\documentclass[12pt]{article}
\pdfoutput=1

\newlength{\abstractwidth}
\usepackage{cancel}
\usepackage{hyperref}

\usepackage{epsf}
\usepackage{color}
\usepackage{graphicx}
\usepackage{tikz}

\usepackage{amsmath}
\usepackage{amssymb}
\usepackage{latexsym}
 \usepackage{setspace}

\numberwithin{equation}{section}

\renewcommand{\thefootnote}{\fnsymbol{footnote}}
\renewcommand{\thanks}[1]{\footnote{#1}}
\newcommand{\starttext}{
\setcounter{footnote}{0}
\renewcommand{\thefootnote}{\arabic{footnote}}}
\newcommand{\bea}{\begin{eqnarray}}
\newcommand{\eea}{\end{eqnarray}}
\newcommand{\be}{\begin{eqnarray}}
\newcommand{\ee}{\end{eqnarray}}

\def\ie{\begin{equation}\begin{aligned}}
\def\fe{\end{aligned}\end{equation}}
\def\half{{\scriptstyle \frac 12}}
\def\quart{{\scriptstyle \frac 14}}

\def\threeh{{\scriptstyle \frac 32}}
\def\fiveh{{\scriptstyle \frac 52}}
\def\eighth{{\scriptstyle \frac 18}}
\def\nineh{{\scriptstyle \frac 92}}

\def\cG{{\cal A}}
\def\cB{{\cal B}}

\def\cD{{\cal D}}
\def\cE{{\cal E}}
\def\cF{{\cal F}}
\def\cG{{\cal G}}

\def\cI{{\cal I}}

\def\cL{{\cal L}}
\def\cM{{\cal M}}
\def\cN{{\cal N}}
\def\cO{{\cal O}}

\def\cT{{\cal T}}

\def\cX{{\cal X}}

\def\ms{\mathfrak{s}}

\def\muu{\mathfrak{u}}

\def\Z{{\mathbb Z}}
\def\M{{\mathbb M}}
\def\ZZ{{\mathbb Z}}
\def\RR{{\mathbb R}}

\def\nn{\nonumber}

\def\tr{{\rm tr}}

\def\p{\partial}

\def\a{\alpha}
\def\b{\beta}

\def\ep{\varepsilon}

\def\wt{{\hat\tau}}

\begin{document}

\starttext

\setcounter{footnote}{0}

\begin{flushright}
%\scriptsize 
{\small QMUL-PH-20-23}
\end{flushright}

\vskip 0.3in

\begin{center}

{\large \bf Maximal $U(1)_Y$-violating $n$-point correlators}
\vskip .1in
 {\large in \bf $\cN=4$ super-Yang-Mills theory}

\vskip 0.2in

{Michael B. Green$^{(a)(b)}$  and Congkao Wen$^{(b)}$} 
   
\vskip 0.15in

{ \small (a) Department of Applied Mathematics and Theoretical Physics }\\
{\small  Wilberforce Road, Cambridge CB3 0WA, UK}

\vskip 0.1in

{\small  (b) School of Physics and Astronomy, Queen Mary University of London, }\\ 
{\small  London, E1 4NS, UK}

\vskip 0.15in

{\tt \small  M.B.Green@damtp.cam.ac.uk, c.wen@qmul.ac.uk}

\vskip 0.5in

\begin{abstract}
\vskip 0.1in

This paper concerns a special class of $n$-point correlation functions of operators in the stress tensor supermultiplet of 
$\mathcal{N}=4$ supersymmetric $SU(N)$  Yang--Mills theory.  These are ``maximal  $U(1)_Y$-violating'' correlators that violate the bonus $U(1)_Y$ charge by a maximum of $2(n-4)$ units.  We will demonstrate that such correlators satisfy $SL(2,\mathbb{Z})$-covariant recursion relations that relate $n$-point correlators to $(n-1)$-point correlators in a manner 
 analogous to the soft dilaton relations that relate the corresponding  amplitudes in flat-space type IIB superstring theory.  
These recursion relations are used to determine terms in the large-$N$ expansion of $n$-point maximal  $U(1)_Y$-violating correlators in the chiral sector,  including correlators with four superconformal stress tensor primaries  and $(n-4)$ chiral Lagrangian operators,  starting from known properties of the  $n=4$ case. We concentrate on the first three orders in $1/N$  beyond the supergravity limit.  The Mellin representations of the correlators are polynomials
in Mellin variables, which correspond to higher derivative contact terms  in the low-energy expansion of type IIB superstring theory in $AdS_5 \times S^5$ at  the same orders as $R^4, d^4R^4$ and $d^6R^4$. The coupling constant dependence of these terms is found to be described by non-holomorphic modular forms with holomorphic and anti-holomorphic weights $(n-4,4-n)$ that are  $SL(2, \mathbb{Z})$-covariant derivatives of Eisenstein series and certain generalisations. 
This determines a number of non-leading contributions to $U(1)_Y$-violating $n$-particle interactions  ($n>4$) in the low-energy expansion of type IIB superstring amplitudes in $AdS_5\times S^5$.

 \end{abstract}

\end{center}

\baselineskip=15pt
\setcounter{footnote}{0}

\newpage

\setcounter{page}{1}
\tableofcontents

\newpage
\section{Introduction and overview}

The holographic relationship between the correlation functions of $\cN=4$ supersymmetric Yang--Mills theory ($\cN=4$ SYM) \cite{Brink:1976bc} with gauge group $SU(N)$ and  type IIB superstring  scattering amplitudes in $AdS_5\times S^5$  is the best-studied example of the gauge/gravity correspondence \cite{Maldacena:1997re, Gubser:1998bc, Witten:1998qj}.    
The large-$N$ expansion of the gauge theory correlation functions corresponds to the low-energy expansion of the superstring amplitudes.  This system possesses maximal supersymmetry, which has 32 supersymmetry components.  In the gauge theory these supersymmetries form part of the superconformal symmetry $PSU(2,2|4)$, which is also the super-isometry group of the superspace containing $AdS_5\times S^5$ in which  the string theory is embedded.
   
Both sides of the correspondence are also invariant under the action of the duality group $SL(2,\ZZ)$.  In $\cN=4$ $SU(N)$ SYM this is Montonen--Olive duality \cite{Montonen:1977sn, Witten:1978mh, Osborn:1979tq}, which relates the theory at one value of the  complex coupling constant $\wt= \frac{\theta_{_{\rm YM}} }{2\pi}+ i \frac{4\pi}{g_{_{\rm YM}}^2}$
  to the theory defined at a value of $\wt$ that is  transformed by an element of $SL(2,\ZZ)$\footnote{More generally,  following \cite{Goddard:1976qe}, S-duality relates a theory with a simply-laced  gauge group $G$ and coupling $\tau$ to a theory with the GNO/Langlands dual  gauge group $G^\vee$, and  the duality group is a sub-group of $SL(2,\ZZ)$.  If $G=SU(N)$ the dual group is $G^\vee=SU(N)/\ZZ_N$. The arguments in this paper are insensitive to the global distinction between $SU(N)$ and $SU(N)/\ZZ_N$ and the duality group is $SL(2,\ZZ)$. }
\be
\label{sl2tau}
\wt\to \frac{a\wt+b}{c\wt+d}\,, \qquad\qquad \qquad a,b,c,d\in \ZZ\,, \qquad  ac-bd=1\,.
\ee 

The holographic image of Montonen--Olive duality is manifested as the invariance  of type IIB superstring theory under S-duality \cite {Hull:1994ys}, in which $\tau$  is a complex scalar field which transforms under $SL(2,\ZZ)$ the same way as $\wt$ in \eqref{sl2tau}. We may expand around a constant  background value by setting $\tau=\tau^0+\delta\tau$, with $\tau^0=\chi^0+ i/g_s$ where $g_s$ is the string coupling constant and $\chi$ is an angular variable with period $1$.

The holographic connection between the gauge theory and type IIB supergravity involves identifying the complex gauge theory coupling constant, $\wt$, with the background value $\tau^0$. 
Furthermore, the string scale is related to $N$ by  ${\alpha'}^2 /L^4= 1/(g_{_{\rm YM}}^2N)$, where $L$ is the length scale of the $AdS_5\times S^5$ background. The fluctuation $\delta\tau$ couples to the gauge theory ``chiral Lagrangian'' operator $\cO_\tau$, which is in the same $\cN=4$  supermultiplet as the stress tensor. Fluctuations of the other massless type IIB supergravity fields couple to the other operators in the same supermultiplet.

\subsubsection*{$U(1)_Y$  violation} 

The scalar field $\tau$ of type IIB supergravity parameterises the coset space $SL(2,\RR)/U(1)$, where $U(1)$ is identified with the R-symmetry that rotates the two fermionic supercharges into each other.  However, in the superstring theory the coset space is subject to discrete identifications, which breaks the duality symmetry from $SL(2,\RR)$ to $SL(2,\ZZ)$.  In particular, the $U(1)$  is not a symmetry in the string theory, since the two supercharges move in opposite directions on the world-sheet.  Therefore, they cannot be continuously transformed into each other but are interchanged by the discrete transformation  that flips the orientation of the world-sheet.  Combining this  with the fact that the fermionic variables may also change sign leads to invariance under $\ZZ_4$ -- which is the residual $U(1)$  symmetry 
embedded in $SL(2,\ZZ)$.  This conclusion can also be deduced from the value of the  coefficient of a chiral $U(1)$ anomaly in ten-dimensional type IIB supergravity \cite{Gaberdiel:1998ui}.  
Whereas  $U(1)$ charge is conserved in tree-level supergravity amplitudes, there is a well-defined pattern of $U(1)$ charge violation in  type IIB superstring amplitudes.\footnote{This discussion refers to scattering amplitudes of massless string states, which are defined in terms of fluctuations of supergravity fields with respect to a fixed background geometry and fixed  $\tau=\tau^0$.  The invariance of the full type IIB superstring effective action under $SL(2,\ZZ)$ includes the transformations of the background, which compensates for the $U(1)$ violation.} 

   It was argued in \cite{Green:1997me} that a type IIB scattering amplitude with $n$ external supergravity states could violate the $U(1)$ charge  by $|q_U| \le  2(n-4)$. Amplitudes that violate $U(1)$ by  $(2n-8)$  units are ``maximal $U(1)$-violating'' (MUV) and  have special features  \cite{Boels:2012zr} that were elucidated in \cite{Green:2019rhz}.  Similarly, ``minimal $U(1)$-violating'' amplitudes ($\overline{\rm MUV}$)  violate $U(1)$ by  $-(2n-8)$.\footnote{All properties concerning maximal $U(1)$ violation have minimal $U(1)$ violation counterparts and so they do not need separate consideration in the following.}

The origin of these statements in the four-dimensional gauge theory was studied in \cite{Intriligator:1998ig} and \cite{Intriligator:1999ff} where it was suggested that there is a corresponding ``bonus'' $U(1)_Y$ symmetry. This was interpreted as an automorphism of $PSU(2,2|4)$, which  extends the symmetry to $U(1)_Y \ltimes PSU(2,2|4)$ where the $U(1)_Y$ factor in the semi-direct product is an R-symmetry that acts on the fermionic generators.  It is again broken to $\ZZ_4$ and  is identified with the centre of the $SU(4)$ R-symmetry group.    In  \cite{Intriligator:1998ig} this was referred to as a ``bonus'' $U(1)_Y$ symmetry and it was conjectured that the pattern of $U(1)_Y$ violation in $\cN=4$ SYM correlators would follow the same pattern as in the corresponding superstring scattering amplitudes.\footnote{We stress that  the ``bonus $U(1)_Y$ symmetry'' is really a $\ZZ_4$ symmetry in both $\cN=4$ SYM and the type IIB superstring.  It  is enhanced to a  $U(1)$ symmetry of the equations of motion in the abelian $\cN=4$ SYM  theory and in the supergravity limit of the type IIB superstring theory.}  Both $SL(2, \ZZ)$ duality and the bonus $U(1)_Y$ have interesting geometric interpretations in terms of the six-dimensional $(2, 0)$  theory when dimensional reduced to four dimensions \cite{Assel:2016wcr}. 

\subsubsection*{Connections between the large-$N$ expansion and the low-energy expansion}

A precise understanding of the connection between the correlation functions of $\cN=4$ SYM  and the scattering amplitudes  in type IIB superstring theory in $AdS_5\times S^5$  has emerged in recent times based largely on the study of the Mellin transform of the  four-dimensional gauge theory position-space correlation functions  \cite{Mack:2009mi}. As emphasised in \cite{Penedones:2010ue} the large-$N$ expansion of these Mellin amplitudes is closely related to the low-energy expansion of the string scattering amplitudes.

Much of this work has considered the four-point correlator and the limit in which $N\to \infty$ with fixed 't~Hooft coupling, $\lambda=g_{_{\rm YM}}^2 N$.   In this limit each order in the $1/N$ expansion corresponds to a particular order in string perturbation theory. A further expansion in powers of $1/\lambda$ corresponds to the low-energy expansion of string perturbation theory in powers of $\alpha's$ (where $s$ represents a Mandelstam invariant) and $\alpha'/L^2$. This limit has been the subject of a considerable amount of research using a mixture of bootstrap and localisation techniques  \cite{Rastelli:2016nze, Aharony:2016dwx, Rastelli:2017ymc, Rastelli:2017udc, Caron-Huot:2018kta, Alday:2017xua, Alday:2018pdi, Aprile:2017bgs, Aprile:2017qoy, Alday:2017vkk, Aprile:2018efk, Binder:2019jwn, Chester:2019pvm, Aprile:2019rep, Drummond:2019odu, Drummond:2019hel, Alday:2019nin, Goncalves:2019znr, Chester:2020dja, Alday:2020dtb, Drummond:2020uni}. 

However, this version of the large-$N$ limit suppresses the contribution of instantons  and obscures $SL(2,\ZZ)$  duality. In order to demonstrate the holographic relationship between instantons in the gauge theory and D-instantons in string theory we need to consider the limit in which $g_{_{\rm YM}}$ is fixed as $N\to \infty$  \cite{Banks:1998nr}.   The $SL(2,\ZZ)$  modular property of the large-$N$ expansion of $\mathcal{N}=4$ SYM four-point correlation function at finite $g_{_{\rm YM}}$ has been recently studied in \cite{Chester:2019jas, Chester:2020vyz} using the powerful tools of supersymmetric localisation \cite{Pestun:2007rz}. 

\vskip 0.3cm
{\it Modular properties of flat-space type IIB amplitudes}
\vskip 0.2cm

The connections between modular functions and the coefficients of low-order terms in the  low-energy expansion of the flat-space graviton scattering amplitudes were obtained by a number of arguments involving the interplay of M-theory dualities and supersymmetry \cite{Green:1997as, Green:1999pu, Green:2005ba}.  These are in some sense to be thought of as BPS coefficients and are expected to be protected by maximal supersymmetry.
Certain modular forms that arise at the same order as $R^4$ were determined in 
\cite{Green:1997me, Green:1999qt}.  Particular emphasis was placed on the situation in which these coefficients were various  Eisenstein series transforming with non-zero modular weights. As shown in  \cite{Green:1998by}, at  the first non-trivial order in the low-energy expansion   supersymmetry together with $SL(2,\ZZ)$ covariance uniquely  constrains the coefficients to satisfy Laplace eigenvalue equations in the hyperbolic plane that parameterises the coset space $SL(2,\RR)/U(1)$ (see also \cite{Sinha:2002zr} for an extension of this result to the first non-leading order).
More recently, an efficient method for determining the Laplace equations satisfied by these coefficients was developed \cite{Wang:2015jna, Wang:2015aua} and has been extended \cite{Green:2019rhz} to determine the modular forms that arise in the low-energy expansion of $U(1)$-violating amplitudes in type IIB superstring theory up to terms with dimension 14.

\subsection {Layout}

The aim of this paper is to extend the study of correlation functions in $\cN=4$ SYM to the large-$N$ expansion of $n$-point correlators of $\half$-BPS operators that violate the bonus $U(1)_Y$ maximally.  This makes contact with the discussion of the corresponding flat-space amplitudes considered in  \cite{Green:2019rhz} and will shed some light on  properties of the corresponding scattering amplitudes in type IIB superstring theory in $AdS_5 \times S^5$.  The method will make use of the differential equations relating correlators of different $U(1)_Y$ charge that were discussed in  \cite{Intriligator:1998ig, Intriligator:1999ff} and developed in \cite{Basu:2004dm, Basu:2004nt}.   We will be able to make precise contact with the large-$N$ expansion of holographic amplitudes in $AdS_5 \times S^5$ by explicit Mellin transform following  \cite{Mack:2009mi, Penedones:2010ue}, as well as modern bootstrap approaches that have been developed for the study of holographic correlators. We will also make use of the harmonic superspace formalism \cite{Galperin:1984bu} that was used to obtain analogous differential equations in \cite{Eden:2011we}.\footnote{The aim of  \cite{Eden:2011we}  was quite different  from ours since it concerned the development of efficient methods for evaluating multi-loop perturbative  contributions to correlators  and did not study the non-perturbative $SL(2,\ZZ)$ properties.}  In this formalism, which is particularly convenient for describing the $\half$-BPS representations of $\cN=4$ SYM, the $SU(4)$ flavour symmetry is described in terms of $SU(2)\times SU(2)'\times U(1)$ subgroup.  This is implemented in terms of a superspace that includes  bosonic coordinates that parameterise the coset space  $S(U(2)\times U(2)') \backslash SU(4)$, and the Grassmann variables are charged under the bonus $U(1)_Y$ that is broken to $\ZZ_4$ (the centre of $SU(4)$ R-symmetry).\footnote{Comments in \cite{Intriligator:1998ig}  questioned whether there  might be problems with the harmonic superspace approach, but these questions were answered in \cite{Eden:1999gh, Howe:1999hz}.} The $\ZZ_4$ imposes selection rules on the correlation functions in $\cN=4$ SYM.

In section~\ref{maxu} we will review the structure of correlators of $\half$-BPS operators in the stress tensor supermultiplet with particular emphasis on the structure of MUV correlators.  The harmonic superspace formalism used in \cite{Eden:2011we} (see also \cite{Eden:2011yp, Eden:2011ku}), is  particularly useful for discussing ``chiral correlators''.  These  are correlators  of chiral operators in the stress tensor multiplet, where a chiral operator has R-symmetry quantum numbers in one $SU(2)$ sub-group of $SU(2)\times SU(2)' \in  SU(4)$ and space-time symmetry in one $SU(2)$ sub-group of the Lorentz group.  Chiral MUV correlators are a subset of all MUV correlators that  possess special features that will simplify our analysis.
 
  In section~\ref{modrel}  we will discuss recursion relations that relate  a $SL(2,\ZZ)$-covariant derivative of a  $(n-1)$-point correlator to a $n$-point correlator in which there is an insertion of the integral of the chiral Lagrangian operator,  $\int d^4x_{n}  \cO_\tau(x_{n})$. 
 These modular covariant recursion relations are analogous to the soft dilaton relations that play an important r\^ole in determining the structure of $U(1)$-violating amplitudes in flat-space  type IIB superstring theory \cite{Green:2019rhz}. 
    The fact that $\cO_\tau$  is an operator in the stress tensor multiplet, which also contains the supersymmetry and R-symmetry currents,  is a special feature of the $\cN=4$ theory. That means that its OPE's can be related to superconformal Ward identities  \cite{Basu:2004nt, Basu:2004dm}.

   In section \ref{softoper}  we will see that these recursion relations are particularly simple for chiral MUV $n$-point correlators. For this particular class of correlators the dependence on the species of the $n$ operators is contained in an overall prefactor which is fixed by the symmetries,  and the non-trivial dependence of the correlator on the space-time coordinates as well as the coupling constant is common to all correlators with a given value of  $n$.

We  will determine those $n$-point correlators that are obtained recursively starting from the four-point correlator, focussing on terms   in the large-$N$ expansion of order $c^\quart$, $c^{-\quart}$ and  $c^{-\half}$, where $c=(N^2-1)/4$ is the conformal anomaly for $SU(N)$ $\cN=4$ SYM.  These are the first three non-vanishing orders beyond the leading order, which is the supergravity limit of order $N^2$. Our analysis  will make use of recent results  \cite{Chester:2019jas, Chester:2020vyz},  based on supersymmetric localisation methods  that have determined the exact $SL(2,\ZZ)$-invariant coefficients  of terms at these orders  in the  $1/N$ expansion of  the Mellin transform of this correlator.\footnote{The integrated correlators, which are averaged over the space-time dependence, have been computed to much higher orders in the $1/N$ expansion \cite{Chester:2019jas, Chester:2020vyz}. }
Inputting this information into the recursion relations  uniquely determines the $n$-point MUV correlator of the chiral sector up to order $N^{-1}$ (although we also need to input a piece of information concerning the flat space superstring six-particle scattering amplitudes obtained in \cite{Green:2019rhz}). 
  
 The space-time dependence of such correlators can be expressed in terms of the specific $AdS_5\times S^5$ Witten diagram ``$D$-functions'' and their coupling constant-dependant  coefficients are specific $SL(2,\ZZ)$-covariant modular forms.
 Making use of the Mellin transforms of the correlation functions  together with the AdS/CFT dictionary, these results give precise expressions for interactions in the low-energy expansion of MUV $n$-point amplitudes of type IIB string theory in  $AdS_5\times S^5$, generalising the flat-space $U(1)$-violating interactions that were studied in  \cite{Green:2019rhz}.  An important general feature of these MUV amplitudes is that they do not possess any poles corresponding to massless intermediate states.  As a consequence their low energy expansions are sums of  higher derivative contact interactions.

In section~\ref{nonmin}, our  focus will be on  demonstrating how the semi-classical evaluation of instanton contributions (which keeps only the  leading order terms in the large-$N$ expansion and in the $g_{_{\rm YM}}\to 0$ limit) reproduces the anticipated form of such correlators that we obtained using recursion relations. Specific examples that will be discussed in section~\ref{instchiral} include $\langle\cO_2(1) \dots \cO_{2} (4)\,   \cO_\tau(5)  \dots \cO_\tau (4+m) \rangle$,  $\langle\cE(1 )\dots \cE(8) \, \cO_\tau(9)\dots \cO_\tau(8+m) \rangle$ and $\langle\Lambda (1)\dots \Lambda(16) \, \cO_\tau(17) \dots \cO_\tau(16+m) \rangle$, where $\cO_2$, $\cE$, $\Lambda$ and $\cO_\tau$ are operators  in the stress tensor multiplet   that are defined in \eqref{Lamnewdef} and have $U(1)_Y$ charges $0$,  $1$, $\threeh$, $2$, respectively.  

We do not have a general  expression for the structure of non-chiral MUV correlators.  However, the semi-classical instanton calculations in such cases are very similar to the chiral cases and their structure will be discussed in section~\ref{nonchiral}. 

 The semi-classical instanton contributions to the large-$N$ limit of the correlators considered in sections~\ref{instchiral} and \ref{nonchiral} are particularly simple since only the 16 superconformal fermionic moduli are of relevance in the instanton profiles of the $\half$-BPS operators.\footnote{ The instanton profile of an operator  is its  value in an instanton background, which depends on the bosonic and fermionic moduli. }
We will end section~\ref{nonmin} with a brief discussion of instanton  contributions to more general classes of correlators.  These examples include  MUV correlation functions with non-zero Kaluza--Klein charges and specific non-MUV correlators.  In these cases the semi-classical instanton contribution requires an understanding of extra fermionic moduli that enter into the ADHM construction of instanton moduli space in $SU(N)$ $\cN=4$ SYM. 

 Finally, in section~\ref{discuss} we  discuss  these results and possible future directions.  A number of technical details concerning $\cN=4$ SYM,  Mellin amplitudes, as well as $SL(2,\ZZ)$ modular forms and the $\alpha'$-expansion of the flat-space superstring amplitudes are given in the appendices.

\section{ $U(1)_Y$-violating correlators of stress tensor multiplet}
\label{maxu}
 
$\cN=4$ SYM  is invariant under sixteen supersymmetries (eight of each chirality) and sixteen conformal supersymmetries.  Its field content may be described in terms of a superfield that is a function of sixteen Grassmann coordinates, $\theta^A_\alpha$ and $\bar\theta^{\dot \alpha}_A$, where $A=1,2,3,4$ labels a $\bf 4$ or $\bar {\bf 4}$ of the R-symmetry group $SU(4)$ and $\alpha,\dot \alpha$ are chiral and anti-chiral spinor labels.

Correlators of  $\half$-BPS states can efficiently be expressed  by use of  the  $\cN=4$ harmonic superspace formalism \cite{Galperin:1984bu, Hartwell:1994rp, Andrianopoli:1999vr}  in terms of a superfield that depends on a total of eight anti-commuting coordinates, which is half the number of odd coordinates in  $\cN=4$ 
  super Minkowski space.   This can be made manifest by decomposing the Grassmann coordinates according to $SU(2)\times SU(2)' \times U(1)$ so that\footnote{For the most part our conventions follow those of \cite{Eden:2011we}. } 
\bea
\theta^A_\alpha = (\rho^a_\alpha,\theta_\alpha^{a'})\,,\qquad \bar\theta_A^{\dot \alpha} = (\bar\rho_{a'}^{\dot \alpha},\bar \theta^{\dot \alpha}_a)\,, 
\eea
where the indices  $\alpha$, $\dot \alpha$ label  chiral and anti-chiral two-component space-time spinors and $a,a'=1,2$ label the ${\bf (2,1)}_1$ and  ${\bf (1,2)}_{-1}$ representations.
The Grassmann variables $\rho^a_\alpha$ and $\bar \rho_{a'}^{\dot \alpha}$ are defined by
  \bea
\label{rhodef}
\rho^a_\alpha=\theta^a_\alpha+\theta^{a'}_\alpha\, y_{a'}^a\,,\qquad\bar \rho_{a'}^{\dot \alpha}=        \bar\theta_{a'}^{\dot \alpha}+\bar \theta_a^{\dot \alpha}\, \bar y_{a'}^a\, ,
\eea
where the  (complex) bosonic coordinate $y$ parameterises the eight-dimensional coset space $S(U(2)\times U(2)') \backslash SU(4)$.  From these definitions we see that  $ y_{a'}^a$ and   $\bar y_{a'}^a$  are bifundamentals of $SU(2)\times SU(2)'$ and have $U(1)$ charges equal to $+2$ and $-2$, respectively.

 The on-shell $\cN=4$ Yang--Mills Field strength multiplet of \eqref{YMcomp} may be identified with the components of a scalar superfield, $W(x,\rho,\bar\rho,y)$  that satisfies certain constraints and which takes values in the Lie algebra $\ms\muu(N)$.
The $\half$-BPS gauge-invariant superconformal primary  $\cO_p$ is the lowest component of the superfield  $\cT^p= \tr(W^p)$.  We will concentrate on the  stress tensor supermultiplet, which is the $p=2$ case $\cT=\tr(W^2)$ (dropping the superscript when  $p=2$), for which the primary $\cO_2$ is the $\bf 20'$  of $SU(4)$.  This may be defined in terms of the component fields by
 \bea
 \label{O2}
 \mathcal{O}_2(x, y) = Y_I Y_J \cO^{IJ}_{\bf 20'}(x) =\frac{1}{g^2_{_{\rm YM}}}   Y_I Y_J\,  {\rm tr}(\varphi^I \varphi^J)(x)  %-\frac{1}{6} \delta^{IJ} \,{\rm tr}(\varphi^K\varphi^K)(x) 
  \,,
\eea
where $I,J=1,\dots, 6$ and $Y_J$ is a fixed null vector satisfying $Y\cdot Y=0$. This can be expressed in terms of $y_{a'}^a$ introduced in \eqref{rhodef}, 
\ie
Y_I = {1 \over \sqrt{2} } (\Sigma_I)^{AB} \epsilon_{ab} g^a_A g^b_B \, ,
\fe
where 
\bea
g_A^b = (\delta^b_a, y^b_{a'}) \,,
\label{gdef}
\eea
and where $(\Sigma_I)^{AB}$ are Clebsch--Gordan coefficients that couple a ${\bf 6}$ to two ${\bf 4}$'s  of $SO(6)$.  These satisfy $\sum_{I=1}^6 (\Sigma_I)^{AB} (\Sigma_I)^{CD} = {1\over 2} \epsilon^{ABCD}$, which implies 
$(Y_i)_I  \, (Y_j)_I= \half \epsilon^{a'b'} \epsilon_{ab} \, (y_{ij})^a_{a'} (y_{ij})^b_{b'} \equiv  (y_i - y_j)^2 $.  It follows from this that $Y_I$ has $U(1)$ charge equal to $+2$ and  $\cO_2$ has $U(1)$ charge equal to $+4$.

The operator $\cO_2$ is  annihilated by half the supersymmetries, which we will take to be the eight 
supersymmetry  components, $Q_\alpha^{a'}$ and $\bar Q^{\dot \alpha}_a$. The super-descendant  states in the short $\half$-BPS stress tensor supermultiplet are then generated by the action of the remaining eight supersymmetry components on $\cO_2$.  The structure of the stress tensor supermultiplet  is expressed in an efficient manner by 
 \bea
 \label{O21}
\mathcal{T}(x, \rho, \bar{\rho}, y)=  \exp\left(\rho^a_{\alpha} Q_a^{\alpha} + \bar{\rho}_{a'}^{\dot \alpha} \bar{Q}_{\dot \alpha}^{a'}   \right)  \cO_2(x, y)  \, .
 \eea
The expansion of  $\cT$ in powers of $\rho$ and $\bar \rho$ generates the super-descendants in the $\cO_2$ multiplet.
A special feature of the $p=2$ multiplet, which has components listed in \eqref{LitM},
 is that  terms of order $\rho^r \, \bar\rho^s$ with $r+s\geq 5$ vanish,\footnote{This follows upon using the equations of motion.}

Since $Q_a^{\alpha}$ and  $\bar{Q}_{\dot \alpha}^{a'}$ have $U(1)$ charges $-1$ and $+1$, respectively, $\mathcal{T}(x, \rho, \bar{\rho}, y)$ has a charge $+4$.  As we will see, correlators of $\mathcal{T}(x, \rho, \bar{\rho}, y)$ are polynomials in $\rho_i,  \bar\rho_i, y_i$,that are strongly constrained by requiring $U(1)$ 
 invariance at each operator position. 
 The bonus $U(1)_Y$ R-symmetry only acts on the fermionic coordinates, $\rho$ and $\bar \rho$ (and not on $y$ or $x$).    Since it is the holographic dual of the $U(1)\to \ZZ_4$  in the type IIB superstring, we will assign  $U(1)_Y$ charges $-\half$ to $\rho$ and $+\half$ to $\bar  \rho$.\footnote{In the convention used in \cite{Intriligator:1998ig}  these $U(1)_Y$ charges are $\pm 1$.  }  This leads to charge assignments for the operators in the $\cO_2$ multiplet  (the stress tensor multiplet) that are equal in magnitude and opposite in sign to those of the supergravity fields that act as sources for the operators according to the holographic relationship between the $\cN=4$ SYM  and type IIB superstring theory.
 For example, the operator $\cO_\tau$  has $U(1)_Y$ charge 2 while the conjugate supergravity field, which is the fluctuation of the complex type IIB dilaton (the field called $Z$ in \cite{Green:2019rhz}) has charge $-2$. 

The $U(1)_Y$ charges of all  the operators in the stress tensor supermultiplet are summarised in \eqref{LitM}. These are correlated with the powers of $\rho$ and $\bar \rho$.   In the following a general  $U(1)_Y$ charge of a correlator will be denoted $q_U$  and the $U(1)_Y$ charge of a field $\Psi$ will be denoted $q_\Psi$.  The dimension of such a field will be denoted $\Delta_\Psi$.

\subsection{Correlation functions of the stress tensor multiplet}
 
We are interested in properties of the correlation function of $n$ operators in the stress tensor multiplet.  We may associate the coordinates $(x_i,\rho_i,\bar\rho_i,y_i)$ with each operator $\cT$ in the correlator so that all possible  $n$-particle correlators are generated as coefficients in the expansion in powers of the Grassmann coordinates, 
\bea
 G_n = \langle  \mathcal{T}(1)  \mathcal{T}(2) \cdots  \mathcal{T}(n) \rangle
 =  \sum_{\underset{  |k-\ell |\,    \leq\, n-4 } { \{ k_r,\ell_r\}=0}}^4  \widehat G_{n; k,\ell}(j_1, j_2, \cdots, j_n) \,  \rho_1^{k_1} \bar\rho_1^{\ell_1}\dots \rho_n^{k_n} \bar\rho_n^{\ell_n}  \, ,
\label{stresscorr}
\eea
where the correlator $ \widehat G_{n; k,\ell}(j_1, j_2, \cdots, j_n)$ is an expectation value of the form 
 \bea
 \widehat G_{n; k,\ell}(j_1, j_2, \cdots, j_n)= \langle \cO_{j_1}(x_1,y_1) \dots \cO_{j_r}(x_r,y_r) \dots  \cO_{j_n}(x_n,y_n)  \rangle\,,
 \label{correx}
 \eea
and $\langle \dots \rangle$ is defined by the functional integral 
\bea
\langle \prod_{r=1}^n {\mathcal{O}}_{j_r} (x_r,y_r) \rangle = \int [d \Phi] \, e^{i \int d^4x \cL[\Phi]} \prod_{r=1}^n {\mathcal{O}}_{j_r} (x_r,y_r) \,.
\label{corredef}
\eea
The dependence of $\hat G$ on $x_i$ and $y_i$ has been suppressed and the
index $j_i$ on  an operator in the stress tensor supermultiplet   labels its $U(1)_Y$ charge and its dimension, i.e. $j_i=(q_i ,\Delta_i)$.\footnote{The operators in the stress tensor supermultiplet are uniquely specified by $j=(q,\Delta)$, apart from the two-fold degeneracy of the  operators with $q=1\, , \Delta=3$ and  $q=-1\, , \Delta=3$ as shown in  \eqref{LitM}.}
The sums in \eqref{stresscorr}  are subject to the following restrictions. Each component operator must lie in the $p=2$ superconformal multiplet shown in \eqref{LitM}, which requires
\ie
k_r+\ell_r\le 4\,, \qquad \quad (k_r,\ell_r) \ne (1,3) \ {\rm or} \ (3,1)\,, \qquad\quad  {\rm where}\    1\leq r \leq n
\label{klcon}
\fe
and we have defined $k$ and $\ell$ by
\ie
\sum_{r=1}^n k_r =4k\,, \qquad\quad  \sum_{r=1}^n \ell_r =4\ell\,.
\label{kldef}
\fe
Furthermore, as explained in \cite{Eden:2011we}, supersymmetry and superconformal symmetry imply that one can gauge away $16$ $\rho$'s (and $16$ $\bar{\rho}$'s), from which it follows that 
\bea
\bigg|\sum_{r=1}^n k_r - \sum_{r=1}^n  \ell_r\bigg| =| 4k-4\ell |  \le 4n-16\,.
\label{sumkl}
\eea
The quantity   $\widehat G_{n; k,\ell}(j_1, j_2, \cdots, j_n)$ in \eqref{stresscorr} is a polynomial in $y_i$ of the form $\sum  y_1^{p_1} y_2^{p_2} \cdots y_n^{p_n}$.
The fact that $\mathcal{T}(r)$ has $U(1)$ charge $+4$ implies a restriction $2p_r+k_r-\ell_r =4 \, .$

As discussed in \cite{Intriligator:1998ig}, the correlation functions that are dual to type IIB supergravity contributions are invariant under $U(1)_Y$, which is a symmetry of type IIB supergravity.   In particular, the leading terms in the large-$N$ expansion,   which are of order $N^2$, correspond to tree-level supergravity contributions that preserve $U(1)_Y$.   However, in general,  the correlation functions are not  $U(1)_Y$ invariant. Nevertheless the unbroken discrete $\ZZ_4$ subgroup imposes constraints on the structure of the polynomials in $\rho, \bar\rho$.  This restricts the polynomials to terms of the form  $\rho^{4u}\, \bar \rho^{4v}\, (\rho\bar\rho)^w$ ($u,v,w\in \ZZ$), which  are invariant under  the $\ZZ_4$, transformations $\rho \to \omega \rho$, $\bar\rho \to \omega^{-1} \bar\rho$, where $\omega^4=1$ \cite{Eden:1999gh}.   The total $U(1)_Y$ charge of the correlator $ \widehat G_{n; k,\ell}$ is given by $\sum_{r=1}^n q_{r}=2k-2\ell$.
 The maximum $U(1)_Y$ violation arises when $k-\ell=n-4$ and the $U(1)_Y$ charge is violated by $q^{max}_{U} = 2n-8$ units.  For chiral correlators this condition becomes  $\ell_r=0\,\ \forall r$ and $\sum_r k_r=4n-16$,  but there are many examples of non-chiral MUV correlators.  Similarly, the minimum $U(1)_Y$ violation is $q^{min}_{U}=-2(n-8)$,  which arises when $\ell-k=n-4$.

Although we would ultimately like to describe any of these correlation functions, we are here focussing on the components of \eqref{correx} that violate $U(1)_Y$ maximally, namely the MUV correlators. 

\subsection{Correlation functions of chiral operators}
\label{chircorr}

For technical reasons (soon to become apparent) many of our considerations  will be restricted to the ``chiral'' sector, which is obtained by setting $\bar \rho= 0$  (equivalently, setting $\ell_i=0$).  This gives the expression
 \bea
 \label{O22}
\cT^C(x,\rho,y)  \equiv \cT(x, \rho, 0, y)=  \cO_2(x, y) + \cdots + \rho^4 \cO_\tau(x)\,,
\eea
which contains the operators connected by the red arrows in \eqref{LitM} (the expressions for these operators in terms of the elementary fields are given in appendix~\ref{stressops}).

A general $n$-point  chiral correlator can be expanded in terms of  Grassmann variables $\rho$ in the form,  
 \bea
 \label{corn}
 G_n = \langle  \mathcal{T}^C(1)  \mathcal{T}^C(2) \cdots  \mathcal{T}^C(n) \rangle
 =  \sum_{k=0}^{n-4}  \sum_{\underset{\sum_{k_l=4k}}{k_i\ge 0}} \widehat{G}_{n; k}(j_1, j_2, \cdots, j_n) \,  \rho_1^{k_1}\dots \rho_n^{k_n}   \,, 
 \eea
where $\cT^C(i)\equiv \cT^C(x_i,\rho_i,y_i)$ and  the values of $j_i$ are determined in  terms of the values of $k_i$ (and we have set $\widehat G_{n;k} \equiv \widehat G_{n;k,0}$).  The expression $\widehat G_{n; k}$ has total degree $4k$ in the Grassmann variables  $\rho_i$, and as a consequence it describes a correlator that violates the $U(1)_Y$ charge by $2k$.  
Here we will be interested in the special cases of  MUV  corrrelators, which have $k=n-4$ and violate $U(1)_Y$ by the maximum value of $q_U=2(n-4)$.\footnote{These were called ``maximal nilpotent"  correlators in \cite{Eden:2011we}. The focus of  \cite{Eden:2011we} is on the perturbation theory and the issue of $U(1)_Y$ violation was not considered in that reference.}    The only $U(1)_Y$-conserving example of such correlators has $n=4$ and is given by\footnote{Two and three-point correlators  also preserve $U(1)_Y$,  and are known to be independent of the coupling constant, $\wt$, so we will only consider the correlators with $n\geq 4$.}
\ie
\widehat G_{4; 0}(j_1,j_2,j_3,j_4)=\langle \cO_2(x_1,y_1)\, \cO_2(x_2,y_2)\, \cO_2(x_3,y_3)\, \cO_2 (x_4,y_4)  \rangle\,,
\label{fourex}
\fe
which is the component in \eqref{corn} with $n=4$ and $k_i=0\ \forall i$.  
      
The chiral MUV  correlators have special properties.  In particular, the dependence on $y_i$ and $\rho_r$  can be factored out into a prefactor $\cI_n(\{x_i,\rho_i, 0,y_i\}) $  \cite{Eden:2011we} so that we can write a chiral MUV $n$-point correlator  in the form
\bea
 \label{corn0}
 \sum_{\underset{\sum_{k_i=4n-16}}{k_i\ge 0}} \widehat{G}_{n; k}(j_1, j_2, \cdots, j_n) \,  \rho_1^{k_1}\dots \rho_n^{k_n} = \cI_n(\{x_i,\rho_i, 0,y_i\})  \, \cG_n (x_1, \cdots , x_{n}; \wt) \, .
\eea
Importantly, the ``reduced correlation function'' $\cG_n$ depends only on the $x_i$ and not on the particular species of operators in the MUV correlator so it has the same form for any  chiral MUV correlator with a given value of $n$.
The prefactor  $\cI_n(\{x_i,\rho_i, 0,y_i\})$, which is fixed by the superconformal symmetry, was explicitly constructed in  \cite{Eden:2011we}, and takes the following form, 
\bea
 \label{eq:In}
\cI_n(\{x_i,\rho_i, 0,y_i\})= \int d^4 \epsilon\, d^4 \epsilon' \, d^4  \bar{\xi}\,  d^4  \bar{\xi}' \prod_{i=1}^n 
 \delta^4 \left( \rho_{i, \alpha}^a - \left(\epsilon^a_{\alpha} + \epsilon^{a'}_{\alpha} y_{i, a'}^a  \right) 
 - x_{i, \alpha}^{\dot \alpha} \left(\bar{\xi}^a_{\dot \alpha} +\bar{\xi}^{a'}_{\dot \alpha} y_{i, a'}^a  \right)  \right) \, . \nn\\ 
\eea
 This is a homogeneous polynomial in $\{\rho_i\}$ of degree $4(n-4)$  (using the fact that superconformal symmetry allows 16 of the $\rho^a_{i,\alpha}$ to be set equal to zero). It has $U(1)$ charge $4$ at each point, which implies that $\cG_n$ has no dependence on $y_i$, as we emphasised earlier.  Furthermore, $\cI_n$ is $S_n$ symmetric and has conformal dimension $-2$ at each position $x_r$,   which implies that the dynamical factor $\cG_n(x_1, \cdots , x_{n}; \wt)$ is also $S_n$ symmetric, and has conformal dimension $+4$ at  each position to match the conformal weight of  $\cT^C(j)$, which is $+2$.\footnote{ 
When translated into Mellin space the prefactor  $\cI_n(\{ x_r, \rho_r, 0, y_r\})$ behaves as $(\gamma_{ij})^4$, where  $\gamma_{ij}$ is the Mellin variable. The prefactor plays the r\^ole that the supersymmetry factor $\delta^{16}(Q)$ played in the structure of MUV  type IIB superstring amplitudes   \cite{Boels:2012zr, Green:2019rhz}. The structure of type IIB superstring amplitudes is reviewed in Appendix \ref{holosol}.},  The reduced correlation function  $\cG_n(x_1, \cdots , x_{n}; \wt)$ plays a prominent r\^ole in the remainder of this paper.\footnote{As we will  see later, the  large-$N$ expansion of the Mellin transform of the MUV correlator $\widehat G_{n;n-4}$ is a sum of contact terms. It will follow that \eqref{corn0} implies that  the large-$N$ expansion of the Mellin transform of $\cG_n$  is also a sum of contact terms.}

A particular example is the chiral MUV correlator of four $\cO_2$  operators and $(n-4)$ $\cO_\tau$ operators
 \bea
 \label{relevant}
\langle   \cO_2(x_1, y_1) \cdots  \cO_2(x_4, y_4)\, \cO_\tau(x_5)\dots \cO_\tau(x_n)\rangle \, ,
\eea
for which the prefactor has the form 
 \bea
 \label{I0}
 \cI_n(\{x_i,\rho_i, 0,y_i\})   \big{|}_{\rho_1= \cdots =\rho_4=0}= \left(\prod_{1 \leq i<j \leq 4} x_{ij}^2\right) \times R(1,2,3,4) \times (\rho_5)^4 \cdots (\rho_n)^4\, ,
\eea
where $R(1,2,3,4)$ is the usual R-symmetry invariant of the four-point correlator \cite{Eden:2000bk, Nirschl:2004pa}, and is given by 
   \begin{align}
  \label{Rdef}
 R(1,2,3,4) &= \frac{y^2_{12}y^2_{23}y^2_{34}y^2_{14}}{x^2_{12}x^2_{23}x^2_{34} x^2_{14}}(x_{13}^2 x_{24}^2-x^2_{12} x^2_{34}-x^2_{14} x^2_{23})
+\frac{ y^2_{12}y^2_{13}y^2_{24}y^2_{34}}{x^2_{12}x^2_{13}x^2_{24} x^2_{34}}(x^2_{14} x^2_{23}-x^2_{12} x^2_{34}-x_{13}^2 x_{24}^2) \nn \\
& +\frac{y^2_{13}y^2_{14}y^2_{23}y^2_{24}}{x^2_{13}x^2_{14}x^2_{23} x^2_{24}}(x^2_{12} x^2_{34}-x^2_{14} x^2_{23}-x_{13}^2 x_{24}^2) +
\frac{y^4_{12} y^4_{34}}{x^2_{12}x^2_{34}}   + \frac{y^4_{13} y^4_{24}}{x^2_{13}
x^2_{24}} + \frac{y^4_{14} y^4_{23}}{x^2_{14}x^2_{23}} \,.
\end{align} 

There are many other chiral MUV correlators that can be extracted from \eqref{corn0} by considering various powers of $\rho_i$.    These are $n>4$ correlation functions with a power $\prod \rho_i^{k_i}$  with $\sum_{i=1}^n k_i = 4n-16$, so they violate the $U(1)_Y$ charge by $2(n-4)$ units.   We will call  these ``super-descendant chiral MUV correlators'' since they  involve products of  less than four $\cO_2$ with  a number of superconformal descendant (but conformal primary) operators.    Some examples will be described in section~\ref{nonmin}, one of which is $\langle \Lambda(x_1)\dots \Lambda(x_{16})\rangle$, which has $n=16$, $q_{_{\Lambda^{16}}}=24$.   Since the operator $\Lambda$ (defined in \eqref{Lamnewdef}) is the component of $\cT$ that is cubic in $\rho$ this correlator involves the terms proportional to $ \rho_1^3\dots \rho_{16}^3$ in the expression for $\cI_n$ in  \eqref{eq:In}.  
 
These properties were used to construct  $\mathcal{G}(x_1, \cdots , x_{n}; \wt)$ at Born level in \cite{Eden:2011we}. The aim in that reference was to use the structure of these particular correlators in order to study $\cN=4$  SYM   perturbation theory by relating the correlator of four $\cO_2$'s at $\ell$ loops to the correlator with $\ell$ insertions of the chiral Lagrangian at the Born level.   Here our aim is different -- we are interested in the large-$N$ expansion of $\mathcal{G}(x_1, \cdots , x_{n}; \wt)$, and its non-perturbative $SL(2, \ZZ)$ modular properties.  For $n > 4$ we anticipate that  the large-$N$ expansion has the form \footnote{The $n = 4$ case, which will be discussed in the next section, also has terms that are holographically related to the supergravity four-graviton scattering amplitude, which starts at order $O(c)$. MUV correllators with $n > 4$ do not arise in supergravity since it  conserves $U(1)_Y$. Furthermore, as we will see, the 
$\alpha=1$ term (of order $c^0$)  vanishes.} 
\ie
 \cG_n(x_1, \cdots , x_n;\wt) & =     c^\quart \, \cG_n^{(0)} (x_1, \cdots , x_{n};\wt )   + c^{-\quart} \cG_n^{(2)} (x_1, \cdots , x_{n};\wt)\cr
 &~~~  + c^{-\half  \, }\cG_n^{(3)} ( x_1, \cdots , x_{n};\wt) 
  +O(c^{-\threeh})
\cr
 & =    \sum_{\alpha=0,2,3}  c^{\frac{1-\alpha}{4}}\, \cG_n^{(\alpha)}(x_1,\cdots, x_n;\wt)  +O(c^{-\threeh})\, ,
\label{NN}
 \fe
 where $c=(N^2-1)/4$.  The terms with $\alpha=0,2,3$ are the ones that are expected to be determined by supersymmetry -- they are holographically dual to  terms in the low-energy expansion of the MUV amplitudes of the type IIB superstring theory in $AdS_5 \times S^5$ of the same dimensions as $R^4$, $d^4 R^4$ and $d^6R^4$. The flat-space limits of these superstring amplitudes were studied in \cite{Green:2019rhz}. 
 
 \section{Modular differential relations}
 \label{modrel}

 We now want to study the detailed form of the differential relations between MUV  correlation functions.  This will make use of techniques suggested in  \cite{Intriligator:1998ig} and \cite{Basu:2004dm}.  We will again  restrict most of our considerations to correlation functions of chiral operators, for which we have the most detailed understanding, although the results apply to the wider class of non-chiral MUV correlators.

\subsection{$SL(2,\Z)$ covariance of $\half$-BPS operators and correlation functions}

Before discussing relations between correlation functions with different $U(1)_Y$ charge  violation we will summarise some notational conventions.  Under a $SL(2,\Z)$ transformation 
\bea
\tau \to  \frac{a\tau+b}{c\tau+d}\,, \qquad \qquad \left|\begin{matrix} a&b \cr c &d \cr    \end{matrix}\right|=1,\qquad a,b,c,d\in \Z\,,
\label{sl2def}
\eea
a modular form  with holomorphic and anti-holomorphic modular weights $(w,\hat w)$ transforms as 
\bea
f^{(w,\hat w)} (\tau) \to  (c\tau+d)^w\, (c\bar \tau + d)^{\hat w}\,  f^{(w,\hat w)}(\tau) \, .
\label{modform}
\eea
When $\hat w=-w$ the transformation is a $U(1)_Y$ transformation by an angle $\phi$, defined by 
\bea
e^{2i w\phi} = \left(\frac{c\tau+d }{c\bar \tau + d}\right)^w  \,.
\label{angdef}
\eea

The modular covariant derivatives  
\be
\label{covderiv}
 \cD_w = i\left(\tau_2 \frac{\partial}{\partial \tau} -\frac{iw}{2}\right),
\qquad  \bar{\cD}_{\hat{w}} = -i\left(\tau_2
\frac{\partial}{\partial \bar\tau} +\frac{i\hat{w}}{2}\right)\, ,
 \ee
act on a $(w,-w)$  modular form as follows 
\bea
\label{dacts}
\cD_w f^{(w,-w)} (\tau) = f^{(w+1,-w-1)}(\tau)\,,\qquad\qquad \bar \cD_{-w} f^{(w,-w)} (\tau) = f^{(w-1,-w+1)}(\tau)\,.
\eea
We may consider two distinct Laplace operators acting on a $(w,-w)$ modular forms, which are defined by
\bea
\Delta_{(-)w}= 4\, \cD_{w-1}\, \bar \cD_{-w}\,,\qquad \Delta_{(+)w}=4\, \bar \cD_{-w-1} \,\cD_w\,,
\label{lapdef}
\eea
so that when acting on a $(0,0)$ form we have $\Delta_{(\pm) 0} = 4 \tau_2^2 \partial_\tau\partial_{\bar\tau}$, which is the standard laplacian on  functions.  Noting that $\tau_2=(\tau-\bar\tau)/2i$ we see that 
\bea
\Delta_{(-)w} - \Delta_{(+)-w}= 2 w\,.
\label{comlap}
\eea

We will consider correlation functions of operators in the stress tensor supermultiplet,
 $\cO_{j_r}$, with conformal dimensions
$\Delta_{j_r}$ and  $U(1)_Y$ charges $q_{j_r}$, and the total modular weight $k = \sum_r w_{j_r}=n-4$.  The total $U(1)_Y$ charge is twice the holomorphic weight, or $q_U=2k$. 
%In \cite{Basu:2004dm}  \cite{Basu:2004nt}

\subsection{$SL(2,\ZZ)$-covariant differential relations between correlators}
 \label{susysl}

 We  begin by recalling the renormalisation of $\half$-BPS operators, such as $\cO_2$ defined in \eqref{O2}. This is the $p=2$ example of the more general superconformal primaries, $\cO_p$, which are defined by  
 \bea
 \label{normop}
 \cO_p    = N \left(\frac{\wt_2}{4\pi N} \right)^{\frac{p}{2}} [\tr \, \varphi^p]_{[0,p,0]}  \,.
 \eea
 With this normalisation a connected vacuum diagram  that can be drawn on a surface  of genus $g$  behaves as $N^\chi$, where the Euler character is $\chi=2-2g$.   Furthermore, in the case of $p=2$, which we will focus on,  the two-point function is independent of $\wt$.  It follows from supersymmetry that every super-descendant  operator in a supermultiplet has a linear dependence on $\wt_2$.
  any operator in the multiplet satisfies
 \bea
 \label{normdef}
 \wt_2\,\frac{\p}{\p \wt}\cO_j  = -{i\over 2}\cO_j \, , \qquad\qquad   \wt_2\,\frac{\p}{\p\bar\wt}\cO_j = {i\over 2}\cO_j\,,
  \eea
 since $\wt_2=(\wt-\bar\wt)/2i$.

 Properties of the operator product expansions (OPE's) of $\cO_\tau$ and $\bar \cO_{\bar \tau}$ with other $\half$-BPS operators  were considered in \cite{Basu:2004nt,Basu:2004dm} (see also \cite{Eden:1999gh, Baggio:2012rr}).  These were obtained by performing a number of supersymmetry transformations on the OPE's of the energy-momentum tensor, which  are highly constrained by Ward identities.
In \cite{Basu:2004nt} it was argued  that these OPE's have the schematic form
\bea
\cO_\tau(z)\, \cO_j(x_j) = a_j \cO_j(x_j)\, \delta^4(z-x_j)+a_{j'}'\frac{1}{(z-x)^4} \cO_{j'}(x_j) + \dots
\label{opes}
\eea
where $j=(q_j, \Delta_j)$ and $j'=(q_j+2,\Delta_j)$ and the ellipsis corresponds to less singular terms involving conformal descendants, long operators and double trace operators.  The  first term on the right-hand side is a contact term that vanishes when the points $z$ and $x_j$ are separated, but it affects the integrated  correlation function in a crucial manner.  The second term  conserves the $U(1)_Y$ symmetry and can be determined from the three point correlator of 1/2-BPS operators, $\langle \cO_\tau(z) \cO_j(x) \cO_{j'} (y)\rangle$. In much of  the following we will take  $O_j$ to be one of the operators in the chiral sector in \eqref{O22}, which have  $q_j=\Delta_j-2$.  In this case the operator $\cO_{j'}$ has $j' =(\Delta_j,\Delta_j)$ and  does not correspond to any operator in the stress tensor multiplet and so  it is not a short operator and the coefficient $a_{j'}' = 0$.  
 
A similar equation to \eqref{opes}  applies  when  $\cO_\tau$ is replaced by  $\bar\cO_{\bar \tau}$.  The coefficients in that case will be denoted by $\bar a_j$ and $\bar a'_{j'}$.  

The coefficients $a_j$ and  $\bar a_j$   of the contact terms can be determined from the requirement that correlation functions transform covariantly under $SL(2,\Z)$ \cite{Basu:2004dm}.  This is seen by considering the  action of $\partial/\partial \bar\wt$ and $\partial/\partial \wt$ on the correlation functions 
$\widehat G_{n; n-4} (j_1\dots j_n)=\langle \cO_{j_1}(x_1) \dots \cO_{j_n}(x_n)\rangle$.  This is a correlation function that violates the $U(1)_Y$ charge by $2(n-4)$ units and therefore transforms as a $(w,-w)$  modular form, with $w=n-4$.
 The differentials act on the explicit factor of $\tau_2$  in the definition of each operator as in  \eqref{normdef} and in addition they act on $e^{i\int d^4 z\cL(z)} = e^{\frac{1}{2\wt_2}\int d^4 z (\wt \cO_\tau(z)-\bar\wt \bar\cO_{\bar\tau}(z))}$ inside the expectation value. The net effect is to give expressions of the form  (after transforming to euclidean space)
\bea
\label{relations1}
i  \wt_2\,  \frac{\partial}{\partial \wt}  \, \langle \prod_{r=1}^n   \cO_{j_r}(x_r) \rangle  = \frac{n}{2}  \langle \prod_{r=1}^n   \cO_{j_r}(x_r) \rangle  + \frac{1}{2}\int d^4z
\langle \cO_\wt(z) \prod_{r=1}^n\cO_{j_r}(x_r) \rangle  \,,
\eea
\bea
\label{relations2}
- i \wt_2\,  \frac{\partial}{\partial \bar\wt} \, \langle \prod_{r=1}^n   \cO_{j_r}(x_r) \rangle  =  \frac{n}{2} \, \langle \prod_{r=1}^n   \cO_{j_r}(x_r) \rangle + \frac{1}{2}\int d^4z \langle \bar\cO_{\bar \tau} (z) \prod_{r=1}^n\cO_{j_r}(x_r) \rangle  \,,
\eea
where the inhomogeneous term proportional to $n/2$ arises from differentiation of the $n$ operators in the correlation function. The second term on the right-hand side of these equations comes from differentiating the  factor of   $e^{-\int d^4 z\cL^E(z)}$, where $\int d^4z \cL^E(z)$ is the euclidean action.

We now separate the contribution of the contact terms from the integrals in the second terms on the  right-hand sides of \eqref{relations1} and \eqref{relations2}.  This gives terms with coefficients $a_j$ and $\bar a_j$.  
The value of $a_j$ is determined by requiring that the  appropriate covariant derivatives act  on the weight-$(w,-w)$ correlation function.  We have
\begin{align}
\label{aeter}
&\left(i \wt_2 \frac{\partial}{\partial \wt} +\frac{w}{2}\right)\,   \langle \prod_{r=1}^n   \cO_{j_r}(x_r) \rangle   =  \frac{n+w}{2} \, \langle \prod_{r=1}^n   \cO_{j_r}(x_r) \rangle + \frac{1}{2} \int d^4z  \langle \cO_\wt(z) \,  \prod_{r=1}^n   \cO_{j_r}(x_r) \rangle 
 \nonumber\\
&=\left(  \frac{n+w +  \sum_r a_r\, }{2}\right)   \, \langle \prod_{r=1}^n   \cO_{j_r}(x_r) \rangle   + \frac{1}{2} \int_\epsilon d^4z  \langle \cO_\wt(z) \,  \prod_{r=1}^n   \cO_{j_r}(x_r) \rangle     \,.
\end{align}
The symbol $\int_\epsilon$ in the second line indicates that the integration region excludes a small ball around each $x_r$, where $\epsilon \to 0$.   This allows us to  isolate the contact terms that gives the terms proportional to  proportional to $a_r$.
From now on we may suppress the $\epsilon$ with the understanding that the integration is performed by first evaluating the integrand  at separated points. Therefore, the net effect of the final result is that we will consider correlators only at distinguished points.

The left-hand side of \eqref{aeter} is equal to $\cD_w    \langle \prod_{r=1}^n   \cO_{j_r}(x_r) \rangle $. 
 In order for this equation to transform covariantly under $SL(2,\Z)$ it therefore follows that the coefficients of the contact terms must be given by
 \bea
 a_r= -1 - \frac{q_r}{2} \, ,
 \label{apone}
 \eea
 which implies
\be\label{Ares}
 \sum_{r=1}^n a_r= -n-w\,.
\ee
We then see that \eqref{aeter} becomes
\bea \label{aetertwo}
\cD_w \,   \langle \prod_{r=1}^n   \cO_{j_r}(x_r) \rangle   =  \frac{1}{2} \int%{_\epsilon}
 d^4z  \langle \cO_\tau(z) \,  \prod_{r=1}^n   \cO_{j_r}(x_r) \rangle     \,.
\eea

In summary, we see that a simple derivative on a correlator with respect to $\tau$ leads to the  correlator with the insertion of  the integral of the marginal operator $\cO_{\tau}$.  The presence of contact terms requires  regularisation consistent with $SL(2,\ZZ)$ covariance, which turns simple $\tau$ derivatives into $SL(2, \ZZ)$-covariant derivatives.  The correlator in the integrand on the right-hand side of \eqref{aetertwo} is then defined with separated points and so is UV finite.\footnote{ The r\^ole of contact terms and covariant derivatives has also been discussed  recently in \cite{Niarchos:2019onf} in the content of four-dimensional $\mathcal{N}=2$ supersymmetric CFTs.} 

Similarly, the action of $\bar\cO_{\bar \tau}$ constrains the sum of the values of $\bar a_r$. 
In this case  $SL(2,\Z)$-covariance leads to the constraint 
\be\label{Ares-2}
\sum_{r=1}^n \bar a_r =  -n+w \, .
\ee
However,  the correlation function  in the second term on the right-hand side of \eqref{relations2} is not a MUV correlator (and, in particular, is not a chiral correlator if all the $\cO_j$ are chiral).  In that case its structure is more complicated, and we will comment on this further in section \ref{nonmin}. 

We will utilise the relation  \eqref{aetertwo} as a recursion relation that imposes non-trivial constraints on the $(n+1)$-point correlator with a chiral Lagrangian inserted once we know the $n$-point correlator \footnote{An analogous relation was explored in  \cite{Eden:1999gh, Basu:2004nt, Baggio:2012rr} to prove the non-renormalisation theorems of three-point correlators using the fact that the four-point correlators cannot violate the $U(1)_Y$ symmetry. }.  The relation \eqref{aetertwo}, which will be applied to MUV correlators in the next sections, mirrors the soft dilaton theorems in superstring amplitudes \cite{Green:2019rhz}.

Relations  of the general nature of \eqref{aetertwo} have arisen in related contexts.  In particular, \cite{Intriligator:1998ig,Intriligator:1999ff} discussed consequences of the differential relation for correlators with the emphasis on cases with $n\le 4$, rather than on the $SL(2,\Z)$-covariance of the $n>4$ cases.  Although $SL(2,\Z)$ covariance was the main focus in \cite{Basu:2004dm}, the arguments  given there were incomplete.  A similar relation  to \eqref{aetertwo}, but with the covariant derivative replaced by a simple derivative was used in \cite{Eden:2011we} to construct loop integrands in perturbation theory, where  $SL(2,\Z)$ plays no r\^ole.  This is equivalent to considering the combination of  covariant derivatives that inserts the Lagrangian is $\cD_w + \bar \cD_{-w}$, which does not depend on the modular weight, $w$.  As was discussed in  \cite{Green:2019rhz}, a similar issue arises in the study of  the soft dilaton theorem in superstring  perturbation theory  \cite{Ademollo:1975pf, Shapiro:1975cz, DiVecchia:2015jaq}.

 \section{Recursion relations between correlators}
 \label{softoper}

 From \eqref{corn0} and \eqref{aetertwo} the recursion relation for chiral MUV correlators reduces to 
 \ie
 \cI_{n-1}(\{x_r,\rho_r, 0,y_r\})  \cD_w\, \cG_{n-1}(x_1, \cdots , x_{n-1}; \wt)=  {1\over 2} \int {d^4 x_n}\cI_n(\{x_r,\rho_r, 0,y_r\})\,\cG_n(x_1, \cdots , x_{n}; \wt)  \, . 
 \fe
Since the operator at $x_{n}$ is $\cal{O}_{\tau}$,  it follows that  $\cI_n(\{x_r,\rho_r, 0,y_r\}) \big{|}_{\rho_n^4}= \cI_{n-1}(\{x_r,\rho_r, 0,y_r\}) $  and  therefore  the factors of $\cI_n$ and $\cI_{n-1}$  cancel in this equation,  resulting in a relation for $\cG$, 
  \ie
   \label{eq:soft}
 \cD_w\, \cG_{n-1}(x_1, \cdots , x_{n-1}; \wt)=  {1\over 2}   \int {d^4 x_n } \,\cG_n(x_1, \cdots , x_{n}; \wt) \, ,
 \fe
where the modular weight of the $( n-1)$-point correlator is $w=n-5$.\footnote{This relationship  is the analogue of  soft dilaton relations in superstring amplitudes in flat space, where the supersymmetry factor $\delta^{16}(\sum_{i}Q_i)$ also cancels out when a dilaton is taken to be soft. See further comments in section~\ref{nonchiral} and appendix~\ref{holosol}. }

 We will now demonstrate how knowledge of the large-$N$ expansion of the correlator of four $\cO_2$ operators  obtained in \cite{Chester:2019jas} and \cite{Chester:2020vyz}  determines the large-$N$ expansion of the MUV  $n$-point correlators with $n>4$ by using the recursion relation \eqref{eq:soft} in the large-$N$ expansion.   Our arguments mirror those in \cite{Green:2019rhz}, which concerned the low-energy expansion of flat-space MUV amplitudes in type IIB superstring theory.
  
Among  the key properties that  determine the structure of MUV correlators is  their OPE structure  
\ie \label{eq:OPE}
\langle \cO_{j_1}(x_1)  \cO_{j_2}(x_2)\dots \cO_{j_{n+1}}(x_{n+1}) \rangle \sim \sum_{j'}  C_{j_1 j_2 \bar{j'}} (x_1 - x_2) \langle \cO_{j'} (x_2)  \cO_{j_3}(x_3)   \dots \cO_{j_{n+1}}(x_{n+1}) \rangle + \cdots
\fe
where the ellipsis indicates a sum of non-BPS long operators and double-trace operators. In the large-$N$ limit with finite $g_{_{\rm YM}}$  the long operators develop large anomalous dimensions. The $U(1)_Y$ charge and dimension of  $\cO_{j'}$ are given by $j'=(q_{j_1} + q_{j_2},\Delta_{j'})$, and its conjugate has $\bar{j'}=(-q_{j_1} - q_{j_2},\Delta_{j'})$.
This follows because the three-point function $C_{j_1 j_2 \bar{j'}}  \sim \langle \cO_{j_1} \cO_{j_2} \cO_{\bar{j'}}  \rangle$ conserves $U(1)_Y$ charge \cite{Intriligator:1998ig}.  Therefore,  the $n$-point correlator with  $\cO_{j'}(x_{2})$ on the right hand side of \eqref{eq:OPE}  vanishes because it violates the $U(1)_Y$ charge by $2(n-3) >2(n-4)$, which is beyond the maximal $U(1)_Y$ charge of a $n$-point correlator. As a result, a MUV correlator $\widehat{G}_{n;n-4} (j_1,\dots,j_n) $  is dual to a contact amplitude  that has no single-trace operator poles.  This implies that the large-$N$ expansion of  {$\widehat{G}_{n;n-4} (j_1,\dots,j_n) $  is dual to the $\alpha'$-expansion of an $AdS_5\times S^5$ type IIB superstring amplitude that does not have any intermediate poles due to  light states.  The Witten diagrams  associated with terms in  the large-$N$ expansion of $\widehat{G}_{n;n-4} (j_1,\dots,j_n) $ are given by a sum of  $D$-functions multiplied by polynomials in $x_{ij}$.   $D$-functions and their Mellin transforms are reviewed in appendix~\ref{mellind}
  
  We will  now argue that  the large-$N$ expansion of the reduced correlator,  $\cG_n(x_1,\dots,x_n)$ (defined in \eqref{corn0}) is likewise a sum of $D$-functions multiplied by polynomials in $x_{ij}$, rather than multiplied by more general rational functions that have singularities in $x_{ij}$.   To prove this we note that any such singularities in  $\cG_n(x_1,\dots,x_n)$ would have to be cancelled by zeroes in coeffiicient of $\rho_1^{k_1}\dots \rho_n^{k_n}$ in the expansion of the prefactor $ \cI_n(\{x_i,\rho_i, 0,y_i\})$ given in \eqref{eq:In} in order for $\hat G_{n,n-4} (j_1,\dots,j_n) $ to be free of such  singularities .
    However,  this is not possible as is demonstrated by  the $(n>4)$-point  correlator  \eqref{relevant}.  In  this case the prefactor, which is  given in \eqref{I0}, is quadratic in powers of $x_{ij}^2$ with $i,j\le 4$, and is therefore not symmetric in $x_{ij}$.  Since  $\cG(x_1,\dots,x_n)$ is permutation symmetric, if it had singularities these would arise in all channels and they would not all be cancelled by the  prefactor,  which would be necessary for $\widehat{G}_{n; n-4}(j_1,\dots,j_n)$ to be free of such poles.
 
 This suggests that the Mellin amplitude derived from $\cG_n(x_1,\dots,x_n)$ should be a polynomial in the  Mellin variables $\gamma_{ij}$, and the degree of the polynomial is determined by the number of derivatives.  The Mellin amplitudes corresponding to contact interactions are discussed in appendix~\ref{mellinflat} following closely the discussion in  \cite{Penedones:2010ue}. Indeed, the poles of Mellin amplitudes correspond to exchange of single-trace operators and as we argued they are not present in MUV correlators in the large-$N$ expansion. This is consistent with the fact that the terms in  the low-energy expansion of flat-space MUV amplitudes of type IIB superstring theory do not have massless poles. The Mellin amplitude corresponding to a contact vertex with derivatives acting on it is displayed in \eqref{eq:Mellin1}.  This property will play an important r\^ole in the following discussion. In particular, it allows us to write down an ansatz for a given MUV correlator in the large-$N$ expansion, with a few unknown coupling constant dependent coefficients, which are then fixed by the recursion relation \eqref {eq:soft}. 

\subsubsection{A note on notation}

In the following we will use the notation $\cG_n^{(\alpha)}(x_i;  \wt)$  to denote the contribution to the $n$-point correlator at order $c^{(1-\alpha)/4}$ in the large-$N$ expansion,\footnote{We will use the economic notation $\cG_n^{(\alpha)}(x_i; \wt) \equiv   \cG_n^{(\alpha)}(x_1,\dots,x_n; \wt)$,  $A_n^{(m)}(x_i) \equiv A_n^{(m)}(x_1,\dots,x_n)$, etc.},
\bea
\label{anote}
\cG_n^{(\alpha)}(x_i; \wt) = \sum_{m=0}^{\alpha }  \cF^{(\alpha)}_{n, m}(\wt)\, A_n^{(m)}(x_i) \, ,
\eea
where $\cF^{(\alpha)}_{n, m}(\wt)$ is an appropriate modular form with holomorphic and anti-holomorphic modular weights $(n-4, 4-n)$ for the MUV correlators we are considering,\footnote{This notation makes contact with the notation used in \cite{Green:2019rhz}  for the coefficients of the low energy expansion of the holographically dual scatttering amplitudes in flat-space type IIB superstring theory.} $\wt$ is the complex coupling constant and $A_n^{(m)}(x_i)$ form the kinematic basis, which we will construct explicitly, corresponding to the contact interaction with $2m$ derivatives acting on it.  We will consider the cases where $\alpha$ takes values $0, 2, 3$\footnote{Note, for the case $\alpha=1$, $\cF^{(1)}_{n, m} (\wt) =0$.  This follows from the fact $\cF^{(1)}_{4, m} (\wt) =0$ together with the recursion relation \eqref{eq:soft}. } for the interactions of the same dimension as $R^4, d^4R^4, d^6R^4$, respectively.  Note that in general when $m \geq 3$ there is  more than one kinematic invariant.  This will be of particular relevance when $\alpha=3$ and $n \geq 6$. For such cases, we will have to introduce additional indices to distinguish a two-fold degeneracy of kinematic factors.

In order to make contact with the $AdS_5\times S^5$ amplitudes we will discuss the Mellin transforms of these correlators, which have the form
\bea
\cM^{(\alpha)}_n (\gamma_{ij}; \wt)=   \sum_{m=0}^{\alpha } \cF^{(\alpha)}_{n, m}(\wt)\,  M_{n}^{(m)}(\gamma_{ij})\,,
\label{mellincon}
\eea
where $M_{n}^{(m)}(\gamma_{ij})$ is the Mellin transform of $A_{n}^{(m)}(x_i)$, and is a symmetric polynomial in the Mellin variables $\gamma_{ij}$ with weight $m$. Some general properties of  the Mellin representation of holographic correlators are reviewed in appendix \ref{mellinflat}.

\subsection{The four-point correlator}

The correlation function of four $\mathcal{O}_{2}$'s provides initial data for determining $n$-point correlators using the recursion relation \eqref{eq:soft} and so we will begin with a brief review of some of its properties.
Its form is determined by a function of two independent cross-ratios $U$ and $V$ defined by
\bea
\label{crossr}
U= \frac{x_{12}^2x_{34}^2}{x_{13}^2 x_{24}^2} \,,\qquad V= \frac{x_{14}^2x_{23}^2}{x_{13}^2 x_{24}^2} \, .
\eea 
After stripping off the prefactor $R(1,2,3,4) \prod_{1 \leq i<j \leq 4} x^2_{ij}$,  the correlator can be conveniently expressed in terms of an inverse Mellin transform 
\bea \label{eq:Mellin}
\cG_{4}(x_i; \wt) = {1 \over x_{12}^4x_{34}^4x_{13}^4x_{24}^4} \int   {ds dt \over (4\pi i)^2 } \, U^{{s\over 2}} V^{{t\over 2}-2}  \Gamma \left(2- {s\over 2} \right)^2\Gamma\left(2- {t\over 2} \right)^2 \Gamma \left({s+t\over 2} \right)^2  \cM_4 (s, t; \wt)  \, ,
\eea
where $\cM_4 (s, t; \wt)$ is the Mellin amplitude, and  $s, t$ are the Mellin variables. Some general properties of Mellin amplitudes are reviewed in appendix~\ref{mellind}.

The large-$N$ expansion of $\cM_4 (s, t; \wt)$ (or equivalently $\cG_{4}(x_i;\wt )$) has recently been determined up to order $c^{-1/2}$ using supersymmetric localisation \cite{Chester:2019jas, Chester:2020vyz}, and takes the following form, \footnote{ Our  normalisation differs from that of  \cite{Chester:2020vyz}, by an overall factor of $c^2$  so that the leading term in the large-$c$ (i.e. large-$N$) limit, which corresponds to tree-level supergravity, is of order $c\sim N^2$.   Furthermore, we have corrected a typo in the prefactor of the $c^{-\half}$ term. }
 \ie \label{eq:4pt-Mellin}
  \cM_4(s, t; \wt) =&\, c\, \frac{8 }{(s - 2) (t - 2) (u - 2)}  
  + c^{1/4} \frac{15\, E(\threeh, \wt) }{4 \sqrt{2 \pi^3}} +  { \cM_{\text{1-loop}}(s, t)}
  \cr
  & +c^{-1/4} \frac{315\, E(\fiveh, \wt)}{128 \sqrt{2 \pi^5}}  \left[ (s^2 + t^2 + u^2) -3\right]
    \\
   & + c^{-1/2} \frac{945 \,   \cE(3,\threeh,\threeh, \wt)}{64\pi^3}\left[stu-\frac14 (s^2 + t^2 + u^2) -4\right] + O(c^{-3/4}) \, ,
 \fe
 where $u= 4-s-t$.  The leading term of order  $c \sim N^2$  is the classical supergravity contribution, which is independent of the coupling constant.  Therefore its derivative with respect to the coupling vanishes, which is consistent with \eqref{eq:soft} and the fact that there are no $U(1)_Y$-violating $n$-point   correlators in the supergravity limit for $n>4$.  A similar statement applies to the one-loop supergravity contribution, $\cM_{\text{1-loop}}(s, t)$. We will be interested in the higher-derivative terms that correspond to the string corrections.

 The leading string correction arises at order $c^\quart\sim N^\half$  (this is the $\alpha=0$ term in \eqref{NN})  and is associated with the higher-derivative interaction $R^4$, which has Mellin amplitude that is simply a constant. In order to apply the recursion relation  \eqref{eq:soft} it is more convenient to express the correlators in space-time coordinates, which are given by $D$-functions 
  \ie \label{eq:4ptnew}
\cG_4^{(0)} (x_i;\wt) = \frac{15E(\threeh, \wt)}{4 \sqrt{2 \pi^3}}  \,A_{4}^{(0)}(x_i)\, , \quad {\rm with} \quad
A^{(0)}_{4}(x_i) = D_{4444}(x_i) \,.
\fe
The definitions of $D_{4444}(x_i)$ (and $D$-functions with general conformal dimensions) and their Mellin transforms are reviewed in appendix~\ref{mellind}.
The modular  function $E(\threeh,\wt)$ is a non-holomorphic Eisenstein series with properties that are reviewed in appendix \ref{holosol}.  More generally, in the following we will encounter non-holomorphic Eisenstein series $E_w(\threeh,\wt)$  that are weight $(w,-w)$ modular forms, 
\ie \label{eq:Ew}
E_{w} (s,\wt)= \frac{2^w \Gamma(s) } { \Gamma(s+w)}\, \cD_{w-1} \cdots \cD_0\, E(s,\wt)\, .
\fe
The properties of $E_{w} (s,\wt)$ are also reviewed in appendix \ref{holosol}. In the special case $w=0$ we will drop the subscript ${}_0$ and set  $E_0(\threeh,\wt) \to E(\threeh,\wt)$.

The $\alpha=2$  term (of order $c^{-\quart}\sim N^{-\half}$) in \eqref{eq:4pt-Mellin} that is proportional to $s^2+t^2+u^2$  corresponds, in the flat-space limit,  to the higher derivative  $\quart$-BPS  interaction, $d^4R^4$,  of the  $AdS_5 \times S^5$ type IIB superstring.       The other $c^{-\quart}$ term is independent of $s,t,u$ and corresponds to a correction to the $R^4$ interaction proportional to $(\alpha' )^2/L^4$.   The $c^{-\quart}$ contribution can be expressed in coordinate space in terms of the linear combination of $D$-functions, 
\ie \label{eq:A4s2-2}
\cG_4^{(2)} (x_i; \wt)=  \frac{315\, E(\fiveh, \wt )}{32 \sqrt{2 \pi^5}}  \left[  A_{4} ^{(2)} (x_i)- {19 \over 4}A_{4 }^{(0)} (x_i)\right] \, ,
\fe
where $A_{4}^{(0)}(x_i)$ is defined in \eqref{eq:4ptnew}
and $A_{4} ^{(2)}(x_i) $ is given by
\ie \label{eq:4pts2-3}
A_{4 } ^{(2)}(x_i) =  \left( x^2_{12} x^2_{34}+ x^2_{13} x^2_{24}+ x^2_{14} x^2_{23} \right) {D}_{5555}(x_i)  \, ,
\fe
where the  prefactors of  $x^2_{ij} x^2_{kl}$ result from the higher powers of $s$, $t$, $u$ in the second line of  \eqref{eq:4pt-Mellin}.

 This is seen by  expressing  $\cG_4^{(2)} (x_i; \wt)$ as an inverse  Mellin transform by using \eqref{eq:Mellin0} and \eqref{eq:Mellin1}
\ie
\label{melltwo}
\cG_4^{(2)} (x_i; \wt) &=  \frac{315\, E(\fiveh, \wt )}{32 \sqrt{2 \pi^5}} {1\over x_{12}^4x_{34}^4x_{13}^4x_{24}^4 }  \int   {d \gamma_{12} \over 2\pi i }  {d \gamma_{13} \over 2\pi i } U^{2-\gamma_{12}} V^{\gamma_{12}+\gamma_{13}-4}  \Gamma (\gamma_{12})^2 \cr
& \times \Gamma (4-\gamma_{12}-\gamma_{13})^2 \Gamma (\gamma_{13})^2  \left(\gamma_{12} \gamma_{34} +\gamma_{13} \gamma_{24} + \gamma_{14} \gamma_{23} - {19 \over 4}  \right) \, ,
\fe
with the constraints $\sum_{j\neq i} \gamma_{ij} = 4$ for all $i$. The integrand  translates into a function of $s,t,u$ by using the definitions in $\eqref{deltak}$, which imply  $\gamma_{12} = -s/2+2$ and  $\gamma_{13} =s/2+t/2$.  With this change of variables the integrand in \eqref{melltwo}  matches the coefficient of the $c^{-\quart}$ term in \eqref{eq:4pt-Mellin}.

Similarly,  the $\alpha=3$ term (of order $c^{-\half}\sim N^{-1}$)  that is proportional to $stu$ in  the last line of  \eqref{eq:4pt-Mellin} corresponds to the $\eighth$-BPS  interaction $d^6R^4$ in the flat-space limit of  $AdS_5 \times S^5$.  The $c^{-\half}$ term proportional to $s^2+t^2+u^2$ corresponds to a correction to the $d^4R^4$ interaction of order $\alpha'/L^2$.   The $c^{-\half}$ term that  is independent of $s,t,u$ corresponds to a $(\alpha')^3/L^6$ correction to the $R^4$ interaction.  After an inverse Mellin transform these terms of order $c^{-\half}$ package into the correlator 
 \ie \label{eq:4pt-3s}
\cG_4^{(3)}  (x_i;\wt)=  -   \frac{945\, \cE(3,\threeh,\threeh, \wt)}{32\pi^3}  \left[ A_4^{(3)} (x_i)  + {9\over 2} A_4^{(2)} (x_i) -  32 A_4^{(0)} (x_i) \right] \, ,
\fe
where $ \mathcal{E}(3, \threeh, \threeh, \wt)$ is the modular function that arises as the coefficient of $d^6R^4$ in the ten-dimensional  type IIB effective action \cite{Green:2005ba}\footnote{This function was denoted $\cE_({\threeh,\threeh)}(\wt)$ in \cite{Green:2005ba} and has often also been denoted $\cE_{(0,1)}(\wt)$.}. This is a generalised Eisenstein series that satisfies an inhomogeneous Laplace eigenvalue equation, as reviewed in appendix~\ref{holosol}.  The functions $A_4^{(0)}(x_i) $ and $A_4^{(2)}(x_i) $ were defined previously and we have also introduced  
\ie
A_4^{(3)} (x_i)  &= x^2_{14} x^2_{24} x^2_{34} D_{5557}(x_i)  + x^2_{13} x^2_{23} x^2_{34} D_{5575} (x_i) \cr
&+ x^2_{12} x^2_{23} x^2_{24} D_{5755}(x_i) 
+ x^2_{12} x^2_{13} x^2_{14} D_{7555} (x_i) \, ,
\label{a4def}
\fe
where $A_4^{(3)}(x_i)$ corresponds to a six derivative term in the Mellin transform.\footnote{All the terms in $A_4^{(3)}(x_i)$ in   \eqref{a4def}   are in fact equal to each other. The expression as a sum makes the permutation symmetry manifest.} 

The terms in the large-$N$ expansion that we have described up to this order are those that correspond to BPS interactions in the flat-space limit where they are completely fixed by supersymmetry together with $SL(2,\Z)$ duality.  They are also the terms that have been determined by holography in $AdS_5\times S^5$ starting from the localised integrated correlation function. In this paper we will not consider terms of higher order than $c^{-\half}$ although a number of higher-order terms have been strongly motivated by the localisation arguments \cite{Chester:2019jas, Chester:2020vyz}.  Since  the localisation analysis is based on the structure of integrated correlation functions it produces averaged information and can only determine linear combinations of the higher-order interactions in string theory.    

Given the results of the first few terms in the $1/N$ expansion of the four-point correlator  we will now determine higher-point MUV correlators using the recursion relation \eqref{eq:soft}, with the four-point function as the initial data.

\subsection{$n$-point correlators at order $c^\quart$}

We start with the $\alpha=0$ terms in \eqref{NN} that arise at  order  $c^\quart$  (or $N^\half$) in the large-$N$ expansion of a MUV $n$-particle correlator.  These  terms correspond to contact interactions in the low-energy expansion of the type IIB theory in $AdS_5\times S^5$  of the form $R^4Z^{n-4}$ (and its supersymmetry completion), where $Z$ is the fluctuation of the complex scalar field $\tau$  that carries $U(1)_Y$ charge $q_Z=-2$.\footnote{ The scalar field $Z$ is a reparameterisation of $\tau$ that is defined by $Z=(\tau-\tau^0)/(\tau-\bar\tau^0)$ where $\tau^0$ is the constant background value of $\tau$ \cite{Green:2019rhz}. } In order to illustrate the idea, we will begin by determining the five-point correlator from the known four-point result in the previous section. Dimensional analysis suggests that the five-point Mellin amplitude at this order should be a constant.  Furthermore as discussed earlier  $\cG_5^{(0)}  (x_i;\wt)$ should have conformal dimension $+4$ at each operator position, which implies it is proportional to $D_{44444}(x_i)$. These properties together imply that it is given by
\ie \label{eq:5pt0}
\cG_5^{(0)}  (x_i;\wt)=  \cF^{(0)}_{5}(\wt)   A_5^{(0)}(x_i) \, ,  \quad {\rm with} \quad A_5^{(0)} (x_i)=  D_{44444} (x_i)\, ,
\fe
where $\cF^{(0)}_{5}(\wt)$  will be determined from the recursion relation \eqref{eq:soft}. 
Using \eqref{eq:soft} and the four-point result \eqref{eq:4ptnew}, we have the relation,\footnote{Here and in the following we are using a condensed notation for the integral  over one variable of a function that depends on $n$ variables.  The integrated variable is displayed explicitly  so we write $A(x_i,x_{n}) \equiv A(x_1,\dots,x_{n-1}, x_{n})$, (where $i=1,\dots, n{-}1$) as an example in which the variable $x_{n}$ is singled out. }
\ie
{1 \over 2} \cF^{(0)}_{5}(\wt)  \int {d^4 x_5 } A_5^{(0)}(x_i, x_5 ) = \frac{15 }{4 \sqrt{2 \pi^3}}  \cD_0 E(\threeh, \wt) A_4^{(0)}(x_i)  \, .
\fe
The $D$-functions can be integrated explicitly using the following general integral identity
\begin{align}  \label{eq:integral1}
\int d^4y \left[ (y-y_1)^2 \right]^m \left({z_0 \over  z_0^2 + (y-z)^2} \right)^{m+4}  &={\pi^2  \over (m+2)(m+3)}  \left( {z_0 \over  z_0^2 + (y_1-z)^2} \right)^{-m} \, ,
\end{align}
where the $m=0$ case is relevant for current considerations.  Using this  identity, we obtain,  
\ie
\label{fourfive}
  \int {d^4 x_5  } A_5^{(0)} (x_i, x_5) =  {\pi^2 \over 6} \times  \mathcal{N}_{44444; 4444} A_4^{(0)} (x_i)\, , 
\fe
where the factor $\mathcal{N}_{44444; 4444}$ is defined as 
\ie
\label{normalise}
\mathcal{N}_{\Delta_1 \Delta_2 \cdots \Delta_{n}; \Delta'_1 \Delta'_2 \cdots \Delta'_{n-1}} =  
 \frac{\prod_{i=1}^n \Gamma(\Delta_i) }{ \prod_{i=1}^{n-1} \Gamma(\Delta'_i ) } 
 \frac{ \Gamma\left({\sum_{i=1}^{n-1} \Delta'_i \over 2}-2 \right) }{\Gamma\left({\sum_{i=1}^{n} \Delta_i \over 2}-2 \right)}  \, ,
\fe
which is correlated with the normalisation of the $D$-functions defined in  \eqref{eq:D-function}.

It follows from \eqref{fourfive} that the  coefficient modular form $\cF^{(0)}_{5}(\wt)$ is determined in terms of the four-point modular function,
\ie
{\pi^2 \over 12} \times  \mathcal{N}_{44444; 4444} \,\cF^{(0)}_{5}(\wt)  =  \frac{15 }{4 \sqrt{2 \pi^3}}  \cD_0 E(\threeh, \wt) \, , 
\fe
which implies,  
\ie
 \cF^{(0)}_{5}(\wt)  =  \frac{945 }{ 4  \sqrt{2 \pi^7}} \,   E_1(\threeh, \wt) \, , 
\fe
where we have used $\mathcal{N}_{44444; 4444} = {1 \over 7}$, and the definition for $E_{w} (s,\wt)$ given in \eqref{eq:Ew}.

The discussion  of the $c^{\quart}$ correlators for general $n>4$  is analogous. This gives
\ie \label{eq:npt0}
\cG^{(0)}_{n}(x_i; \wt) =  \cF^{(0)}_{n}(\wt)   A_n^{(0)}(x_i) \, ,  \quad {\rm with} \quad A_n^{(0)} (x_i) =  
D_{ \underbrace{\scriptstyle  44\dots 4}_{n}}  (x_i) \, ,
\fe
where, as indicated, the $D$-function has $n$ indices. Again, the space-time dependence  is determined by the fact that the Mellin amplitude is constant and the correlation function should have conformal dimension $4$ at each $x_i$. The integral given in \eqref{eq:integral1}  leads to 
\ie \label{eq:int0}
 \int d^4 x_n  A_n^{(0)} (x_i, x_n)  =  {\pi^2 \over 6}  \mathcal{N}_{ \underbrace{\scriptstyle  44\dots 4}_{n}; \underbrace{\scriptstyle  44\dots 4}_{n-1}}  A_{n-1}^{(0)} (x_i)\, ,
\fe
where the normalisation factor is given by
\ie
 \mathcal{N}_{ \underbrace{\scriptstyle  44\dots 4}_{n}; \underbrace{\scriptstyle  44\dots 4}_{n-1}} = \frac{3}{(n-2) (2 n-3)} \, .
 \fe
From \eqref{eq:int0} and \eqref{eq:soft}, we determine  that the modular form coefficients of MUV $n$-point correlators at leading order in the large-$N$ expansion have the form
\ie
\cF^{(0)}_{n}(\wt) %= \left(- {24 \over \pi^2} \right)^{n-4}  \frac{15 }{4 \sqrt{2 \pi^3}} D_{n-5} E_{n-5}(\threeh, \tau) 
=   {\Gamma (2 n-2)  \Gamma( n-\fiveh) \over 16 \sqrt{2}\, \pi^{2n-6}  }  E_{n-4}(\threeh, \wt) \,, \quad\quad n\geq 4\,.
\fe

\subsection{$n$-point correlators at order $c^{-\quart}$}

At this order  the correlators we are considering correspond to local interactions  of  supergravity fields of the form $d^4R^4 Z^{n-4}$ (and its supersymmetry completion)\footnote{The tensor contractions of the curvature $R$ and the derivatives, $d$,   have been suppressed in this expression,  as has the precise indication of which fields the derivatives act on.}  in $AdS_5 \times S^5$.  Therefore we expect the Mellin amplitude to be proportional to a polynomial  of weight $(\gamma_{ij})^2$,  as in the $\alpha=2$, $n=4$ case reviewed previously. We begin with the five-point correlator, which involves two independent kinematic invariants of order  $(\gamma_{ij})^0$ and $(\gamma_{ij})^2$. The former corresponds to the  position space correlator $A_5^{(0)}(x_i)$, which appeared at order $c^\quart$ in the previous section. The higher-derivative terms correspond to a new position-space expression,
\ie \label{eq:A52}
A_5^{(2)} (x_i)= (x^2_{12} x^2_{34}+ x^2_{13} x^2_{24}+ x^2_{14} x^2_{23}) {D}_{55554} (x_i) +  {\rm perm} \, ,
 \fe
 where ``perm" indicates a sum over all independent permutations. The function $A_5^{(2)}(x_i)$ is constructed so that it has conformal dimension $4$ at each point, and has permutation symmetry. Note that, up to terms that have less  derivatives that will be included later, the choice of $A_5^{(2)} (x_i)$ is unique. This can be seen by noting that the Mellin amplitude $\gamma_{12} \gamma_{34} + {\rm perm}$ is the unique symmetric polynomial at order $(\gamma_{ij})^2$, again, up to lower-derivative terms.

 Therefore the five-point MUV correlator at order $c^{-\quart}$ should take the following form, 
 \ie
\cG_5^{(2)}(x_i; \wt)  =   \cF^{(2)}_{5, 2}(\wt) \,A_5^{(2)} (x_i)  + \cF^{(2)}_{5,0}(\wt) \, A_5^{(0)} (x_i)  \, ,
 \fe
 where the $\wt$-dependent functions $\cF^{(2)}_{5,2}(\wt)$ and $\cF^{(2)}_{5,0}(\wt) $ will be shown to be uniquely determined by the recursion relation \eqref{eq:soft} and the four-point result.  To utilise  \eqref{eq:soft}, we need to integrate both $ A_5^{(0)} (x_i)$ and $ A_5^{(2)} (x_i)$. The integral $\int d^4 x_5 A_5^{(0)}(x_i,x_5)$ reduces to $A_4^{(0)}(x_i)$ as discussed previously.  The integral of   $A_5^{(2)}(x_i,x_5)$ is evaluated by using  \eqref{eq:integral1} for $m=0$ and $m=1$, which gives 
\ie \label{eq:5pts2}
 \int {d^4 x_5 } A_5^{(2)} (x_i,x_5) &=  {\pi^2 \over 6}\! \left[ \cN_{45555;5555}  (x^2_{12} x^2_{34}+ x^2_{13} x^2_{24}+ x^2_{14} x^2_{23}) {D}_{5555} (x_i) +\cN_{45555;4455} \! \left(  x^2_{12}   {D}_{5544} (x_i) \right.  \right. \cr 
&
\left. \left. +\, x^2_{13}  {D}_{5454} (x_i)  +x^2_{14}  {D}_{5445}  (x_i) + x^2_{23}  {D}_{4554}   (x_i) + x^2_{24} {D}_{4545}  (x_i)+ x^2_{34} {D}_{4455} (x_i) \right) \right] \cr
&=  {\pi^2 \over 72}  \left[  A_4^{(2)}(x_i)+ \frac{128}{7} A_4^{(0)} (x_i)\right] \, .
\fe
In arriving at this result we have used 
\ie
 \mathcal{N}_{45555;5555}  = {1 \over 12} \, , \qquad \mathcal{N}_{45555;4455}  = {4 \over 21} \, ,
\fe
and the following identity satisfied by  $D$-functions, 
\ie \label{eq:5pt-id}
x^2_{12}   {D}_{5544}(x_i)  &+ x^2_{13}  {D}_{5454}(x_i)   + x^2_{14}  {D}_{5445} (x_i)  + x^2_{23}  {D}_{4554} (x_i) \cr
&  + x^2_{24} {D}_{4545} (x_i) + x^2_{34} {D}_{4455} (x_i) = 8  {D}_{4444} (x_i) \, .
\fe
This identity can be obtained  by noting that each side of the equation has the same Mellin transform, which is simply equal to  $8$. 

Therefore, using \eqref{eq:5pts2}, we have
\ie
{1 \over 2} \int d^4 x_5\,  \cG_{5}^{(2)} (x_i,x_5; \wt)  = & \, \frac {\pi^2}{ 144}  \left[ \cF^{(2)}_{5,2}(\wt) A_4^{(2)}  (x_i)  + \frac{128}{7}  \cF^{(2)}_{5,2}(\wt)    A_4^{(0)}   (x_i)  \right.  \cr
& \left.   + \,  \frac{12}{7} \cF^{(2)}_{5,0}(\wt)  A_4^{(0)}  (x_i)   \right] \, .
\fe
This expression is related by the recursion relation \eqref{eq:soft} to the covariant derivative acting on the four-point correlator.  Using the expression \eqref{eq:A4s2-2} leads to following relations
\ie
\cF^{(2)}_{5,2}(\wt) =   \frac{14175}{8 \sqrt{2 \pi^9}}   E_1(\fiveh, \wt) \, , \qquad
  \cF^{(2)}_{5,0}(\wt) =  - {215 \over 16} \cF^{(2)}_{5,2}(\wt)  \, ,
\fe
where we have used \eqref{eq:Ew} for the definition of $E_1(\fiveh, \wt)$. In summary, we find that  the five-point MUV correlator at order  $c^{-\quart}$ is given by
\ie
\cG_{5}^{(2)} ( x_i; \wt) = \cF^{(2)}_{5,2}(\wt) \left[   A_5^{(2)} (x_i) - { 215 \over 16 } A_5^{(0)} (x_i) \right] \, .
\fe

More generally, we may consider the $n$-point correlator with any $n\ge 4$.  Again, the Mellin amplitude at  order $c^{-\quart}$ has two independent kinematic invariants.  One of them is simply $A_n^{(0)}(x_i)$, and the other takes the following form in terms of $D$-functions, 
\bea
A_n^{(2)}(x_i) = (x^2_{12} x^2_{34} + x^2_{13} x^2_{24} +x^2_{14} x^2_{23}) \, D_{ \underbrace{\scriptstyle  55554\dots 4}_{n}}(x_i) + {\rm perm}\, ,
\eea
which is the generalisation of $A_5^{(2)}(x_i)$ in \eqref{eq:A52}.  The correlator $\cG_n^{(2)} (x_i; \wt)$ is then a linear combination of $A_n^{(0)}  (x_i) $ and $A_n^{(2)}  (x_i)$,\footnote{Due to identity \eqref{eq:5pt-id} and its $n$-point generalisation, the function $A_n^{(1)} (x_i) = x^2_{12} D_{ \underbrace{\scriptstyle 554\cdots4 }_n}(x_i) +{\rm perm}$ is not independent of $A_n^{(0)} (x_i)$. }
\bea \label{eq:npt-s2}
\cG_n^{(2)}  (x_i;\wt) = \cF^{(2)}_{n,2}(\wt) \, A_n^{(2)} (x_i) + \cF^{(2)}_{n, 0}(\wt)\, A_n^{(0)} (x_i)  \, . 
\eea
To perform the integration over $x_n$, we use \eqref{eq:int0} and the generalisation of the five-point integral relation \eqref{eq:5pts2}, which takes the following form, 
\bea \label{eq:nptsIn2} 
 \int {d^4 x_n }  A_n^{(2)}(x_i,x_n)  =   \frac{\pi^2}{2(n-1) (2 n-1)}
\left( A_{n-1}^{(2)}(x_i) +  \frac{16 (n-3) (n-1)}{2 n-3}   A_{n-1}^{(0)} (x_i)\right)\, ,
\eea
where we have used 
\ie
\cN_{\underbrace{\scriptstyle  55554\dots 4}_{n}; \underbrace{\scriptstyle  55554\dots 4}_{n-1} }= \frac{3}{(n-1) (2 n-1)}\, , \quad 
\cN_{\underbrace{\scriptstyle  55554\dots 4}_{n}; \underbrace{\scriptstyle  554\dots 4}_{n-1} }=\frac{48}{(n-1) (2 n-3) (2 n-1)}\, . 
\fe
Therefore, 
\ie
{1 \over 2} &  \int  {d^4 x_n}  \cG_{n}^{(2)} (x_i,x_n; \wt) =   \frac{\pi^2}{ 4(n-1) (2 n-1)}  \left[  \cF^{(2)}_{n,2}(\wt)    A_{n-1}^{(2)}(x_i) \right. \cr
& \left. + \,  \left(  \frac{16 (n-3) (n-1)}{2 n-3}    \cF^{(2)}_{n,2}(\wt) +  \frac{(n-1) (2 n-1) }{(n-2) (2 n-3)}   \cF^{(2)}_{n,0}(\wt) \right) A_{n-1}^{(0)} (x_i)  \right] \, .
\fe
According to the recursion relation \eqref{eq:soft}, the right hand side should match with a covariant derivative acting on the $(n-1)$-point correlator, namely,  
\ie
\cD_{n-5} \, \cG_{n-1}^{(2)}  ( x_i; \wt) =  \cD_{n-5} \cF^{(2)}_{n-1,2}(\wt) A_{n-1}^{(2)}(x_i) + \cD_{n-5} \cF^{(2)}_{n-1,0}(\wt)A_{n-1}^{(0)}(x_i) \, .
\fe
This leads to recursion relations for the coefficients $ \cF^{(2)}_{n-1,2}(\tau)$ and $ \cF^{(2)}_{n-1,0}(\tau)$.  Solving the recursion relations, and using the four-point initial data in \eqref{eq:A4s2-2},   
\ie
\cF^{(2)}_{4,2}(\wt)= \frac{315E(\fiveh, \wt )}{32 \sqrt{2 \pi^5}}  \, , \qquad \cF^{(2)}_{4,0}(\wt)= - {19\over 4}\cF^{(2)}_{4,2}(\wt)\, ,
\fe 
we obtain,  
\ie
\cF^{(2)}_{n,2}(\wt) &=   \frac {  \Gamma (2n) \Gamma(n-\threeh)} { 384 \sqrt{2} \pi ^{2 n-5}   } \,  E_{n-4}(\fiveh, \wt) \, , \cr
\cF^{(2)}_{n,0}(\wt) &=- \frac{(2 n-5) \left(4 n^2-12 n+3\right)}{4 (n-1)} \, \cF^{(2)}_{n,2}(\wt)    \, .
\fe

Therefore, the $n$-point MUV correlator at order $c^{-\quart}$ is given 
\ie
\cG^{(2)}_{n} ( x_i; \wt)=  \cF^{(2)}_{n,2}(\wt)   \left[  A_n^{(2)} (x_i) - \frac{(2 n-5) \left(4 n^2-12 n+3\right)}{4 (n-1)}  A_n^{(0)}(x_i) \right] \, ,
\fe
and in Mellin space it takes the following form, 
\ie
\cM^{(2)}_{n} ( \gamma_{ij}; \wt)=  \cF^{(2)}_{n,2}(\wt)   \left[  (\gamma_{12} \gamma_{34} + {\rm perm}) - \frac{(2 n-5) \left(4 n^2-12 n+3\right)}{4 (n-1)}   \right] \, ,
\fe
with the constraints $\sum_{  j\neq i }\gamma_{ij} = 4$\  $\forall i$ .

\subsection{Correlators at order $c^{-\half}$}

In this sub-section, we consider the correlator at order $c^{-\half}$, corresponding to local interactions  of  supergravity fields of the form $d^6R^4 Z^{n-4}$ (and its supersymmetry completion) in $AdS_5 \times S^5$. It contains three independent kinematic invariants.  Two of these are $A_n^{(0)}(x_i)$ and $A_n^{(2)}(x_i)$ that we encountered earlier. We will introduce a new structure, which is a polynomial of degree 3 in the Mellin variables $\gamma_{ij}$. We will see that there are in fact two independent kinematic invariants of degree $3$ when $n\geq 6$, which we will denote by $A^{(3)}_{n,1}(x_i)$ and $A ^{(3)}_{n,2}(x_i)$.  This closely resembles properties of MUV type IIB superstring amplitudes with $6$ or more massless external states \cite{Green:2019rhz}.  Since the cases with  $n\geq6 $ are special, we will discuss the cases with $n=5$, $n=6$ and general $n$ separately in the following. 

\subsubsection{$5$-point correlator}

We begin with the five-point correlator, which is uniquely determined by the four-point correlator via the recursion relations. Besides $A_5^{(0)} (x_i) $ and $A_5^{(2)} (x_i)$, the new structure $A^{(3)}_{5}(x_i)$ takes the following form  
\ie
A_5^{(3)}  (x_i) = x^2_{14} x^2_{24} x^2_{34} D_{55574}  (x_i)+ {\rm perm} \, .
\fe
It is constructed so that $A_5^{(3)}  (x_i)$ has the correct conformal dimension at all $x_i$ and furthermore generates six derivatives which are implemented by the three pre-factors of $x_{ij}$. The full correlator is given by the following linear combination, 
\ie
\cG_5^{(3)}(x_i;\wt) = \cF_{5, 3}^{(3)}(\wt) \, A_5^{(3)}(x_i)  + \cF_{5, 2}^{(3)}(\wt) \, A_5^{(2)}(x_i) + \cF_{5, 0}^{(3)}(\wt) \, A_5^{(0)}(x_i)   \, .
\fe
Again, the recursion relation \eqref{eq:soft} relates the above five-point correlator to the four-point correlator. To perform the integral of $A_5^{(3)}  (x_i)$, in addition to the identity \eqref{eq:integral1}, we will use another integration identity, 
\begin{align} \label{eq:integral3}
& \int d^4y\, (y-y_1)^2(y-y_2)^2(y-y_3)^2 \left( {z_0 \over  z_0^2 + (y-z)^2} \right)^7  \cr
%=&\,  {z_0^7 \over 6!} \int d^4 y \, (y-y_1)^2  (y-y_2)^2 (y-y_3)^2  \int_0^{\infty} du\, u^6 \, e^{- u (z_0^2 + (y-z)^2)} \cr
=& \left[ \frac{\pi ^2}{30}  \left( {z_0 \over  z_0^2 + (y_1-z)^2} \right)^{-1} \left( {z_0 \over  z_0^2 + (y_2-z)^2} \right)^{-1} \left( {z_0 \over  z_0^2 + (y_3-z)^2} \right)^{-1} \right. \cr
& \left. - \, \frac{\pi ^2}{5!}  \left( \left( {z_0 \over  z_0^2 + (y_1-z)^2} \right)^{-1}  (y_2 -y_3)^2 +
\left( {z_0 \over  z_0^2 + (y_2-z)^2} \right)^{-1}  (y_1 -y_3)^2 \right. \right. \cr
& \left. \left. + \, \left( {z_0 \over  z_0^2 + (y_3-z)^2} \right)^{-1}  (y_1 -y_2)^2   \right)  \right]   \, ,
 \end{align}
which leads to 
\ie
\int {d^4 x_5 } A_5^{(3)}(x_i,x_5) &= {\pi^2 \over6} \left[ \cN_{55574;5557} A_4^{(3)}  (x_i)  + {1\over 2}  \cN_{55574;5546}  \left( x^2_{14} x^2_{24} D_{5546} (x_i) +{\rm perm} \right) \right. \cr 
& \left.  +\, {4 \over 5} \cN_{55574;4444}  D_{4444} (x_i)  -  {1\over 10} \cN_{55574;5544} \left( x^2_{12} D_{5544} (x_i) +{\rm perm} \right) \right]   \, .
\fe
Using the identity \eqref{eq:5pt-id} the terms on the second line in this equation can be expressed in terms of $A_4^{(0)}(x_i) $, and similarly, we have another type of $D$-function identity, 
\ie
 x^2_{14} x^2_{24} D_{5546}(x_i)  +{\rm perm}  &= - 2 \left( x^2_{12} x^2_{34} D_{5555} (x_i) +{\rm perm} \right) + 32 D_{4444} (x_i) \cr 
 & = -2 A_4^{(2)}(x_i)  + 32  A_4^{(0)}(x_i)  \, . 
\fe
This means that altogether we have 
\ie
 \int {d^4 x_5 } A_5^{(3)}(x_i,x_5) =  {\pi^2 \over 90} \left[ A_4^{(3)}(x_i)  - 3 A_4^{(2)} (x_i) +{416 \over 7}  A_4^{(0)}(x_i)   \right] \, ,
\fe
where we have used the form of the normalisation factor $\cN_{\Delta_1 \cdots \Delta_5; \Delta'_1 \cdots \Delta'_4}$ in \eqref {normalise}.  Therefore, 
\ie
 {1\over 2}  \int {d^4 x_5}  & \, \cG_{5}^{(3)}  (x_i,x_5; \wt)  =  {\pi^2 \over 180}   \left[ \cF_{5, 3}^{(3)}(\wt) \left(  A_4^{(3)} (x_i) - 3 A_4^{(2)} (x_i) +{416 \over 7}  A_4^{(0)} (x_i)  \right)  \right. \cr
& \left. +\, \cF_{5, 2}^{(3)}(\wt) \left( {5 \over 4}  A_4^{(2)} (x_i) + {160 \over 7}   A_4^{(0)}(x_i)  \right) +  {15 \over 7}  \cF_{5, 0}^{(3)}(\wt)A_4^{(0)} (x_i)   \right]  \, .
\fe
The recursion relation implies that this should match with a covariant derivative acting on the four-point correlator given in \eqref{eq:4pt-3s}, which leads to
\bea
\cG_{5}^{(3)} (x_i; \wt) = - \frac{42525}{16 \pi ^5}  \mathcal{E}^{(3)}_{1, 1} ( \wt)  \left[A_5^{(3)} (x_i) + 6 A_5^{(2)}(x_i)  - {320 \over 3} A_5^{(0)}(x_i)  \right]   \, ,
\label{g5def}
\eea
where we have defined, 
\ie \label{eq:cE31}
\mathcal{E}^{(3)}_{w, 1} ( \wt)  = 2^w \, \cD_{w-1} \cdots \cD_0 \mathcal{E}(3, \threeh, \threeh, \wt) \, ,
\fe
with $\mathcal{E}^{(3)}_{0, 1} ( \wt)=\mathcal{E}(3, \threeh, \threeh, \wt)$.  The properties of the modular form $\mathcal{E}^{(3)}_{w, 1} ( \wt)$ and its applications to the low-energy expansion of flat-space MUV superstring amplitudes are reviewed in appendix~\ref{mod14}.

\subsubsection{$6$-point correlator}

As anticipated earlier, when $n\geq 6$ and $\alpha=3$ there are two distinct $x_i$-dependent structures that can contribute to the MUV correlator.     This reflects the fact that there are two independent kinematic factors in flat-space MUV $n$-particle scattering  amplitudes with $n\geq 6$, as was discussed in \cite{Green:2019rhz} (these are the combinations of Mandelstam invariants in $\mathcal{O}^{(3)}_{6,1} (s_{ij})$ and  $\mathcal{O}^{(3)}_{6,2} (s_{ij})$ in \eqref{eq:kinn}).
 The corresponding independent  structures that contribute to the  gauge theory position-space correlators are proportional to
\ie
\label{twost}
(a) \  \  x^2_{14} x^2_{24} x^2_{34} D_{555744}  (x_i) \,, \quad\qquad  \qquad  (b)\  \   x^2_{12} x^2_{34} x^2_{56} D_{555555} (x_i)\, . 
\fe
As we will shortly see, these terms have interesting  properties: 
\begin{itemize}
\item[(1)]
Term (a) is a straightforward generalisation of the $4$ and $5$-point cases. The integral of term (a) over $x_6$ reproduces the form of $A_5^{(3)}(x_i)$ (together with terms with  $\alpha <3$).
\item[(2)]
Term (b) has a form that does not exist for $n<6$.  Furthermore, its integral over $x_6$ has the form of a five-point correlator with $\alpha=2$ rather than $\alpha=3$.   
\end{itemize}

Although there is arbitrariness in the choice of two independent linear combinations of these structures, there is one particularly natural choice of basis motivated by our knowledge of the type IIB amplitudes in the flat-space limit.   This suggests one special  linear combination  should be chosen to be 
\ie
A_{6,1}^{(3)}(x_i)  = \left(x^2_{14} x^2_{24} x^2_{34} D_{555744}(x_i) + {\rm perm} \right) - {3\over 4} \left( x^2_{12}  x^2_{34} x^2_{56} D_{555555}(x_i) + {\rm perm} \right) \, ,
\fe
which has a  Mellin transform that matches the corresponding flat-space amplitude given in \cite{Green:2019rhz} (up to an overall constant factor).  In other words,  in the flat-space limit the Mellin transform of $A_{6,1}^{(3)} (x_i)$ reduces  to 
\ie
M_{6,1}^{(3)} (\gamma_{ij}) \big{|}_{\gamma_{ij} \rightarrow \infty} = {5\over 12} \sum_{i<j} \gamma_{ij}^3 + {1\over 8} \sum_{i<j<k} \gamma_{ijk}^3 \, ,
\fe 
where we have defined $\gamma_{ijk} = \gamma_{ij} + \gamma_{ik}+ \gamma_{jk}$.  One may also translate $\gamma_{ij}$ to $s_{ij}$ using \eqref{deltak}, which implies $\gamma_{ij} \rightarrow -s_{ij}/2$ in the flat-space limit.  This expression is in agreement  with the $s_{ij}$-dependence of the flat-space amplitude proportional to $\mathcal{O}^{(3)}_{6,1} (s_{ij})$ in  \eqref{eq:kinn}.

 The contribution to the $n=6$, $\alpha=3$  correlator that contains $A_{6,1}^{(3)}(x_i)$  will be denoted $\cG_{6,\, 1}^{(3)}$,  and it has  the   following structure, 
\ie \label{eq:6pt-s31}
\cG_{6, 1} ^{(3)}( x_i;\wt) =   \cF_{6, 3, 1}^{(3)}(\wt) \,  A_{6,1}^{(3)} (x_i) + \cF_{6, 2, 1}^{(3)}(\wt)  \, A_6^{(2)}(x_i) + \cF_{6, 0, 1}^{(3)}(\wt)  \,A_6^{(0)} (x_i) \, .
\fe
This expression is required to satisfy the recursion relation, \eqref{eq:soft}  which requires
\bea
\label{recurmore}
{1 \over 2}  \int {d^4 x_6 }\, \cG_{6, 1}^{(3)} ( x_i,x_6;\wt)   =  \cD_1 \cG_5^{(3)} (x_i;\wt) \, .
\eea
The integral on the left-hand side can be evaluated using  \eqref{eq:integral1} and \eqref{eq:integral3}, which leads to
\ie
  \int {d^4 x_6 }\,  A_{6,1}^{(3)}(x_i, x_6)  =    {\pi^2  \over 132} \left[ A_5^{(3)}(x_i) -3 A_5^{(2)}  +\frac{256}{3} A_5^{(0)}  (x_i)  \right] \, .
\fe
This corresponds to item (1) above.
Inputting the expression \eqref{g5def} for $\cG_5^{(3)} (x_i;\wt)$ into the right-hand side of   \eqref{recurmore}  determines all the coefficients in \eqref{eq:6pt-s31} and the result is 
\ie
\cG^{(3)}_{6, 1} (x_i; \wt) =  - \frac{1403325}{4 \pi ^7} \mathcal{E}^{(3)}_{2, 1} ( \wt) \! \left[  A_{6,1}^{(3)}  (x_i) + {15 \over 2}  A_{6}^{(2)}  (x_i)- {2592 \over 11}  A_{6}^{(0)} (x_i)\right]  \, .
\fe

Since $\cG^{(3)}_{6, 1} (x_i; \wt)$  already obeys the recursion relation, and reduces to the five-point correlator upon integration,  the second kinematic invariant must vanish upon integration. This determines $A_{6,2}^{(3)}(x_i)$ to be
\bea
A_{6,2}^{(3)}(x_i)  =  x^2_{12} x^2_{34} x^2_{56} D_{555555}   (x_i)+ {\rm perm} \, ,
\eea
that is term (b) in \eqref{twost}. 
We see that after integration, $A_{6,2}^{(3)}(x_i)$ reduces to a five-point correlator  with a Mellin transform with $\alpha<3$, as was mentioned in item (2) above.  Explicitly,  we have
\bea
 \label{eq:6pt-s32}
 \int {d^4 x_6} A_{6,2}^{(3)}  (x_i) =  {\pi^2 \over 165} \left(  x^2_{12} x^2_{34} D_{55554}  (x_i) + {\rm perm} \right)  =  {\pi^2 \over 165}  A_{5}^{(2)} (x_i) \, ,
\eea
and indeed the right-hand side is of order $\gamma_{ij}^2$ in Mellin space.   With \eqref{eq:6pt-s32} at hand, it is straightforward to see that the combination
\bea
\label{newform}
\cG_{6,  2}^{(3)}  (x_i;\wt) =   \cF_{6, 3, 2}^{(3)}(\wt)  \left[ A_{6,2}^{(3)} (x_i) - \frac{2}{3}  A_6^{(2)} (x_i)+  {128 \over 11}A_6^{(0)} (x_i) \right] \, ,
\eea
has the property that its integral vanishes, namely
\bea
 \int {d^4 x_6  }\, \cG_{6,  2}^{(3)}  (x_i;\wt) =  0 \, .
 \eea

Since $\cG_{6,  2}^{(3)}  (x_i;\wt)$  integrates to zero, we cannot determine the form of the coefficient $\cF_{6, 3, 2}^{(3)}(\wt)$ from the recursion relation. But we can deduce it from knowledge of the flat-space limit. In particular, the Mellin transform of  $\cG_{6,  2}^{(3)}  (x_i;\wt)$ is equivalent to $\gamma_{12}\gamma_{34} \gamma_{56} + {\rm perm}$ in the flat-space limit, which matches precisely with the flat-space kinematic invariant $\cO_{6,2}^{(3)}(s_{ij})$ given in \eqref{eq:kinn2}. The associated modular form is denoted by $\mathcal{E}^{(3)}_{2,2}(\wt)$, which has weights $(2,-2)$ and is reviewed in appendix \ref{holosol}. Therefore the requirement that the flat-space results should be reproduced determines that the  unknown modular form in \eqref{newform} is given by \footnote{ In \cite{Green:2019rhz}, the modular form $\mathcal{E}^{(3)}_{2,2}(\wt)$ was determined apart from  an  overall constant, so the normalisation of $\cF_{6, 3, 2}^{(3)}(\wt)$ in this equation is arbitrary.  However, we will see that once  the six-point correlator is given, the higher-point correlators are uniquely determined.}
\ie
\cF_{6, 3, 2}^{(3)}(\wt)  = \mathcal{E}^{(3)}_{2,2}(\wt) \, . 
\fe

\subsubsection{$n$-point correlators}
 
  In this sub-section, we will show that with the results of the six-point correlators at order $c^{-\half}$, the recursion relation \eqref{eq:soft} completely determines all the higher-point correlators at this order without appealing to knowledge of the flat-space limit. 

It is straightforward to see that $n$-point generalisations of $A_{6,1}^{(3)}(x_i)$ and $A_{6,2}^{(3)}(x_i)$ should take the following forms, 
\ie
A_{n,1}^{(3)}(x_i) =& \, \left( x^2_{14} x^2_{24} x^2_{34} D_{\underbrace{\scriptstyle  55574\dots 4}_{n}} (x_i)  + {\rm perm} \right) - {3\over 4}  \left(x^2_{12} x^2_{34} x^2_{56}  D_{\underbrace{\scriptstyle  5555554\dots 4}_{n}} (x_i)+ {\rm perm} \right) \, , \cr
A_{n,2}^{(3)}(x_i) =&\,  x^2_{12} x^2_{34} x^2_{56} D_{\underbrace{\scriptstyle  5555554\dots 4}_{n}} (x_i) + {\rm perm} \, .
\fe
Therefore, the two independent correlators at order $c^{-\half}$ with $n \geq 6$ are given by expressions of the form 
\ie \label{eq:npt-s31}
\cG_{n,1}^{(3)} (x_i;\wt)=   \cF_{n, 3, 1}^{(3)}(\wt) A_{n,1}^{(3)}(x_i)  + \cF_{n, 2, 1}^{(3)}(\wt) A_n^{(2)} (x_i)+ \cF_{n, 0, 1}^{(3)}(\wt) A_n^{(0)}  (x_i) \, ,
\fe
and 
\ie \label{eq:npt-s32}
\cG_{n, 2}^{(3)}(x_i;\wt) =  \cF_{n, 3, 2}^{(3)}(\wt) A_{n,2}^{(3)}(x_i)  + \cF_{n,2, 2}^{(3)}(\wt) A_n^{(2)} (x_i)+ \cF_{n, 0, 2}^{(3)}(\wt) A_n^{(0)}  (x_i) \, .
\fe

All the coefficients in the above equations can be determined by using \eqref{eq:soft} and the $n=6$ results. To utilise the recursion relation \eqref{eq:soft}, we first note the $x_n$ integral of $A_{n,1}^{(3)}(x_i, x_n)$ is given by
 \ie
\int {d^4 x_n }   A_{n,1}^{(3)} (x_i, x_n) = 
 \frac{\pi^2}{ 2 n (2 n-1)}\! \left[ A_{n-1,1}^{(3)}  (x_i)  - 3 A_{n-1}^{(2)} (x_i) 
+ \frac{16 (n-3) (3 n-2)}{2 n-3}  A_{n-1}^{(0)}   (x_i)   \right] \, .
 \fe
Using the above result, we obtain, 
  \ie
&  { 1 \over 2}  \int {d^4 x_n} \,  \cG_{n, 1}^{(3)}  (x_i,x_n; \wt)   = \cr
&   \frac{\pi^2}{ 4n (2 n-1)}  \left[  \cF_{n, 3, 1}^{(3)}(\wt)  \left( A_{n-1,1}^{(3)}  (x_i)  - 3 A_{n-1}^{(2)} (x_i) 
+ \frac{16 (n-3) (3 n-2)}{2 n-3}  A_{n-1}^{(0)}   (x_i)   \right) \right.  \cr
&  \left.  + \, \cF_{n, 2,1}^{(3)}(\wt)  \left( {n\over n-1} A_{n-1}^{(2)} (x_i)  +  \frac{16 (n-3) n}{2 n-3}  A_{n-1}^{(0)}(x_i)   \right) + \cF_{n, 0, 1}^{(3)}(\wt) \frac{n (2 n-1)}{(n-2) (2 n-3)} A_{n-1}^{(0)}(x_i)    \right] \, . 
\fe
The recursion relation identifies this integrated result with $\cD_{n-5} \cG^{(3)}_{n-1, 1}(x_i; \wt)$, which leads to the relations
\ie
\cF_{n, 3, 1}^{(3)}(\wt) &=  - { 3\, \Gamma (2 n+1) \over 4096 \pi ^{2 n-5}}   \mathcal{E}_{n-4, 1}^{(3)}( \wt) 
 \, , \quad
\cF_{n, 2, 1}^{(3)}(\wt) = \frac{3 (n-1)}{2}  \cF_{n, 3, 1}^{(3)}(\wt) \, , \cr
 \cF_{n, 0, 1}^{(3)}(\wt) &=-\frac{4 (n-3) (n-2) n (n+3)}{2 n-1} \cF_{n, 3, 1}^{(3)}(\wt)  \, ,
\fe
with $ \mathcal{E}_{w, 1}^{(3)}( \wt) $ defined in \eqref{eq:cE31}.  Therefore, the first contribution to the $n$-point correlator at order $c^{-\half}$   is given by
\ie  
\cG^{(3)}_{n, 1}( x_i; \wt)  &= \cF_{n, 3, 1}^{(3)}(\wt) \! \left[ A_{n,1}^{(3)} (x_i)   +  \frac{3 (n-1)}{2}   A_n^{(2)} (x_i)  
-\frac{4 (n-3) (n-2) n (n+3)}{2 n-1}  A_n^{(0)}(x_i)   \right] \, .
\fe

Similarly, using the following integral identity (with $n>6$),
\bea
 \int {d^4 x_n} A_{n,2}^{(3)} (x_i, x_n)    =   \frac{\pi^2}{2n (2 n-1)}
 \left[ A_{n-1,2}^{(3)} (x_i)   + \frac{4 (n-5)}{n-1}  A_{n-1}^{(2)}   (x_i)  \right] \, ,
 \eea
and the result of the six-point correlator \eqref{newform}, the second independent contribution to the $n$-point correlator at order $c^{-\half}$ is given by
\ie
\cG^{(3)}_{n, 2}(x_i; \wt)  &= \cF^{(3)}_{n,3,2}(\wt)   \left[ A_{n,2}^{(3)} (x_i)   - \frac{2 (n-4) (n-5)}{n} A_n^{(2)} (x_i)  \right. \cr
& \left. ~~ + \, \frac{16 (n-5) (n-4) (n-3) (n-2)}{3 (2 n-1)}  A_n^{(0)}(x_i)   \right] \, ,
\fe
where the coefficient $\cF^{(3)}_{n,3,2}(\wt)$ is given by
\ie
\cF^{(3)}_{n,3,2}(\wt) =  \frac{ \Gamma (2 n+1)}{ \Gamma(13) \pi^{2n-12} }   \mathcal{E}_{n-4,2}^{(3)} (\wt) \, ,
\fe
and we have defined 
\ie \label{eq:cE32}
\mathcal{E}_{w,2}^{(3)} (\wt) = 2^{w-2} \, \cD_{w-1} \cdots \cD_2 \mathcal{E}_{2,2}^{(3)} (\wt) \, , \quad {\rm with} \quad w \geq 2 \, .
\fe
Note $2^{w-2} \, \cD_{w-1} \cdots \cD_2 \mathcal{E}_{2,2}^{(3)} (\wt)$ is interpreted as $\mathcal{E}_{2,2}^{(3)} (\wt)$ when $w=2$. 

In Mellin space, the correlators are given by  
\ie
{\cM}^{(3)}_{n, 1} (\gamma_{ij}; \wt)  &= \cF^{(3)}_{n,3,1}(\wt) \left[  \left( \gamma_{14} \gamma_{24} \gamma_{34}  +{\rm Perm} \right)    -  {3\over 4} \left( \gamma_{12} \gamma_{34} \gamma_{56} +{\rm Perm} \right)   \right. \cr 
& \left. + \, {3(n-1)  \over 2} \left( \gamma_{12} \gamma_{34}  +{\rm Perm}  \right) -\frac{4 (n-3) (n-2) n (n+3)}{2 n-1}  \right]\,,
\fe
and
\ie
{\cM}^{(3)}_{n, 2} ( \gamma_{ij}; \wt)  &= \cF^{(3)}_{n,3,2}(\wt)  \left[ \left( \gamma_{12} \gamma_{34} \gamma_{56} +{\rm Perm} \right)  - \frac{2 (n-4) (n-5)}{n}   \left( \gamma_{12} \gamma_{34}  +{\rm Perm}  \right) \right. \cr
& \left. ~~~ +  \frac{16 (n-5) (n-4) (n-3) (n-2)}{3 (2 n-1)}   \right] \, , 
\fe
where the Mellin variables obey the constraints $\sum_{j\neq i}\gamma_{ij} = 4$ for all $i$. In the flat-space limit,  ${\cM}^{(3)}_{n, 1} (\gamma_{ij}; \wt) $ and ${\cM}^{(3)}_{n, 2} (\gamma_{ij}; \wt)$ are in agreement with the known flat-space superstring amplitudes obtained in \cite{Green:2019rhz}. In particular,  for the first case 
\ie
\cM_{n,1}^{(3)} (\gamma_{ij}; \wt)  \big{|}_{\gamma_{ij} \rightarrow \infty} = {4 \over 3}  \cF^{(3)}_{n,3,1}(\wt) \,  \mathcal{O}^{(3)}_{n,1} (\gamma_{ij}) \, ,
\fe
where $\mathcal{O}^{(3)}_{n,1} (\gamma_{ij})$ is the kinematic invariant that appears in the flat-space superstring amplitude and is given in  \eqref{eq:kinn}.  Similarly, for the second case, we have, 
\ie
\cM_{n,2}^{(3)}(\gamma_{ij}; \wt)  \big{|}_{\gamma_{ij} \rightarrow \infty} =   {1 \over 6} \cF^{(3)}_{n,3,2}(\wt) \,   \mathcal{O}^{(3)}_{n,2} (\gamma_{ij}) \, ,
\fe
where $\mathcal{O}^{(3)}_{n,2} (\gamma_{ij})$ is defined in  \eqref{eq:kinn2}.

\subsection{Summary of main results in this section} 
\label{sec:summary}

We will here give a brief summary of the main results  in this section since the details are reasonably complicated.

 Using the recursion relations \eqref{eq:soft}, we have considered the first few orders of the large-$N$ expansion of a general $n$-point MUV correlator,
 \bea \label{eq:MUV}
\widehat {G}_{n; n-4}(j_1, j_2, \cdots, j_n; \wt)  = \cI_n(\{x_i,\rho_i, 0,y_i\})  \times \sum_{\alpha=0, 2, 3} c^{1-\alpha \over 4} \sum_r \cG^{(\alpha)}_{n, r}(x_i; \wt)  \, ,
\eea
where we are ignoring the supergravity contributions which conserve $U(1)_Y$ so they only contribute to $n=4$ correlators.
The index $\alpha$ labels contributions at order $c^{\frac{1-\alpha}{4}}$ and the index $r$ labels the distinct kinematic invariants at a given value of $\alpha$. This index is omitted for terms with $\alpha=0, \,2$ since in those cases there is a unique structure.  However, when $\alpha=3$ there are two independent kinematic invariants for $n\geq 6$, so the index takes the values $r=1,2$ for these cases.

The prefactor $\cI_n$ is determined by the symmetries of the theory, and is given in \eqref{eq:In}. One can extract different chiral MUV correlators by choosing appropriate powers of  $\rho_i$. The dynamic part of the correlator is contained in $\cG^{(\alpha)}_{n }(x_i; \wt)$ and is independent of the species of operators in the correlator. For $\alpha=0, 2$ we find
\ie \label{eq:alpha02}
\cG^{(0)}_{n}(x_i; \wt) &=  {  \Gamma (2 n-2)  \Gamma(n-\fiveh) \over  16 \sqrt{2} \pi ^{2 n-6} }  E_{n-4}(\threeh, \wt)  A^{(0)}_n(x_i)\, , \cr
\cG^{(2)}_{n} (x_i; \wt) &=   { \Gamma (2 n) \Gamma(n-\threeh) \over  384 \sqrt{2} \pi ^{2 n-5} }   E_{n-4}(\fiveh, \wt) \!  \left[  A_n^{(2)} (x_i)- \frac{(2 n-5) \left(4 n^2-12 n+3\right)}{4 (n-1)} A_n^{(0)}(x_i) \right] \, ,
\fe
where $E_w(s,\wt)$ is a non-holomorphic Eisenstein series with weights $(w,-w)$ as defined in \eqref{eq:Ew}, and $A_n^{(\alpha)}(x_i)$ is a sum of $D_{\Delta_1\dots \Delta_n}(x_i)$ functions  multiplied by powers of $x^2_{ij}$ (listed in \eqref{adefs} below).  

When $\alpha=3$, there are two independent contributions.  One of these is connected to the $n=4$, $\alpha=3$ correlator by iterative use of the recursion relation and has the form
\ie  
\label{sixone}
\cG^{(3)}_{n,1}(x_i; \wt)  &= -    {3\Gamma (2 n+1) \over 4096 \pi ^{2 n-5} }  \mathcal{E}^{(3)}_{n-4,1}( \wt) \left[ A_{n,1}^{(3)} (x_i)  +  \frac{3 (n-1)}{2}  A_n^{(2)} (x_i)  
 \right. \cr
& \left.~~~   -\frac{4 (n-3) (n-2) n (n+3)}{2 n-1}   A_n^{(0)}(x_i)   \right] \, ,
\fe
where $\mathcal{E}^{(3)}_{n-4,1}( \wt)$ is a weight-$(n-4, 4-n)$ modular form as defined in \eqref{eq:cE31}. The second $\alpha=3$ contribution  has the form
\ie
\label{sixtwo}
   \cG^{(3)}_{n, 2} (x_i; \wt)  &=  \frac{\Gamma (2 n+1)}{  \Gamma(13) \pi^{2n-12}}  \mathcal{E}_{n-4,2}^{(3)} (\wt) \left[ A_{n,2}^{(3)} (x_i)    - \frac{2 (n-4) (n-5)}{n} A_n^{(2)} (x_i)  \right. \cr
& \left. ~~~ +\frac{16 (n-5) (n-4) (n-3) (n-2)}{3 (2 n-1)}   A_n^{(0)}(x_i)   \right] \,.
\fe
This contribution only arises when $n \geq 6$, and modular form $\mathcal{E}^{(3)}_{n-4,2}( \wt)$ has weights $(n-4, 4-n)$, and it is given in \eqref{eq:cE32}.
  
Finally, the space-time dependent  functions, $A_n^{(0)}(x_i), A_n^{(2)}(x_i), A_{n,1}^{(3)} (x_i), A_{n,2}^{(3)} (x_i)$ are defined in terms of $D$-functions. They are given by
\ie
\label{adefs}
A_n^{(0)}(x_i) &= D_{44\cdots 4}(x_i)\, ,  \cr
A_n^{(2)}(x_i) &= (x^2_{12} x^2_{34} + x^2_{13} x^2_{24} +x^2_{14} x^2_{23}) D_{5555 44\cdots 4}(x_i) + {\rm perm}\, , \cr
A_{n,1}^{(3)}(x_i) &=  \left( x^2_{14} x^2_{24} x^2_{34} D_{555744\cdots 4}(x_i)  + {\rm perm} \right) - \Theta(n-6) \left( {3\over 4}  x^2_{12} x^2_{34} x^2_{56} D_{5555554\cdots 4}(x_i)+ {\rm perm} \right) \, , \cr
A_{n,2}^{(3)}(x_i) &=\,  x^2_{12} x^2_{34} x^2_{56} D_{5555554\cdots 4}(x_i) + {\rm perm} \, ,
\fe
where the step function $\Theta(x)=1$ if $x \geq 0$ and $\Theta(x)=0$ if $x < 0$, and each $D$-function has $n$ indices. 
All $A_n^{(\alpha)}(x_i)$ have conformal dimension $4$ at each point, $x_i$, and have a simple expression in Mellin space using the formalism in  Appendix \ref{mellinflat}, 
\ie
M_n^{(0)}(\gamma_{ij}) &=1 \, ,  \qquad M_n^{(2)}(\gamma_{ij}) =  \left( \gamma_{12} \gamma_{34} + {\rm perm}  \right) \, , \cr
M_{n, 1}^{(3)}(\gamma_{ij}) &=   \left( \gamma_{14} \gamma_{24} \gamma_{34}   + {\rm perm} \right) -
\Theta(n-6)  \, {3\over 4} \left( \gamma_{12} \gamma_{34} \gamma_{56} + {\rm perm} \right) \, , \cr
M_{n, 2}^{(3)}(\gamma_{ij}) &=  \left( \gamma_{12} \gamma_{34} \gamma_{56} + {\rm perm} \right) \, ,
\fe
where $\gamma_{ij}$ satisfy the constraints $\sum_{j\neq i} \gamma_{ij} = 4$  $\forall i$. The MUV correlators in Mellin space correspond to $n$-point contact terms in $AdS_5 \times S^5$.

In the next section we will make further comments on MUV and non-MUV correlators involving super-descendant operators based on knowledge of instanton contributions to such correlators.

\section{Semi-classical Instanton contributions}
\label{nonmin}

The expression \eqref{eq:MUV} contains information about all possible chiral MUV correlators.  The dependence on the particular operators in a chiral correlator is encoded in  the expansion of $\cI_n(\{x_i,\rho_i, 0, y_i\})$  in powers of $\rho_i$.   
The expression of any of these  correlators  is of the form  \eqref{corn0}, with the same function $\cG_n(x_i;\wt)$ as in the last section.  In this section we will illustrate how semi-classical instanton calculations reproduce detailed information about the form of the instanton sector  of these $n$-point correlators at leading order in the large-$N$ expansion and at leading order in the perturbative expansion in powers of  $g_{_{\rm YM}}$. The semi-classical instanton contributions to these correlators will  be constructed by starting with a lower-point chiral correlator (such as those computed in  \cite{Bianchi:1998nk, Dorey:1999pd}) and appending it with any number of $\cO_\tau(x_i)$ operators.  In the next sub-section we will review specific examples and demonstrate that the results agree with the leading D-instanton contributions to chiral MUV correlators that we obtained using the recursion relations.

We do not have a general expression for the structure of non-chiral MUV correlators, which involve non-zero powers of both $\rho_i$ and $\bar\rho_i$.  However, there is no problem, in principle, in evaluating the semi-classical instanton contributions to such correlators, as will be described in sub-section~\ref{nonchiral}.

Information about the large-$N$ behaviour of other classes of correlators for which we do not have a general expression analogous to  \eqref{eq:MUV} may also be obtained from instanton calculations.  Some examples are   briefly described in section~\ref{others}.  These include the structure of correlators involving $p>2$ $\half$-BPS operators, as well as chiral and non-chiral  correlators that  are non MUV.  These examples are ones in which the instanton profiles of the operators in a correlator depend in important ways on  fermionic moduli beyond those associated with broken superconformal symmetries.   

 Before discussing these examples we will briefly review some general features of the instanton calculations.

\subsubsection*{Generalities concerning  $\cN=4$ SYM  instantons}

Recall that the contributions of Yang--Mills instantons to correlators in  $\cN=4$ SYM involve integration over the 16 exact superconformal fermionic moduli,  $(\eta^A_\alpha,\,  \bar{\xi}^A_{\dot \alpha})$   \cite{Banks:1998nr}.  These correspond to the eight Poincar\'e  supersymmetries and eight conformal supersymmetries that are broken in an instanton background.    While   $(\eta^A_\alpha,\,  \bar{\xi}^A_{\dot \alpha})$ are the only fermionic moduli if the gauge group is $SU(2)$ in the  $SU(N)$  case there are many more fermionic moduli,  almost all of which develop moduli space interactions and so only the 16 superconformal fermionic moduli are exact moduli.  This means that in order for a correlator to have a non-zero instanton contribution, a total of 16 superconformal fermionic  moduli have to be supplied by the instanton profiles of the operators in the correlator. Integration over these results in a functional dependence of MUV  correlation functions on $\{x_i,y_i\}$ and $g_{_{\rm YM}}$ that is independent of $N$.  However,  in order to match the $N$-dependence of the coefficients in the large-$N$ expansion, and to determine properties of multi-instanton contributions, it is important to  consider  the  contribution of instantons in $SU(N)$ gauge theory in the large-$N$ limit. 

The large-$N$ limit of the ADHM construction of the $k$-instanton moduli space was studied in \cite{Dorey:1999pd} using a string-inspired procedure.  This determined the moduli space action for $k$ $SU(2)$ instantons embedded in $SU(N)$.   This  involves large numbers of fermionic and bosonic coordinates associated with the relative orientation and positions in Euclidean space, as well as their relative gauge orientations.    The 16 superconformal fermionic moduli that appear in the $N=2$ case remain exact -- they do not appear in the moduli space action. These are the moduli that are protected by supersymmetry.  As a consequence the instanton measure is independent of these Grassmann coordinates and the instanton contribution will vanish unless the instanton profiles of the operators in the correlator provide the 16 fermionic moduli,   $\eta^A_\alpha$ and  $\bar{\xi}^A_{\dot \alpha}$,  needed to saturate the integral.   
  
  However, the extra super-moduli that appear for $N>2$ do appear in the action and are therefore {\it pseudo}-moduli. Integration over these moduli affects the instanton measure and the expressions for general correlation functions.  
 In a tour de force,  the procedure in  \cite{Dorey:1999pd}  made use of a large-$N$ saddle point technique to integrate over these extra super-moduli.  This demonstrated, among other things, that to leading order in $1/N$ the $k$-instanton moduli space collapses to a point in $AdS_5\times S^5$, which represents a charge-$k$ D-instanton in the holographically dual type IIB superstring. 
 
 The general procedure is reasonably complicated, but simplifies greatly in the  case of a single instanton ($k=1$). In that case the extra  bosonic moduli are those associated with the coset space $SU(N)/SU(N-2)$ that parameterises  the embedding of $SU(2)$ in $SU(N)$.  The fermionic super-partners of these extra bosonic moduli are Grassmann variables  
 $\bar\nu^{A\,u}$ and $\nu^A_u$ where $u=1,\dots, N$, labels the fundamental representation of $SU(N)$. After accounting for constraints satisfied by these coordinates they each have  $4(N-2)$ independent components. 
   Gauge invariant quantities, such as the instanton profiles of composite operators in the stress tensor multiplet,  depend only on the gauge-invariant combination $\sum_{u=1}^N \bar \nu^{Au}\nu^B_u$, which decomposes into two $SU(4)$ representations,
 \ie
 (\bar \nu \nu)^{[AB]}_{\bf 6} = \sum_{u=1}^N \bar \nu^{[Au}\nu^{B]}_u\,, \qquad\quad  (\bar \nu \nu)^{(AB)}_{\bf 10} = \sum_{u=1}^N \bar \nu^{(Au}\nu^{B)}_u\,.
 \label{nudefs}
 \fe 
 
To leading order in $g_{_{\rm YM}}$ only the exact fermionic moduli,   $(\eta^A_\alpha,\,  \bar{\xi}^A_{\dot \alpha})$, in the instanton profiles of operators in the $\cO_2$ multiplet affect MUV correlators. However, when considering non-MUV correlators or correlators of operators in $\cO_p$ multiplets with $p>2$ the extra fermion modes, $\bar\nu^{A\,u}$ and $\nu^A_u$, enter in important ways.
 
These very general comments  are sufficient  for the considerations of this section although this is a  much richer subject 
of intrinsic interest.  We will now present a few examples of instanton contributions to correlators at leading order in perturbation theory (the semi-classical approximation).
  
\subsection{Instanton contributions to  chiral MUV correlators}
\label{instchiral}

In semi-classical instanton contributions of this type there are no contributions from fermionic moduli beyond the 16 broken superconformal  supersymmetries, $\eta^A_\alpha$ and $\bar{\xi}^A_{\dot \alpha}$.

The first example  that  we will consider is the correlator with four $\mathcal{O}_2$ and $m$ $\cO_{\tau}$ insertions.  Following   \eqref{eq:MUV} and \eqref{eq:alpha02},  we find that at the leading power of $c$ (i.e., of $N$), it is given by
\ie \label{eq:4O2-Otau}
& \widehat{G}_{\cO_{2}^{4} \cO_\tau^m}(x_i, y_i; \wt) = \langle \cO_2(x_1,y_1) \cdots \cO_2(x_4,y_4)  \cO_\tau(x_5) \cdots \cO_\tau(x_{m+4}) \rangle \cr
&= c^{\quart} R(1,2,3,4)\, \left(\prod_{1< i<j \leq 4} x_{ij}^2\right) 
\frac{\Gamma(2m+6) \Gamma(m+\threeh) }{ 16 \sqrt{2} \pi^{2m+2} } E_{m}(\threeh, \wt)  D_{\underbrace{\scriptstyle  44\dots 4}_{4+m}} (x_i) + O(c^{-\quart})\,,
\fe
where $ E_{m}(\threeh, \wt)$ is the weight $(m,-m)$ modular Eisenstein series defined in \eqref{zswdef}.
We have only displayed the leading large-$N$ term since that is what emerges from the semi-classical instanton calculus, which we will now review. 

The semi-classical one instanton contribution to this  correlator in the $m=0$ case (the  four-point correlator) was obtained for $SU(2)$ gauge group  in  \cite{Bianchi:1998nk}, and for $k$ instantons to leading order in the large-$N$ limit of  $SU(N)$ in \cite{Dorey:1999pd}.  The latter result used a large-$N$ saddle point method to determine the absolute coefficient in the $k$-instanton sector at leading order in $g_{_{\rm YM}}$.  
After integration over the 16 superconformal fermionic moduli,  the semi-classical $k$ instanton contribution to the four-point correlator  has the form (ignoring an overall numerical factor) \footnote{We have here included the dependence on $N$, $k$ and $\wt_2$ by 
  extending the discussion in  \cite{Dorey:1999pd}, and also the R-symmetry factor $R(1,2,3,4)$ \cite{Bianchi:2013xsa}. } 
\ie
& \widehat{G}_{\cO_2^4}(x_i, y_i; \wt) = \langle \mathcal{O}_2(x_1, y_1)  \mathcal{O}_2(x_2, y_2) \mathcal{O}_2(x_3, y_3) \mathcal{O}_2(x_4, y_4) \rangle \cr
& = c^{\quart}  R(1,2,3,4)  \left(\prod_{1< i<j \leq 4} x_{ij}^2\right) \sigma_{-2}(k) k^{\half} e^{2\pi ik\wt} \! \int\! {d\rho_0 d^4 x_0 \over \rho_0^5}  \prod_{i=1}^4 \left( \rho_0 \over \rho_0^2 + (x_i - x_0)^2 \right)^4 \, .
\fe
Here $(\rho_0, x^{\mu}_0)$ are the bosonic moduli representing the size and the position of the instanton. 

 This  instanton calculation generalises straightforwardly to include $m$ insertions of $\cO_\tau$, each one of which inserts the classical instanton profile,
\ie \label{eq:cOtau}
\cO_{\tau}(x_i) \big{|}_{\rm instanton} \rightarrow {24 \,k\, \wt_2 \over \pi} \left( \rho_0 \over \rho_0^2 + (x_i - x_0)^2 \right)^4 \, .
\fe 
This leads to  
\ie
\widehat{G}_{\cO_2^4 \cO_{\tau}^m}(x_i, y_i; \wt) &= \langle \mathcal{O}_2(x_1, y_1) \cdots  \mathcal{O}_2(x_4, y_4) \cO_{\tau}(x_5) \cdots \cO_{\tau}(x_{m+4}) \rangle  \cr 
&=  c^\quart\,
 k^{\half+m}\,\sigma_{-2}(k)\, ( \wt_2)^m\, R(1,2,3,4)      \left(\prod_{1< i<j \leq 4} x_{ij}^2\right)e^{2\pi ik\wt}   \cr
& ~~~  \int {d\rho_0 d^4 x_0 \over \rho_0^5}  \prod_{i=1}^{m+4} \left( \rho_0 \over \rho_0^2 + (x_i - x_0)^2 \right)^4 \, ,
\label{newcorr}
\fe
which is in agreement with \eqref{eq:4O2-Otau} after using the expansion of $E_m(\threeh, \wt)$ in \eqref{expadef}.

As emphasised in \cite{Green:2002vf} for the $m=0$ case, the contribution of anti-instantons to chiral correlators are suppressed by powers of $g_{_{\rm YM}}$.  This suppression arises from the fact that in an anti-instanton background each operator picks up a number of extra fermionic pseudo-zero modes associated with the non-superconformal moduli. For example, the operator $\cO_\tau$ has an anti-instanton profile with eight fermionic modes.    As a result 
the leading anti-instanton contribution to the  correlator $G_{\cO_{2}^{4} \cO_\tau^m}$ in \eqref{eq:4O2-Otau} 
is of order $(\wt_2)^{-m}$ in contrast to the instanton contribution in \eqref{newcorr}. This is again in agreement with the expansion of the modular form $E_m(\threeh, \wt)$.

Another example of a chiral MUV correlator we will consider is  
\bea
\label{sixteenm}
{\widehat G}_{\Lambda^{16} \cO_\tau^m} (x_i, y_i; \wt) = \langle \Lambda(x_1, y_1)\dots \Lambda(x_{16}, y_{16})  \cO_\tau(x_{17})\dots  \cO_\tau(x_{16+m}) \rangle\, ,
\eea
  where  the operator $\Lambda$ is  the $[0,0,1]_{(\frac{1}{2},0)}$ entry in \eqref{LitM} which is of order $\rho^3$ in $\cT^C(x,\rho,y)$ (and is the holographic dual of the dilatino of the type IIB superstring). When $m=0$, it is an example of higher-point correlators without the insertion of $\cO_{\tau}$.  In this case we need to expand $\cI$ in \eqref{eq:In} to order $\rho_i^3$ for $i=1,\dots,16$ and to order $\rho_i^4$ for $i=17,\dots,  16+m$.   Therefore the correlator is given by 
  \ie
  {\widehat G}_{\Lambda^{16} \cO_\tau^m} (x_i, y_i; \wt) =c^{\quart}\,  \cI_{\Lambda^{16} \cO_\tau^m }(\{x_r,\rho_r, 0,y_r\})  \big{|}_{\prod_{i=1}^{16}\rho^3_i \, \prod_{j=17}^{16+m}\rho^4_j} \, E_{12+m}(\threeh,\wt) \, D_{\underbrace{\scriptstyle  44\dots 4}_{16+m}} (x_i)\, ,
  \fe
where, dropping a multiplicative combinatorial coefficient, the prefactor is given by
\ie
 \label{innew}
 \cI_{\Lambda^{16} \cO_\tau^m }(\{x_r,\rho_r, 0,y_r\})  \big{|}_{\prod_{i=1}^{16}\rho^3_i \, \prod_{j=17}^{16+m}\rho^4_j}  &=\! \int \! d^4 \epsilon\, d^4 \epsilon' \, d^4  \bar{\xi}\,  d^4  \bar{\xi}' \prod_{i=1}^{16} 
  \left(\epsilon^a_{\alpha} + \epsilon^{a'}_{\alpha} y_{i, a'}^a 
 + x_{i, \alpha}^{\dot \alpha} \left(\bar{\xi}^a_{\dot \alpha} +\bar{\xi}^{a'}_{\dot \alpha} y_{i, a'}^a  \right)  \right)\cr
 &=
 \!  \int \! d^8 \eta \, d^8 {\overline \xi}  \prod_{i=1}^{16} 
  \left( \eta^{A_i}_{\alpha_i} 
 + x_{i, \alpha_i}^{\dot \alpha_i}\, \bar{\xi}^{A_i}_{\dot \alpha_i}  \right) g_{A_i}^{a_i}  \, ,
\fe
where we have used the definition of $g_A^a$ in \eqref{gdef} and defined  $\eta_\alpha^A=(\epsilon^a_\alpha,\epsilon^{a'}_\alpha)$ and 
${\overline \xi}^A_{\dot\alpha}= (\bar\xi^a_{\dot\alpha},\bar\xi^{a'}_{\dot\alpha})$, which  encodes the $SU(4)$ representations of the $\Lambda^A_\alpha$ operators.

With $m=0$ this again  has the same form as the semi-classical calculation of the instanton contribution to $\langle \Lambda_1(x_1, y_1) \dots \Lambda_{16}(x_{16}, y_{16})\rangle$ derived in \cite{Bianchi:1998nk} for the $SU(2)$ case and in \cite{Dorey:1999pd} for the $SU(N)$ case at leading order in the large-$N$ limit.  The latter result  determined the absolute coefficient in the $k$-instanton sector at leading order in $g_{_{\rm YM}}$.  This  instanton calculation again generalises very simply to include $m$ insertions of $\cO_\tau$, giving (ignoring an overall numerical factor) 
 \begin{align} 
&	\widehat{G}_{\Lambda^{16}\cO_\tau^m }  (x_i, y_i; \wt) 
    = c^\quart \, (\wt_2)^{12+m} \,  k^{25/2+m}\,  \sigma_{-2} (k)\,  e^{2\pi ik\wt}  \,
    \int \frac{d^{4}x_{0} \, d\rho_0}{\rho_0^{5}}  
	\int d^{8}\eta \, d^{8}{\overline \xi}  \nonumber \\  
	& \prod_{i=1}^{16+m} \left[ 
	\frac{\rho_{0}^{4}}{[\rho_0^{2}+(x_{i}-x_{0})^{2}]^{4}}\right] 
	 \prod_{i=1}^{16}\left[ \frac{1}{\sqrt{\rho_{0}}}
	\left( \rho_0 \eta^{A_{i}}_{\alpha_{i}}+
	{(x_{i}-x_{0})}_{\alpha_{i}{\dot \alpha}_{i}}  
	{\overline \xi}^{{\dot \alpha}_{i}A_{i}} \right)  \right] g_{A_i}^{a_i} \\
	&=  c^\quart  (\wt_2)^{12}   k^{25/2+m}  \sigma_{-2} (k) e^{2\pi ik\wt}    D_{\underbrace{\scriptstyle  44\dots 4}_{16+m}}(x_i)
	\int d^{8}\eta d^{8}{\overline \xi}  \, \prod_{i=1}^{16}  \left( \eta^{A_{i}}_{\alpha_{i}}+
	(x_{i })_{\alpha_{i}{\dot \alpha}_{i}}  
	{\overline \xi}^{{\dot \alpha}_{i}A_{i}} \right) g_{A_i}^{a_i}  \,. \nonumber
	\label{l16-sym} 
\end{align} 
In passing to the second equality we have redefined the eight $\eta$ integration variables to absorb the $\rho_0$ and $x_0$ dependence in the parenthesis involving the Grassmann variables \cite{Bianchi:2013xsa}.  

We see that the fermionic integration of $\widehat{G}_{\Lambda^{16}\cO_\tau^m }  (x_i, y_i; \wt)$  from the instanton calculation is identical to $ \cI_{\Lambda^{16} \cO_\tau^m}$ given in \eqref{innew}, and it also contains the correct $D$-function. The $g_{_{\rm YM}}$-dependent coefficient, as well as its dependence on the D-instanton number $k$, agrees with the expectation from the flat-space amplitude, which has a coefficient proportional to $E_{12+m}(\threeh,\wt)$, \eqref{expadef}, for which the instanton contributions are defined by \eqref{znpddef}.  
 
The last example, considered explicitly in \cite{Bianchi:1998nk}, is the instanton contribution to the correlator
\bea
\label{eightm}
\widehat{G}_{\cE^8 \cO_\tau^m}(x_i, y_i; \wt)= \langle \cE(x_1, y_1)\dots \cE(x_{8}, y_8)  \cO_\tau(x_{9})\dots  \cO_\tau(x_{8+m}) \rangle\, ,
\eea
when $m=0$.   Here  the operator $\cE$ is  the $[0,0,2]_{(0,0)}$ entry in \eqref{LitM}, which is of order $\rho^2$ in $\cT^C(x,\rho,y)$ (and is part of the holographic dual of the complex three-form of the type IIB superstring, where $\cB$ is the other part).
In this case we need to expand $\cI_{\cE^8 \cO_\tau^m}$ in \eqref{eq:In} to give the terms of order $\rho_i^2$ for $i=1,\dots, 8$, and $\rho^4$ for $i=9,\dots, 8+m$.   This gives a factor (again dropping a multiplicative combinatorial coefficient),   
\ie
& \cI_{\cE^8 \cO_\tau^m}  (\{x_r,\rho_r, 0,y_r\})  \big{|}_{\prod_{i=1}^{8}\rho^2_i \, \prod_{j=9}^{8+m}\rho^4_j}   \cr
= &	\int d^{8}\eta\, d^{8}{\overline \xi}  \, \prod_{i=1}^{16}  \left( \eta^{A_{i}}_{\alpha_{i}}+
	x_{i, \alpha_{i}{\dot \alpha}_{i}}  
	{\overline \xi}^{{\dot \alpha}_{i}A_{i}} \right)   \epsilon^{\alpha_i \beta_i}\left( \eta^{B_{i}}_{\beta_{i}}+
	x_{i, \beta_{i}{\dot \alpha}_{i}}  
	{\overline \xi}^{{\dot \alpha}_{i}B_{i}} \right) g_{A_i}^{a_i} g_{B_i}^{b_i} \, .
\label{Ieightm}
\fe
The semi-classical $k$-instanton contribution to the correlator is a straightforward extension to general $m$ of the $m=0$  instanton calculation given  explicitly in  \cite{Bianchi:1998nk} 
\ie
     &{\widehat G}_{{\cE}^8 \cO_\tau^m}(x_i, y_i; \wt)=
    c^{\quart}  \, (\wt_2)^{4+m}\,   k^{\nineh+m} \,   \sigma_{-2} (k)\,  e^{2\pi i k \wt}    \int \frac{d^{4}x_{0} \, d\rho_0}{\rho_0^{5}}    \prod_{i=1}^{8+m} \left[ 
	\frac{\rho_{0}^{4}}{[\rho_0^{2}+(x_{i}-x_{0})^{2}]^{4}}\right]   \\
 &\!\!\!\!\! \!\!\!\!\! 	\int d^{8}\eta d^{8}{\overline \xi}  \prod_{i=1}^{16}  {1\over \sqrt{\rho_0}} \left(\rho_{0} \eta^{A_{i}}_{\alpha_{i}}+
	(x_{i} -x_0 )_{\alpha_{i}{\dot \alpha}_{i}}  
	{\overline \xi}^{{\dot \alpha}_{i}A_{i}} \right)   \epsilon^{\alpha_i \beta_i}  {1\over \sqrt{\rho_0}} \left(\rho_{0}  \eta^{B_{i}}_{\beta_{i}}+
	(x_{i} -x_0)_{\beta_{i}{\dot \alpha}_{i}}  
	{\overline \xi}^{{\dot \alpha}_{i}B_{i}} \right) g_{A_i}^{a_i} g_{B_i}^{b_i}   \,,
  \label{e8-sym} 
\fe
where the prefactor follows implicitly from  \cite{Dorey:1999pd}.  This again agrees with $\cI_{\cE^8 \cO_\tau^m}$ given in \eqref{Ieightm} and the correct $D$-function, as well as the leading instanton contribution obtained from the large-$\wt$  expansion of  the modular form $E_{4+m}(\threeh,\wt)$.

\subsection{Instanton contributions to non-chiral MUV  correlators}
\label{nonchiral}

 There are many non-chiral MUV  $n$-point correlators with $n\geq 4$ that involve products of  chiral,  anti-chiral and mixed chirality $\half$-BPS operators.   All such correlators are holographic duals of scattering amplitudes that are contact interactions.  In the flat-space limit these amplitudes are related by an overall prefactor of $\delta^{16}(\sum_{i=1}^n Q_i)$, where $Q_i$ is the sixteen-component supercharge acting on the $i$th particle (and $\bar Q_i$ would be the conjugate supercharge).\footnote{The definition of the on-shell supercharges $Q_i$ and $\bar Q_i$ in terms of spinor helicity formalism is reviewed in Appendix \ref{holosol}.} Whereas the prefactor $\cI_n(\{x_r,\rho_r, 0, y_r\})$ in \eqref{corn0} plays the same r\^ole for chiral correlators as  $\delta^{16}(\sum_{i=1}^n Q_i)$, we do not know 
 how to generalise this systematically to non-chiral correlators which have non-zero  powers of both $\rho$ and $\bar\rho_r$.   For  that reason
 we do not have a general procedure for determining the coefficients in the large-$N$ expansion

 However, we know that the leading term in the flat-space limit  should reproduce the ten-dimensional super-amplitude, which does not distinguish the chiral from the non-chiral cases.  This  is consistent with the fact that instanton contributions to this subset of correlators  have the same features as in the chiral case, at least in the large-$N$ limit.  In particular, at leading order in $1/N$ and in $g_{_{\rm YM}}$, the only fermionic moduli that are relevant are the 16 superconformal super-moduli, $(\eta^A_\alpha,\,  \bar{\xi}^A_{\dot \alpha})$.   Therefore, the instanton calculations extend straightforwardly to include MUV non-chiral correlators. 
 The only complication is that the instanton profiles of operators that accompany powers of $\bar\rho$  involve a greater number of factors of  $\eta^A_\alpha,$ and $\bar{\xi}^A_{\dot \alpha}$.

 One example, among many,  with $n=4$ is obtained  by  starting from the non-chiral four-point correlator $\langle  \bar\cO_{\bar \tau} \bar\cO_{\bar \tau}  \cO_{\tau}  \cO_{\tau} \rangle$, the structure of which was considered in \cite{Brodie:1998ke, Goncalves:2014ffa}.  Appending it with $m$ $\cO_\tau$ operators leads to the $(4+m)$-point  correlator of order $\rho^{8+4m}\,\bar\rho^8$,
 \ie
\widehat G_{\bar \cO_{\bar \tau}^2 \cO_\tau^{2+m}}  (x_i; \wt) = \langle \bar\cO_{\bar \tau} (x_1) \bar\cO_{\bar\tau}(x_2)  \cO_{\tau}(x_3) \cO_{\tau}(x_4)   \dots  \cO_{\tau} (x_{m+4})\rangle\,.
\label{fourtatus}
\fe

The correlator is related to $\langle  \bar\cO_{\bar \tau} \bar\cO_{\bar \tau}  \cO_{\tau}  \cO_{\tau} \rangle$ by the recursion relations \eqref{aetertwo}. 
Each factor of $\bar\cO_{\bar\tau}$ has an instanton profile containing the product of  eight fermionic superconformal moduli  $(\eta^A_\alpha,\,  \bar{\xi}^A_{\dot \alpha})$ and the leading semi-classical instanton contribution to this correlator may be evaluated in the same manner as for the chiral cases in the last sub-section.

Another  example starts with the non-chiral  $n=5$ MUV correlator  $\langle \cE\,\cE\, \cE\,  \bar\cE \, \cO_2 \rangle$. 
Each $\cE$ operator contains the product of 2 superconformal fermionic moduli, while $\bar \cE$ contains 6 and $\cO_2$ contains 4, so the correlator saturates the total of sixteen fermionic moduli, as in the previous cases.
 Again, adding $m$ $\cO_\tau$ operators  leads to the $(5+m)$-point MUV correlator  of order $\rho^{6+4m}\bar\rho^2$, 
\ie 
\langle \cE (x_1,y_1) \cE(x_2,y_2) \cE(x_3,y_3) \bar\cE(x_4,y_4)\cO_2(x_5,y_5)\, \cO_\tau(x_6)\dots \cO_\tau(x_{5+m})\rangle \, ,
\label{examp2s}
\fe
which is related to  $\langle \cE\,\cE\, \cE\,  \bar\cE \, \cO_2 \rangle$ by  \eqref{aetertwo}.

\subsection{Instanton contributions  to other classes of  correlators}
\label{others}

There are several situations in which the instanton contributions not only involve the 16  superconformal fermionic moduli, but the extra fermionic moduli $\bar \nu^{Au}$, $\nu^A_u$ in the instanton profiles of $\half$-BPS operators also play an important r\^ole.  These will be outlined in this sub-section with few details.  Much of this material is a straightforward extension of  \cite{Green:2002vf}.

\subsubsection{MUV correlators with higher Kaluza--Klein charges}

The spectrum of type IIB supergravity after compactification on $AdS_5\times S^5$ includes Kaluza--Klein excitations of the massless fields.  These correspond holographically to  gauge-invariant operators in multiplets with superconformal primaries $\cO_p$ with $p>2$.
The instanton profiles of such operators involve products of   $\eta^A_\alpha,\,  \bar{\xi}^A_{\dot \alpha}$, and one factor of  the extra-mode bilinear, $(\bar \nu \nu)^{[AB]}_{\bf 6}$ for each Kaluza--Klein charge.  To leading order in $1/N$ the  Kaluza--Klein charges serve to restore the ten-dimensional Poincar\'e invariance of the flat-space theory. Therefore, the leading  large-$N$ behaviour of correlators that are dual to Kaluza--Klein charges should reproduce the corresponding type IIB super-amplitude in flat ten-dimensional space-time.  

For example, the gauge-invariant operator, $\Lambda^*$, that is dual to the dilatino with a single Kaluza--Klein excitation has an instanton profile linear in  superconformal fermionic moduli and has a single factor of $(\bar \nu \nu)^{[AB]}_{\bf 6}$.  Therefore, the   correlator $\Lambda^{14} {\Lambda^*}^2\cO_\tau^m$ has a total of 20 fermionic zero modes and  is related by recursion to the  $m=0$ case.  The leading semi-classical instanton contribution to this correlator matches the flat-space amplitude in the large-$N$ limit, as expected  \cite{Green:2002vf}.

\subsubsection{Chiral non-MUV $n$-point correlators}

The class of  non-MUV correlators of chiral operators is one in which the product of instanton profiles of operators  again involves more that 16 fermionic moduli.  In such cases there are extra factors of $(\bar \nu \nu)^{[AB]}_{\bf 10}$ that are important in determining the properties of the large-$N$ limit.  
Unlike in the case of MUV correlators, in non-MUV cases the large-$N$  limit is dual to scattering  amplitudes that have poles.  

There are many examples of such correlators, two of which are
$G_{\Lambda^{14} \chi^2\cO_\tau^m}$, which involves a total of $20$ fermionic moduli,  and $G_{\Lambda^{16} \cO_2^2\cO_\tau^m}$, which involves a total of  $24$ fermionic moduli.
In these examples the analysis of the instanton behaviour is more complicated, which is associated with the presence of poles in    the holographic dual scattering amplitudes \cite{Green:2002vf}.

\subsubsection{Non-MUV non-chiral  $n$-point correlators}

This class of correlators does not, in  general,  have simplifying  features although special cases share features with the preceding examples.  Of particular interest is the correlator obtained by differentiating a MUV $n$-point correlator with respect to $\bar \tau$.  We saw earlier in \eqref{relations2}  that this inserts a factor of $\int dx_{n+1} \bar\cO_{\bar\tau}(x_{n+1})$  into the   MUV correlator.  The semi-classical instanton calculation we discussed previously can be generalised by taking into account the instanton profile of $\bar O_{\bar\tau}$, that includes a factor of eight fermionic moduli, so such a correlator has a total of 24 fermionic moduli (including the 16 superconformal moduli).   This illustrates how the recursion relation can impose integral constraints on  non-MUV non-chiral correlators by relating them to MUV correlators.

\section{Discussion} 
\label{discuss}

This paper has considered properties of the large-$N$ expansion of $n$-point correlators in  $\cN=4$ SYM  with gauge group $SU(N)$ that violate the bonus  $U(1)_Y$ maximally.  
 The maximal $U(1)_Y$ charge violation for a $n$-point correlator is $2(n-4)$.    These correlators are holographically dual to maximal $U(1)_Y$-violating scattering amplitudes of $n$ massless states in ten-dimensional type IIB superstring theory compactified on $AdS_5\times S^5$.   The large-$N$ expansion of the correlators with fixed Yang--Mills coupling, $\wt$,  corresponds to the expansion of the string theory amplitudes in powers of $\alpha' {\bf s_{ij}}$ and $\alpha'/L^2$ (where $L$ is the $AdS_5$ scale) with fixed background value of the complex 
scalar, $\tau^0=\wt$.

The systematics of $U(1)_Y$-violation is understood on both sides of the holographic correspondence.  In $\cN=4$ SYM it is governed  by the fact that the ``bonus $U(1)_Y$'' is broken to $\ZZ_4$, which is the centre of the R-symmetry $SU(4)$.     Likewise, only a $\ZZ_4$ remnant of the type IIB supergravity $U(1)$ R-symmetry survives in the type IIB superstring. This $\ZZ_4$ is compatible with its embedding in the discrete $SL(2,\ZZ)$ duality group  and invariance under  $\ZZ_4$ restricts the possible $U(1)_Y$-violating terms.

The  $n$-point MUV correlators that we have explicitly studied in this paper are those involving the ``chiral'' operators in the stress tensor multiplet that are related to the superconformal primary $\cO_2$ by successive applications of the four chiral supercharges $Q_\alpha^a$ that lie in one $SU(2)$ subgroup of the R-symmetry $SU(4)$.  This includes, for example, the $n$-point correlators in \eqref{relevant},  \eqref{sixteenm},   and \eqref{eightm}. 
 Chiral MUV correlators have the property that the dependence on $\rho_r$ and $y_r$  factors out as in \eqref{corn0}, where the function $\cI(x_r,\rho_r,0,y_r)$ is given in \eqref{eq:In} (and was derived in \cite{Eden:2011yp}).\footnote{There is an equaivalent class of ``anti-chiral'' correlators obtained by successive supersymmetry transformations of $\cO_2$ by $Q^{a'}_{\dot \alpha}$. }

We used the integrated recursion relation \eqref{eq:soft},  together with recent results concerning four-point correlators of superconformal primaries  \cite{Chester:2019jas,Chester:2020vyz}, to  determine coefficients in the large-$N$ expansion of  $n$--point chiral MUV correlators with any  value of $n\geq 4$.   Explicit results were obtained for the modular covariant coefficients of  the first three orders   in the $1/N$ expansion beyond the supergravity limit. These coefficients, as summarised in section~\ref{sec:summary}, which are $SL(2, \mathbb{Z})$ modular forms  with modular weights $(w,-w)$ (where $2w=q_U=2n-8$),  which determine the exact perturbative and non-perturbative dependence on the Yang--Mills coupling and $\theta$ angle.  These modular forms, which  implement Montonen--Olive duality in the gauge theory, are  the same kind of  modular forms  that entered into the construction of S-dual  MUV amplitudes in flat-space type IIB string theory in \cite{Green:2019rhz}, and the Mellin representation of the correlators reproduces the corresponding flat-space string amplitudes in the flat-space limit.  The present results give a certain amount of information about   $\alpha'/L^2$ corrections to the  low energy expansion of MUV string amplitudes  beyond the flat-space limit of $AdS_5\times S^5$.

Non-chiral MUV correlators involve operators that are obtained by successive transformations of $\cO_2$ by both  $Q^{\alpha}_a$  and $\bar Q^{\dot\alpha}_{a'}$.     Although our understanding of the structure of the chiral cases is quite general
 this is not true of the non-chiral cases.  
 Since non-chiral  MUV correlators are related by 
 superconformal symmetry to chiral MUV correlators,  given the simplicity of chiral MUV correlators, we would expect that one may  directly construct these non-chiral  MUV correlators  from the chiral ones. 
Although this  has not been determined in general, suggestions for addressing this issue were proposed in \cite{Chicherin:2016fbj} for correlators at the Born level.  It would be interesting to  explore whether the methodology of \cite{Chicherin:2016fbj} can be generalised to the holographic correlators. Furthermore, we also expect a factor analogous to $\cI_n$  to factor out in a systematic manner. 
In the absence of a general analysis, the discussion in section~\ref{nonmin} illustrates how semi-classical instanton calculations may be used to determine some features of both chiral and non-chiral MUV $n$-point correlators. In particular, the instanton analysis is useful for analysing non-chiral MUV correlators, such as the examples listed in section~\ref{nonchiral}.  
 
The following are other obvious avenues to explore.

Although we have concentrated  on MUV correlators of operators in the stress tensor supermultiplet the   recursion relation relation applies more generally.  It would be of interest to understand  the extent to which the systematics of bonus $U(1)_Y$-violation applies to more general correlators.  For example, it seems likely that correlators of  $\half$-BPS operators in supermultiplets with $p>2$ have well-defined $U(1)_Y$-violating selection rules, generalising those of the $p=2$ case.  Such operators are holographically dual to higher Kaluza--Klein modes on $S^5$.

Another direction to explore is the extension from maximal $U(1)_Y$-violating to non-maximal $U(1)_Y$-violating correlators, which correspond to Mellin amplitudes that  have poles in $\gamma_{ij}$.  A simple class of such correlators consists of   $n$-point next-to-maximal $U(1)_Y$-violating (NMUV) correlators, which violate $U(1)_Y$ by $2(n-5)$ units. The residue of a pole in a corresponding NMUV Mellin amplitude factorises into the product of a $(n-1)$-point MUV amplitude and a three-point supergravity amplitude.  This is reminiscent of properties of maximal helicity-violating (MHV) amplitudes  in $\cN=4$ SYM, which have played a prominent r\^ole in the modern developments of scattering amplitudes.

Our analysis made  use of recent results concerning the large-$N$ expansion of the supersymmetric localisation  of the correlator of four $\cO_2$ operators  in 
\cite{Binder:2019jwn, Chester:2019jas, Chester:2020dja, Chester:2020vyz} as initial data in the recursion relation.  An alternative  procedure would be to extend the localisation  analysis to directly determine the large-$N$ expansion of correlators  of $U(1)_Y$-violating $n$-point correlators, with four $\cO_2$ and  $(n-4)$ $\cO_\tau$ operators.  This involves  generalising the analysis of the four-point correlator in \cite{Binder:2019jwn, Chester:2019jas, Chester:2020dja, Chester:2020vyz}  by considering higher numbers of derivatives on the $\cN=2^*$ partition function. 

Another particularly interesting  challenge is to directly determine the action of the Laplace operator on the modular form coefficients in the $1/N$ expansion of MUV correlation functions.  This would amount to a more consistent implementation of the procedure in \cite{Basu:2004dm,Basu:2004nt}.  The $\wt$ Laplace operator acting on a $n$-point MUV correlator is  $\Delta_{(+)w}= 4 \bar \cD_{-w-1} \cD_w$ (see \eqref{laplaceplusn}), where the modular weight  is $w=n-4$.  Acting first with 
$\cD_w$ brings down a factor of $\int d^4 z_1 \cO_\tau(z_1)$. So the correlator becomes a $(n+1)$-point MUV correlator integrated over the $(n+1)$th position,  $z_1$.  Next, applying $\bar  \cD_{-w-1}$  inserts a factor of $\int d^4z_2 \bar\cO_{\bar \tau}(z_2)$, which takes it into a ``next-to-next-to-maximal'' $U(1)_Y$-violating (NNMUV) correlator with $n+2$  operators integrated over both $z_1$ and $z_2$.  We know that this must result in the appropriate Laplace equations for the coefficients of terms in the $1/N$ expansion.  However, in order to understand this in detail we need to  understand properties of  NNMUV correlators integrated over $z_1$ and $z_2$.  Such correlators violate $U(1)_Y$ by $2(n-6)$ units.  They correspond to Mellin amplitudes that have poles with residues that factorise into the product of two MUV amplitudes (both with $n\ge 4$ points) or into the product of a three-point supergravity amplitude and a  NMUV $(n-1)$-point amplitude.

Finally, we  have stressed that the recursion relations \eqref{eq:soft} take a form that is reminiscent of   the soft dilaton relation of flat-space type IIB superstring amplitudes studied in  \cite{Green:2019rhz}, which relate a $(n+1)$-particle amplitude with one soft dilaton to a $n$-point amplitude.  However, soft theorems of flat-space amplitudes have much wider applicability  to other theories and in diverse space-time dimensions. They have, for example, played important r\^oles in the bootstrapping flat-space scattering amplitudes \cite{Cheung:2015ota, Luo:2015tat, Cheung:2016drk, Bianchi:2016viy, Elvang:2018dco, Cheung:2018oki}.  It would be of interest to explore analogous recursion relations, for instance, in the six-dimensional flat-space amplitudes of type IIB superstring theory compactified on $K3$, where soft theorems probe the moduli structure of $K3$ surfaces \cite{Heydeman:2018dje}. The corresponding holographic correlators are dual to  $AdS_3 \times S^3$ amplitudes, which have recently been studied in the supergravity limit \cite{Giusto:2018ovt, Giusto:2019pxc, Rastelli:2019gtj, Giusto:2020neo}.

\section*{Acknowledgements}

We would like to thank Shai Chester, Paul Heslop, Silviu Pufu, Rodolfo Russo, and Yifan Wang for insightful discussions.    MBG has been partially supported by STFC consolidated grant ST/L000385/1.   CW is supported by a Royal Society University Research Fellowship No. UF160350.

\appendix

\section{Notation and conventions}
\label{notate}

\subsection{Spinor conventions and the elementary $\cN=4$ SYM fields}
\label{a.1}

The spinor indices, are raised and lowered as follows:
\begin{eqnarray}\label{2.57}
  &&\psi^\alpha = \epsilon^{\alpha\beta}\psi_\beta\,, \qquad \bar\chi^{\dot\alpha} =
  \epsilon^{\dot\alpha\dot\beta}\bar\chi_{\dot\beta}\,, \qquad \psi_\alpha = \epsilon_{\alpha\beta}\psi^\beta\,, \qquad \bar\chi_{\dot\alpha} =
  \epsilon_{\dot\alpha\dot\beta}\bar\chi^{\dot\beta}\,, \label{2.58}
\end{eqnarray}
where the $\epsilon$ symbols have the properties:
\begin{equation}\label{2.59}
  \epsilon_{12} = \epsilon_{\dot 1\dot 2} = -\epsilon^{12} = -\epsilon^{\dot 1\dot 2} =
  1\,, \qquad \epsilon_{\alpha\beta}\epsilon^{\beta\gamma} =
  \delta_\alpha^\gamma\,,  \qquad \epsilon_{\dot\alpha\dot\beta}\epsilon^{\dot\beta\dot\gamma} =
  \delta_{\dot\alpha}^{\dot\gamma}\,.
\end{equation}

Vectors are related to spinors using
\begin{align}\label{spinor-vector1}
&x_{\alpha \dot{\alpha}} =
\sigma^{\mu}_{\alpha \dot{\alpha}} x_{\mu}  \,, \qquad x^{\dot{\alpha}\a} =
\tilde\sigma_{\mu}^{\dot{\alpha}\a} x^{\mu}\nn\\
&  x_{\mu}=\frac{1}{2}(\sigma_\mu)_{\a{\dot \alpha}} x^{{\dot \alpha}\a}=\frac{1}{2} \tilde\sigma_{\mu}^{ \dot{\alpha}\alpha} x_{\alpha
\dot{\alpha}} \,, \qquad x^2 = x_{\mu}
x^{\mu} = \frac{1}{2} x_{\alpha \dot{\alpha}}  x^{
\dot{\alpha}\alpha} 
\end{align}
where $\tilde\sigma_{\mu}^{ \dot{\alpha}\alpha} = \ep^{\a\b} \ep^{{\dot \alpha}{\dot \beta}} (\sigma_\mu)_{\b{\dot \beta}}$. 

The component fields of the Yang--Mills field strength supermultiplet are in $\ms\muu(N)$ and their $U(1)_Y$ charge assignments are as follows,
\bea
\label{YMcomp}
&&\phi^{AB} \qquad \lambda_{A \alpha} \qquad \bar \lambda^A_{\dot \alpha} \qquad F_{(\alpha\beta)} \qquad \bar F_{(\dot \alpha\dot \beta)}\ \nn\\
q_{U}: \quad&& 0 \qquad \quad\, \half \qquad\   - \half \qquad 1 \qquad \  \ -1
\eea
The scalar fields $\phi^{AB}$ satisfy the reality condition
\begin{equation}
\phi_{AB} =  \left(\phi^{AB}\right)^\dagger  = \frac{1}{2} \ep_{ABCD}
\phi^{CD},
\end{equation}
where $\ep_{1234} = \ep^{1234} = 1$.

The Yang-Mills field strength is defined by the commutator of two covariant
derivatives, $[D_\mu, D_\nu] = -i F_{\mu\nu}$, where  $D_\mu = \partial_\mu -iA_\mu$ with $A_\mu\in  \ms\muu(N)$.  In spinor notation we have
\begin{equation}\label{comcor}
 F_{\mu\nu} (\sigma^\mu)_{\alpha\dot\alpha} (\sigma^\nu)_{\beta\dot\beta}
=\epsilon_{\alpha\beta}\bar F_{\dot\alpha\dot\beta} + \epsilon_{\dot\alpha\dot\beta} F _{\alpha\beta}\,,
\end{equation}
where $F_{\a\b} = F_{\b\a}$ and $F_{{\dot \alpha}{\dot \beta}} = F_{{\dot \beta}{\dot \alpha}}$ are given by  
\begin{align}
& F^\beta{}_\alpha=-\frac12 F_{\mu\nu} (\sigma^\mu\bar\sigma^\nu)_\alpha{}^\beta
\,,\qquad \bar F_{\dot\beta}{}^{\dot\alpha} = -\frac12F_{\mu\nu} (\bar\sigma^\mu\sigma^\nu)^{\dot\alpha}{}_{\dot\beta}
\end{align}
which implies
\begin{align}
F_{\mu\nu} F^{\mu\nu} =\frac12 \left(F_{\alpha\beta}F^{\alpha\beta} + 
\bar F_{\dot\alpha\dot\beta} \bar F^{\dot\alpha\dot\beta}\right) \, ,
\qquad 
F_{\mu\nu} \widetilde  F^{\mu\nu}  =\frac i 2 \left(F_{\alpha\beta}F^{\alpha\beta} -
\bar F_{\dot\alpha\dot\beta} \bar F^{\dot\alpha\dot\beta}\right) \, .
\end{align}

\subsection{The $\cN=4$ SYM Lagrangian}
\label{a.2}

The $\cN=4$ SYM Minkowski space Lagrangian  has the form
\begin{align}
\cL =& \tr\bigg\{ \frac1{g_{_{\rm YM}}^2} \bigg( -\frac{1}{4} \left(F_{\alpha\beta}F^{\alpha\beta}+ \bar F_{\dot\alpha\dot\beta}\bar F^{\dot\alpha\dot\beta} \right) 
 + \frac14 D_{\alpha\dot\alpha}\phi^{AB} D^{\dot\alpha\alpha}\phi_{AB} + \frac18   [\phi^{AB},\phi^{CD}][\phi_{AB},\phi_{CD}]
\nn\\  
& 
+i \bar\lambda_{\dot\alpha A}D^{\dot\alpha \alpha}  \lambda_\alpha^A 
-i (D^{\dot\alpha \alpha}\bar\lambda_{\dot\alpha A})  \lambda_\alpha^A-\sqrt{2}  \lambda^{\alpha A} [\phi_{AB},\lambda_\alpha^B] +\sqrt{2}  \bar\lambda_{\dot\alpha A} [\phi^{AB},\bar\lambda^{\dot\alpha}_B] \bigg)
\nn\\ 
&
+\frac{i \, \theta_{_{\rm YM}} }{32 \pi^2}  \left(F_{\alpha\beta}F^{\alpha\beta}- \bar F_{\dot\alpha\dot\beta}\bar F^{\dot\alpha\dot\beta}\right)\bigg\}\,.
\label{lagrangedef}
\end{align} 
 
 \subsubsection*{The chiral and anti-chiral Lagrangian operators}

The Lagrangian in \eqref{lagrangedef} 
can be written as the sum of two complex conjugate parts 
\begin{align}
\cL=&-   \frac{i}{2 \wt_2}  \left( \wt \cO_\tau-  \bar \wt  \bar\cO_{\bar \tau}\right)\,, 
\label{loperdef}
\end{align}
where $\wt = \theta_{_{\rm YM}}/2\pi + 4\pi i/g_{_{\rm YM}}^2$.  The composite operators $\cO_\tau$ and $\bar \cO_{\bar\tau}$ are the chiral and anti-chiral Lagrangians that are defined by 
\begin{align}
 \label{otaudef} 
 \cO_\tau = \frac{ \wt_2}{4\pi}  \,  \tr\bigg\{  -{1\over 2}  F_{\alpha\beta} F^{\alpha\beta} 
+\sqrt{2} \lambda^{\alpha A} [\phi_{AB},\lambda_\alpha^B]   - \frac18  [\phi^{AB},\phi^{CD}][\phi_{AB},\phi_{CD}] \bigg\} \, ,
\end{align}
\be
 \label{obartaudef} \bar\cO_{\bar\tau} = \frac{ \wt_2}{4\pi} \,\tr \bigg\{ - {1\over 2} \bar F_{{\dot \alpha}{\dot \beta}} \bar  F^{{\dot \alpha}{\dot \beta}} 
+\sqrt{2} \lambda^{{\dot \alpha}}_ A [\phi^{AB},\lambda_{\dot \alpha B}]   - \frac18  [\phi^{AB},\phi^{CD}][\phi_{AB},\phi_{CD}]  \bigg\} \, .
\ee
In passing from \eqref{lagrangedef} to \eqref{loperdef}  we have used the field equations and dropped terms that are total derivatives, apart from the topological term proportional to $i \, \theta_{_{\rm YM}} \left(F_{\alpha\beta}F^{\alpha\beta}- \bar F_{\dot\alpha\dot\beta}\bar F^{\dot\alpha\dot\beta}  \right) \sim 2 \theta_{_{\rm YM}} F_{\mu\nu} \tilde  F^{\mu\nu}$, which is non-zero in an instanton background and plays a key r\^ole in the structure of the correlators.  
 
 Note that after substituting the solution of the equations of motion for $\lambda_A^\alpha$,  $\lambda^A_{\dot\alpha}$ and $\phi_{AB}$  into the  Lagrangian \eqref{lagrangedef} the dependence on these fields vanishes apart from boundary terms.  In other words, the dependence on these fields in $\cO_\tau$ and $\bar \cO_{\bar\tau}$  (in \eqref{otaudef} and \eqref{obartaudef}) cancels in the combination \eqref{loperdef} modulo equations of motion.    This means that the condition that \eqref{loperdef} reproduce the on-shell Lagrangian in euclidean space does not uniquely determine the terms involving these fields in $\cO_\tau$ and $\bar \cO_{\bar\tau}$.    However, these expressions are uniquely determined by applying four chiral (or four anti-chiral) supersymmetry transformations to $\cO_2$.

\subsection{Short supermultiplets}
\label{shorts}

The  superconformal primary, $\cO_p$, of a short BPS multiplet is proportional to $[\tr \varphi^{(I_1}\dots \varphi^{I_p)}]_{[0,p,0]}$.   The super-descendants are characterised, following the discussion in  \cite{Dolan:2002zh} and  \cite{Gunaydin:1984fk} 
 by their $SU(4)$ Dynkin labels  $[k,p,q]$  and their spin labels $(j,j')$ under $SO(1,3) \approx SU(2)_L\times SU(2)_R$.  The dimension of such short multiplets is $64p^2(p^2-1)/3$. 
The primary operator $\cO_p$ has  zero $U(1)_Y$ charge.  The superconformal descendant components are obtained by acting with powers of the supercharges $Q^I_\alpha$ (which have $U(1)_Y$ charge $\half$) and 
$\bar Q_{I\dot \alpha}$ (which have $U(1)_Y$ charge $-\half$)  on $\cO_p$. Therefore,  any descendant has a well-defined $U(1)_Y$ charge that is determined by the number of supersymmetry transformations that relate the component to the primary $\cO_p$.  

 The states of the $p=2$ multiplet, which contains the stress tensor as well as the  super-currents and the R-symmetry current  are shown in \eqref{LitM}.   These correspond to terms in the expansion of  $\cT(x, \rho, \bar{\rho}, y)$ (defined in \eqref{O21})  in powers of $\rho$ and $\bar \rho$.    The  sequence of red arrows indicates successive $\rho_\alpha^a\,Q^\alpha_a$ transformations associated with the chiral operators in \eqref{O22}, which are terms in the expansion in powers of $\rho$ with $\bar \rho=0$.  Likewise the blue arrows indicate the sequence of operators generated by successive  applications of   $\bar \rho_{\dot \alpha}^{a'}\,\bar Q^\alpha_{a'}$ to $\cO_2$.  Note that the multiplet terminates after four supersymmetry transformations. Further supersymmetry transformations generate states that vanish upon using the equations of motion.  
 
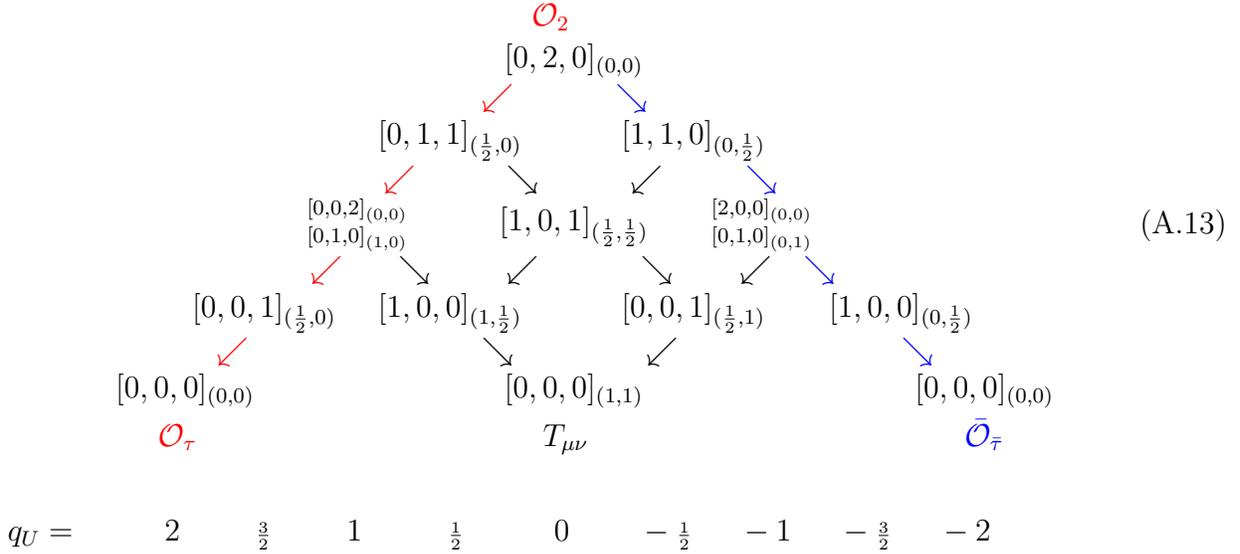
\bea
 \label{LitM}
&&\qquad\qquad\qquad  \qquad\qquad \ \color{red}    \cO_2  \nn\\
&&
\begin{matrix}
 &  &  &  &\!\!\!\!\!\!\!\!\!\!\!\!\!\!\!\!\!\! [0,2,0]_{(0,0)} & & & &  \\

&  &  &{\color{red} \swarrow}  &~~~~ \!\!\!\!    {\color{blue}\searrow} & & & &  \\

& & &\!\!\!\!\!\!\!\!\!\!\!\!\!\!\!\!\!\![0,1,1]_{(\frac{1}{2},0)} &
&\!\!\!\!\!\!\!\!\!\!\!\!\!\!\!\!\!\![1,1,0]_{(0,\frac{1}{2})}&&&  \\

& &\color{red} \swarrow & ~~~~\searrow &~~~~\swarrow &~~~~ {\color{blue}\searrow} & & &  \\

&  &\!\!\!\!\!\!\!\!\!\!\!\!\!\!\!\!{[0,0,2]_{(0,0)} \atop
[0,1,0]_{(1,0)} } & &\!\!\!\!\!\!\!\!\!\!\!\!\!\!\!\!\!\!
[1,0,1]_{(\frac{1}{2},\frac{1}{2})}  & &
\!\!\!\!\!\!\!\!\!\!\!\!\!\!\!\!\!\!\!\!\!\!{ [2,0,0]_{(0,0)} \atop
[0,1,0]_{(0,1)} }  & &  \\

& ~~\color{red} \swarrow& ~~\searrow &~~~~\swarrow & ~~~~~~\searrow &~~\swarrow &
 {\color{blue}\searrow}  & &  \\

&\!\!\!\!\!\!\!\!\!\!\!\!\!\!\!\!\!\![0,0,1]_{(\frac{1}{2},0)} &
&\!\!\!\!\!\!\!\!\!\!\!\!\!\!\!\!\!\!
[1,0,0]_{(1,\frac{1}{2})} &  &
\!\!\!\!\!\!\!\!\!\!\!\!\!\!\!\!\!\!
[0,0,1]_{(\frac{1}{2},1)} & &\!\!\!\!\!\!\!\!\!\!\!\!\!\!\!\!\!\!
~~~~~~[1,0,0]_{(0,\frac{1}{2})}&  \\

\color{red}\swarrow& & & \searrow &~~~~~~~\swarrow & &
& {\color{blue}\searrow}  &\\

\!\!\!\!\!\!\!\!\!\!\!\!\!\!\!\!\!\![0,0,0]_{(0,0)}& &
\!\!\!\!\!\!\!\!\!\!\!\!\!\!\!\!\!\!&
&\!\!\!\!\!\!\!\!\!\!\!\!\!\!\!\!\!\![0,0,0]_{(1,1)}& &
\!\!\!\!\!\!\!\!\!\!\!\!\!\!\!\!\!\!&
&\!\!\!\!\!\!\!\!\!\!\!\!\!\!\!\![0,0,0]_{(0,0)}
\end{matrix}
\\
&& \hskip -.7cm  {\color{red} \cO_\tau}  \hskip 4.6cm T_{\mu\nu}\hskip 5.0cm  {\color{blue} \bar \cO_{\bar \tau}}\nn\\
\nn \\
&& \hskip -  2.7cm   q_U=   \hskip 1.1 cm  2   \hskip 1.0 cm \threeh  \hskip 1.0 cm 1  \hskip 1.0cm  \ \half  \hskip 1.2cm 0   \hskip 0.9cm -\half \hskip 0.6cm -1  \hskip 0.6cm -\threeh \hskip  0.6cm -2  \nn
\eea
The assignment of $U(1)_Y$ charges to the operators in the stress tensor supermultiplet, corresponds to the assignment in  \cite{Intriligator:1998ig} and \cite{Basu:2004dm}.

 \subsubsection*{The chiral stress tensor operators}
 \label{stressops}

  The chiral operators connected by the red arrows in \eqref{LitM}, are the components of the  $\rho$ expansion with $\bar \rho=0$ in \eqref{O22}.  These are gauge-invariant composite operators that are given by the following expressions in terms of the component fields in the $\cN=4$ Yang--Mills super-multiplet \cite{Eden:2011yp}, 
Defining 
\ie
{\phi}^{A a} = \phi^{AB} g^{a}_B \, , \qquad  \lambda^{ a}_\a = \lambda^{A}_{\a}  g^{a}_A \, , \qquad \phi  = - \half g^{a}_A g^{b}_B \epsilon_{ab} \phi^{AB} \, , 
\fe
with $g_A^b = (\delta^b_a, y^b_{a'}) $ as given in \eqref{gdef}, we have  

\begin{align}
\rho^0 \, :\quad\  & \cO_2 =  \frac{1}{g^2_{_{\rm YM}}} {\rm tr} \left(\phi \, \phi  \right)\,, \cr
\rho^1 \, :  \quad \ & \cX^\a_a = \frac{1}{g_{_{\rm YM}}^2} \tr\left( \lambda^{ \a}_a \, \phi \right)\,, \cr
\rho^2 \, : \quad \ & \cE_{(a\, b)}  =  \frac{1}{g_{_{\rm YM}}^2}\tr \left( \lambda^{ \a}_{(a} \lambda_{b) \a}- \sqrt{2} [ \phi^{A}_{(a} , \bar{\phi}_{A, b)} ] \, \phi \right)\,, \cr
\rho^2  \, : \quad \ &  \cB^{(\a\b)}  =\frac{1}{g_{_{\rm YM}}^2} \tr\left( \lambda^{a (\a} \lambda^{\beta)}_{a}
- i \sqrt{2} F^{\a \beta} \phi \right)  \,, \cr
\rho^3 \, : \quad \ &\Lambda^\a_a  = \frac{1}{g_{_{\rm YM}}^2} \tr \left( 
F^{\a}_{\beta} \lambda^{\beta}_a + i \left[{\phi}^{A}_a, \bar \phi_{AB} \right] \lambda^{B \a} \right) \,, \cr
\rho^4  \, :\quad \ &
 \cO_\tau =\frac{1}{g_{_{\rm YM}}^2}   \,  \tr\bigg(  - {1\over 2} F_{\alpha\beta} F^{\alpha\beta} 
 + \sqrt{2} \lambda^{\alpha A} [\phi_{AB},\lambda_\alpha^B]   - \frac18  [\phi^{AB},\phi^{CD}][\phi_{AB},\phi_{CD}] \bigg)\,,
\label{Lamnewdef}
\end{align}
where we have exhibited the power of $\rho$ associated with each operator.

\section{Mellin amplitudes and the flat-space limit }
\label{mellinflat}

The scattering amplitude in $AdS_{d+1}$ is obtained in terms of the Mellin transform of the correlator, which has the form \cite{Penedones:2010ue} 
\bea
\label{mellindef}
\mathcal{G}(x_i) =  \frac{1}{(2\pi i)^{n(n-3)/2}  }\int_C  \prod_{i<j}^n d\gamma_{ij} \, M(\gamma_{ij})  \prod_{i<j} (x_{ij}^2)^{-\gamma_{ij}} \Gamma(\gamma_{ij})  \, ,
\eea
where the integration contours $C$ run parallel to the imaginary $\gamma_{ij}$ axis. The Mellin variables $\gamma_{ij} = \gamma_{ji}$, and are constrained by the conditions
\bea
\sum_{j\neq i}  \gamma_{ij}= \Delta_i \, , \qquad \forall i\,.
\label{gammacons}
\eea
This can be re-expressed as
\bea
\label{deltak}
\gamma_{ij}=k_i\cdot k_j=\frac{\Delta_i+\Delta_j -s_{ij}}{2}\,,
\eea
 with  $s_{ij}= - (k_i+k_j)^2$, where  $k_i^\mu$ are Minkowski vectors satisfying $\sum_i k_i=0$ and   $k_i^2=-\Delta_i$.  
 
 The quantities $s_{ij}$ play the r\^ole of Mandelstam invariants in the context of finite $AdS$ radius.  In taking the flat space limit one has to rescale these variables and recover the scattering amplitude as a function of the conventional Mandelstam invariants ${\bold s}_{ij}$.  More precisely, the flat-space amplitude $T({\bold s}_{ij})$ is related to $M(s_{ij})$ by
 \bea
 M(s_{ij}) \underset {s_{ij}\to \infty}{\approx} {R^{n(1-d)/2+d+1} \over \Gamma \left( {\sum_i \Delta_i -d \over 2} \right)} \int_0^\infty d\beta\,\beta^{\half\sum_i\Delta_i-\frac{d}{2}-1}e^{-\beta} T({\bold s}_{ij})\big|_{{\bold s}_{ij}=2\beta s_{ij}/R^2}\,,
  \label{tdefs}
 \eea
 which may be inverted to give an expression for the flat-space amplitude in terms of a limit of the Mellin transformed correlator,
\bea
\label{Mellinv}
T({\bold s}_{ij})=  \frac{\Gamma \left( {\sum_i \Delta_i -d \over 2} \right)}{R^{n(1-d)/2+d+1}}  \lim_{R\to \infty} \int_{-i\infty}^{i\infty}\frac{d\alpha}{2\pi i}\alpha^{\half(d-\sum_i\Delta_i)}e^\alpha\, M(s_{ij})  \big|_{s_{ij}=\frac{R^2}{2\alpha}{\bold s}_{ij}}\,,
\eea
where the integration contour passes to the right of all the poles of the integrand. Therefore the flat-space amplitude is obtained from the Mellin amplitudes by taking the Mellin variables to infinity.

\subsection{Embedding in $SO(1,d+1)$}

In order to describe a conformally invariant theory, which has $SO(2,d)$ symmetry it is useful,  following Dirac \cite{Dirac:1936fq} , to use coordinates in $(d+2)$-dimensional space of signature $(2,d)$.  In CFT it is conventional to continue to euclidean four-dimensional signature, which results in $SO(1,d+1)$ symmetry and coordinates $X\in \M^{d+2}$ subject to the constraints $X^2=-R^2$, $X^0>0$ that parameterise euclidean $AdS_d$.  The boundary coordinates are efficiently described in terms of a null vector $P\in \M^{d+2}$  satisfying $P^2=0$ and $P\sim \lambda P\, (\lambda\in \RR)$.  The boundary point $P$ and the bulk point can be parametrised as
\bea
P =  (1, x^2, x^\mu)\, ,  \quad X= {1\over z_0} (1, z_0^2 +z^2, z^{\mu}) \, ,
\label{lcP}
\eea
where $x^\mu$ is a $d$-dimensional vector and we have set the radius $R=1$.  We see therefore that
\bea
P_{ij} = (P_i-P_j)^2=-2P_i\cdot P_j =(x_i-x_j)^2\, , \quad -2P \cdot X = {1\over z_0} \left( z_0^2 +(x - z)^2 \right)\, .
\label{moreP}
\eea

\subsection{$D$-functions and Mellin amplitudes}
\label{mellind}

The $D$-functions are defined as\footnote{ The normalisation factor is chosen so that the Mellin transforms of $D$-functions are simple.   }
\ie \label{eq:D-function}
D_{\Delta_1  \Delta_2 \, \cdots \, \Delta_n} (x_i) ={2 \prod_{i=1}^n \Gamma(\Delta_i) \over \pi^h \Gamma \left( {\sum_{i=1}^n \Delta_i  \over 2} -h\right)    } \int_{AdS} dz  K_{\Delta_1} (x_i; z) K_{\Delta_2} (x_i; z) \cdots K_{\Delta_n} (x_i; z)\, , 
\fe
where $h=d/2$, and the bulk-to-boundary propagator is defined as
\ie
K_{\Delta} (x_i ; z)  = \left( {z_0 \over  z_0^2 + (x_i -z)^2} \right)^{\Delta} \,  , 
\fe
which can be expressed  in terms of the six-dimensional embedding space formalism as
\ie
K_{\Delta} (P_i; X)  =  \left( {1 \over  -2 P_i \cdot X} \right)^{\Delta} \, .
\fe
The $n$-point function of a contact Witten diagram without derivatives is given by $D$-functions,
\ie
A_n^{(0)}(x_i) = D_{\Delta_1\, \Delta_2\, \cdots\, \Delta_n} (x_i) \, .
 \fe
In terms of embedding coordinates, the $D$-function can also be expressed as
\ie
D_{\Delta_1 \Delta_2\, \cdots\, \Delta_n} (P_i) 
={2  \over \pi^h \Gamma \left( {\sum_{i=1}^n \Delta_i  \over 2} -h\right)    }\int_0^{\infty} {dt_1 \over t_1} t_1^{\Delta_1} \cdots
\int_0^{\infty} {dt_n \over t_n} t_n^{\Delta_n} \int_{\rm AdS} dX     e^{-2 Q \cdot X} \, ,
\fe
After the integration over $X$, using
\ie
\int_0^{\infty} \prod_i {d t_i \over t_i} t^{\Delta_i } \int _{AdS} dX e^{2 T \cdot X} = \pi^h \Gamma \left( {\sum_{i=1}^n \Delta_i  \over 2} -h\right)  \int_0^{\infty} \prod_i {d t_i \over t_i}  \, e^{T^2} \, ,
\fe
we find
\bea \label{eq:Mellin0}
D_{\Delta_1 \Delta_2 \, \cdots \, \Delta_n} (P_i) &=&2
 \int_0^{\infty} {dt_1 \over t_1} t_1^{\Delta_1} \cdots
\int_0^{\infty} {dt_n \over t_n} t_n^{\Delta_n}  e^{-\sum_{i<j} t_i t_j P_{ij}} \nn\\
&= & \int_C  \prod_{i<j}^n {d \gamma_{ij} \over 2\pi i }  \Gamma(\gamma_{ij} )    (P_{ij})^{-\gamma_{ij}} \, ,
\eea
where we have used the Symanzik formula \cite{Symanzik:1972wj} in the last step.  From the definition of the Mellin transform in \eqref{mellindef} we conclude that the Mellin transform  of  a D function is simply given by  $\cM(\gamma_{ij})=1$.   

Now consider terms with derivatives, for instance $(x_{12}^2)^{\alpha} D_{\Delta_1  \Delta_2 \, \cdots \, \Delta_n} (x_i)$, which can be expressed as  
\ie \label{eq:Mellin1}
A_{n}^{(\alpha; 12)}(P_i) &=2\, (P_{12})^{\alpha}  \int_0^{\infty} {dt_1 \over t_1} t_1^{\Delta_1} \cdots
\int_0^{\infty} {dt_n \over t_n} t_n^{\Delta_n}   \, e^{-\sum_{i<j} t_i t_j P_{ij}} \cr 
&=  \int_C \prod {d \gamma_{ij} \over 2\pi i } \,  \prod_{i<j}^n \Gamma(\gamma_{ij} )    (P_{ij})^{-\gamma_{ij} } (P_{12})^\alpha \cr 
&=  \int_C \prod {d \gamma_{ij} \over 2\pi i } \,  \prod_{i<j}^n \Gamma(\gamma_{ij} )  (P_{ij})^{-\gamma_{ij} } {  \Gamma(\gamma_{12} + \alpha )  \over  \Gamma(\gamma_{12})} \, ,
\fe
where the last equality follows after the change of integration variables $\gamma_{12}\to \gamma_{12}+\alpha$, and the contours $C$ run parallel to the imaginary axis.  So the Mellin amplitude is $M_{n}^{(\alpha;12)}(\gamma_{ij}) = \Gamma(\alpha) (\gamma_{12})_{\alpha}$  (where $(x)_\alpha$ is the Pochhammer symbol), with the following conditions, 
\ie
\sum_{j \neq i} \gamma_{ij}  =\Delta'_i \, , \qquad \Delta'_{i} =  \Delta_{i} - (\delta_{i1} + \delta_{i2}) \alpha \, .
\fe

\section{Type IIB superstring amplitudes and modular forms}
\label{holosol}

Low-energy expansions  of  MUV $n$-particle amplitudes in type IIB superstring theory were studied in \cite{Green:2019rhz}.  The terms up to dimension $14$ (the same order as $d^6R^4$) are BPS interactions and their coefficients are $SL(2,\ZZ)$ modular forms as we will briefly review  in this appendix. The amplitudes can be conveniently expressed as
\ie  \label{eq:superamp}
{\cal A}^{(\alpha)}_{n, r} (x_i; \tau^0)=F_{n, r}^{(\alpha)}(\tau^0)\, \delta^{16}( \sum_{i=1}^n Q_i) \, {\cal O}^{(\alpha)}_{n, r}(s_{ij}) \, .
\fe
The prefactor $\delta^{16}(\sum_{i=1}^n Q_i)$ builds in type IIB supersymmetry.  The supercharge $Q_i$ (and $\bar Q_i$)  associated with the particle $i$ can be expressed in the ten-dimensional spinor helicity formalism by
\ie \label{eq:Qi}
Q_{i}^{A} = \lambda_{i, a}^A \eta_i^a \, , \qquad \bar Q_i^A=  \lambda_{i, a}^A \frac{\partial }{\partial \eta_i^a} \, ,
\fe
where the spinor variables are related to massless momenta by 
\ie
 \quad \lambda_{i, a}^A \lambda_{i}^{B, a} = p_{i, \mu} (\Gamma^{\mu} )^{AB} \,,
\label{lamdef}
\fe
with $A, B=1, 2, \cdots, 16$ labels a chiral  ${\it SO(1, 9)}$ spinor index and $a=1,2, \cdots, 8$ labels a vector of the little group, $SO(8)$. The ten-dimensional Minkowski-space momentum of a massless state,  $p_{i, \mu}$, is a null vector and $\eta_i^a$ is a Grassmann variable.  The on-shell massless states can be packaged into an on-shell  ``linearised superfield" 
\bea
\Phi(\eta) =Z +\eta^a  \Lambda_a  + \eta^a \eta^b \phi_{ab}+\dots+\frac{1}{8!} (\eta)^8 \bar Z\, ,
\label{phiexp}
\eea
where $Z\sim - i (\tau-\tau^0)/(2\tau_2^0)$ is the linearised complex dilaton.\footnote{The nonlinear definition is $Z = (\tau-\tau^0)/(\tau- \bar{\tau}^0)$, which has $SL(2,\ZZ)$ modular weights $(-1,1)$ and therefore transforms with $U(1)$ charge equal to $-2$ \cite{Green:2019rhz}.}

The factor  $\delta^{16}( \sum_{i=1}^n Q_i)$ is of order  $p^8$  and ${\cal O}^{(\alpha)}_{n, i}(s_{ij})$ is a symmetric degree-$\alpha$ monomial of Mandelstam variables $s_{ij}$. Here we will consider the cases of $\alpha=0, 2, 3$. Note that, ${\cal O}^{(1)}_{n, r}(s_{ij})=\sum_{i<j}^n s_{ij} =0$. The coefficient $F_{n, r}^{(\alpha)}(\tau^0)$ is a $SL(2, \ZZ)$ modular form with holomorphic and anti-holomorphic modular weights $(n-4, 4-n)$. 

The subscript $r$ labels the independent  monomials ${\cal O}^{(\alpha)}_{n, r}(s_{ij})$  that arise at order $s^\alpha$.  When $\alpha=0,\, 2$ there is only one independent kinematic invariant and so the subscript $r$ has one value and we will drop it for simplicity.  This is also true when $\alpha=3$ and $n\le 5$.   In those cases we will simply drop the subscript $r$.  For $\alpha=3$ and $n\ge 6$ there are two independent kinematic invariants so that in those cases $r=1,2$.
The explicit expressions  when $\alpha=0, 2$ are
\ie
\cO^{(0)}_{n}(s_{ij}) = 1 \, , \qquad\quad  \cO^{(2)}_{n}(s_{ij}) = \frac{1}{2}  \sum_{1\leq i<j \leq n} s^2_{ij} \, .
\fe
With $\alpha=3$, and $n=4, 5$, the unique structure has the form
\ie
 \cO^{(3)}_{n}(s_{ij}) = \frac{1}{2} \sum_{1\leq i<j \leq n} s^3_{ij} \, ,
\fe
whereas the two independent structures for $n \geq 6$ are given by 
\ie 
\label{eq:kinn}
\mathcal{O}^{(3)}_{n,1} (s_{ij})  &= {1\over 32} \left(  (28-3n) \sum_{i<j}   s_{ij}^3 + 3 \sum_{i<j<k}  s_{ijk}^3 \right) \, , 
\fe
and 
\ie 
\label{eq:kinn2}
\mathcal{O}^{(3)}_{n,2} (s_{ij}) &= (n-4) \sum_{i<j} s_{ij}^3 -\sum_{i<j<k}  s_{ijk}^3 \, , 
\fe
with  $s_{ijk}=s_{ij} + s_{ik} + s_{jk}$.

An important feature that distinguishes these expressions is their behaviour in the limit that one of the momenta becomes soft (i.e., is taken to zero).  For $n>6$ they each reduce in the soft limit to the $n-1$ expressions,\footnote{Clearly,  for $\alpha=0, 2$, we also have $\cO^{(\alpha)}_{n} (s_{ij}) \underset{p_n \to 0}   {\to}    \cO^{(\alpha)}_{n-1} (s_{ij})$. }
\ie
\cO^{(3)}_{n,1} (s_{ij}) \underset{p_n \to 0}   {\to}    \cO^{(3)}_{n-1,1} (s_{ij}) \,, \qquad\quad
O^{(3)}_{n,2} (s_{ij})  \underset {p_n \to 0}  { \to}   \cO^{(3)}_{n-1,2} (s_{ij}) \,,
\label{softns}
\fe
but the soft limit limit in the $n=6$ case is 
\ie
\cO^{(3)}_{6,1} (s_{ij}) \underset{p_6 \to 0}   {\to}    \cO^{(3)}_{5} (s_{ij}) \,, \qquad\quad
O^{(3)}_{6,2} (s_{ij})  \underset {p_6 \to 0}  { \to}   0\,.
\label{softnsw}
\fe

The modular forms $F_{n, i}^{(\alpha)}(\tau)$ were determined by using  $SL(2, \ZZ)$-symmetric versions of  soft-dilaton relations (namely we take the particle $Z$ to be soft) that relate $n$-point MUV amplitudes to $(n-1)$-point MUV amplitudes, 
\ie 
\label{eq:soft-amp}
{\cal A}^{(\alpha)}_{n, r}(s_{ij}; \tau^0) \big{|}_{p_n \rightarrow 0} = 2\, \cD_w  {\cal A}^{(\alpha)}_{n-1,r} (s_{ij}; \tau^0)\, ,
\fe
where $w=n-5$ labels the modular weight, which is related to the $U(1)_Y$ charge violation.  As we commented in \eqref{softns} that the kinematic invariants $\mathcal{O}^{(3)}_{n,r} (s_{ij})$ defined in \eqref{eq:kinn} are consistent with \eqref{eq:soft-amp}. The structure of the superfield \eqref{phiexp}, together with \eqref{eq:superamp}, ensures that the factor of $\delta^{16}(\sum_{i} Q_i)$ in the superamplitudes cancels from  \eqref{eq:soft-amp}. This relation 
has a  form that is similar to the recursion relation \eqref{eq:soft} for  MUV correlators in the gauge theory. 

\subsection{Modular form coefficients of  dimension $8$ and $12$ terms }

These are the coefficients of higher derivative interactions  in the low energy expansion of MUV amplitudes that have the same dimension as  $R^4$ and $d^4R^4$ (which are $\half$-BPS and $\quart$-BPS,  respectively) that arise in the expansion of the four-graviton amplitude, which is an example of a $n=4$ MUV amplitude.
In these  $n=4$  cases  with $\alpha=0,\, 2$  the coefficients  $F_{4}^{(\alpha)}(\tau)$ satisfy Laplace eigenvalue equations  
\bea
\left(\Delta_{\tau} - s_{\alpha}(s_{\alpha}-1) \right) F_{4}^{(\alpha)}(\tau) =0\,,
\label{eiseneq}
\eea
where the hyperbolic laplacian is $\Delta_{\tau} = 4 \tau_2^2(\partial_\tau \, \partial_{\bar \tau})$ and  $s_{0} = \threeh$, $s_{2} = \fiveh$. The function $F_{4}^{(\alpha)}(\tau)$  is  a $SL(2,\ZZ)$ modular function  that satisfies the boundary condition  $\lim_{\tau_2\to \infty} F_{4}^{(\alpha)}(\tau)< \tau_2^a$, where $a$ is a real number.  The solution to this equation with appropriate boundary  conditions  is a non-holomorphic Eisenstein series, which has the form
\bea
\label{eisendef}
E(s,\tau) = \sum_{(m,n) \neq\,(0,0)}\frac{\tau_2^s}{|m+n\tau|^{2s}}     =   \sum_{k \in\ZZ} \cF_k(s,\tau_2) \, e^{2\pi i k \tau_1}\,,
\eea
where the zero mode consists of two power behaved terms,
\bea
\cF_0(s,\tau_2)  = 2\zeta(2s)\,  \tau_2^s \  +  \ \frac{2\sqrt \pi \,\Gamma(s-\frac{1}{2}) \zeta(2s-1)}{\Gamma(s)}\, \tau_2^{1-s} \,,
\label{eisenzero}
   \eea
and the non-zero modes  are D-instanton contributions, which are proportional to $K$-Bessel functions,
\bea
\cF_k(s,\tau_2)  =   \frac{4\,\pi^s}{\Gamma(s)}\,  |k|^{s-\half} \, \sigma_{1-2s}(|k|)
\sqrt{\tau_2}\,K(s-\half, 2\pi |k|\tau_2) \,, \  \ \  k \neq 0\,,
\label{nonzeroeisen}
\eea
where the divisor sum is defined by  
\bea \label{eq:divisor-sum}
\sigma_p(k)=\sum_{d>0, {d|k}}   d^{p}  \, , \quad {\rm for} \quad k>0 \, .
\eea

%The lowest order example of such a modular invariant coefficient is $ E(\threeh,\tau)$, the coefficient of the $R^4$ interaction, in the flat-space type IIB theory.  This has a zero mode that contains two power-behaved terms given by \eqref{eisenzero} with $s=3/2$.  Taking into account the power of $\tau_2^{1/2}$  that arises in transforming to the string frame, these powers are $\tau_2^2$  and $\tau_2^0$, which correspond to tree-level and one-loop perturbative superstring contributions.  The $p=2$ term of order $d^4 R^4$ has a coefficient $E(\fiveh,\tau)$ that has  tree-level and two-loop perturbative contributions.

We are generally interested in amplitudes with $n>4$, for these cases $F_{n}^{(\alpha)}(\tau)$ are modular forms with non-trivial weights $(w,- w)$, with $w=n-4$. For these cases the coefficients  $F_{n}^{(\alpha)}(\tau)$  with $\alpha=0,2$ are determined by \eqref{eq:soft-amp}  to be  non-holomorphic Eisenstein modular forms. These can be defined by  
\be
\label{dzsn}
\cD_w\, E_{w} (s,\tau)= {s+w\over 2}\, E_{w+1}(s,\tau)\, ,
\ee
and
\be
\label{dbarzsn}
\bar \cD_{-w}\, E_{w}(s,\tau) = {s-w\over 2}\, E_{w-1}(s,\tau)\, \,,
\ee
using the definition of modular covariant derivatives given in (\ref{covderiv}).  Iterating   \eqref{dzsn} leads to the expression 
\be
E_{w} (s,\tau)= \frac{2^w \Gamma(s) } { \Gamma(s+w)}\, \cD_{w-1} \cdots \cD_0\, E(s,\tau)\, ,
\label{zwdef}
\ee
where $E_{0}(s,\tau) := E (s,\tau)$.  It is straightforward to show that this implies
\be
\label{zswdef}
E_{w}(s,\tau) =\sum_{(m,n)\ne (0,0)} \left({m+n\bar\tau\over m+n\tau}\right)^w \, {\tau_2^s \over |m + n \tau|^{2s}} \, .
\ee

These modular forms satisfy the Laplace equations
\be
\label{laplaceplusn}
\Delta_{(+)}^{(w)} E_{w}(s,\tau) := 4\bar \cD_{-w-1} \cD_w E_{w} (s,\tau)= (s+w)(s-w-1) \, E_{w}(s,\tau)\, ,
\ee
and
\be
\label{laplaceminusn}
\Delta_{(-)}^{(w)}  E_{w} (s,\tau):=4\cD_{w-1} \bar \cD_{-w} E_{w} (s,\tau)= (s-w)(s+w-1) \, E_{w}(s,\tau)\, .
\ee
The two laplacians acting on a weight-$(w,-w)$ modular form satisfy
\be
\label{difflap}
\Delta^{(w)}_{(+)} - \Delta^{(w)}_{(-)} = -2w\,, \qquad\quad   \Delta^{(0)}_{(+)}=  \Delta^{(0)}_{(-)}=\Delta_{\tau}  \, .
\ee 

These modular forms are periodic in $\tau_1$ and have interesting expansions as Fourier series. 
In the $s=3/2$ case that is relevant for the coefficients of the terms of the order $R^4$, this has a Fourier expansion of the form 
\bea
\label{expadef}
E_{w}(\threeh,\tau)  = 
2\zeta(3)\, \tau_2^{{3\over 2}} + \frac{4 \zeta(2)}{1-4w^2} \, \tau_2^{-{1\over
2}} +  
\sum_{k =1}^\infty \left( \cF_{k,4-w} (\threeh,\tau_2)e^{2\pi i k\tau_1} + 
\cF_{k,4+w}  (\threeh,\tau_2)e^{-2\pi i k \tau_1} \right)\,.\nn\\
\eea
The first two terms in (\ref{expadef}) have the interpretation of contributions that should arise in string perturbation theory at   tree-level and one loop, while the charge-$k$ D-instanton and charge-($-k$) anti D-instanton terms are contained in the sum over $k$, where
\bea
\label{znpddef}
\cF_{k,4-w}  (\threeh,\tau_2)
 = (8\pi)^{\half} \, \sigma_{-2}(k)\,   (2\pi k)^{\half} \,  \sum_{j=-w}^\infty    \frac{a_{4- w, j}}   {(2\pi k \tau_2)^j}\,
e^{-2\pi  k\tau_2} \, ,
\eea
and
\begin{equation}
\label{ckrdef}{a_{n,j} = {(-1)^n \over 2^j (j-n+4)!}   \frac {\Gamma(\threeh) }{ 
\Gamma(n-\fiveh)} \frac{\Gamma (j -\half)}{ \Gamma(- j -\half)}}\,.
\end{equation} % 
The instanton sum in (\ref{znpddef}) 
begins with the power $\tau_2^w$ for  D-instantons  (which have phases $e^{2\pi i k \tau_1}$)   
while the series of corrections to the anti D-instanton  (with phases $e^{-2\pi i k \tau_1}$)    starts with the power $\tau_2^{-w}$.  These powers are consistent with the requirement of saturating the fermionic zero modes that are present in the D-instanton background.

\subsection{Modular form coefficients of  dimension $14$ terms }
\label{mod14}

These are the coefficients of higher derivative interactions  with $\alpha=3$  that have the same dimension as   $d^6 R^4$, which is $\eighth$-BPS.  The modular forms  that arise in these cases satisfy inhomogeneous Laplace eigenvalue equations so they will be called ``generalised Eisenstein modular forms".

\subsubsection{Four-point interaction}  
\label{App:4pts}

The $n=4$, $\alpha=3$  coefficient is a modular function (it has modular weight $w=0$) that is the coefficient of the $d^6R^4$ interaction in the low-energy expansion of the four-graviton amplitude. In \cite{Green:2005ba} this function was shown to satisfy the  inhomogeneous Laplace equation\footnote{As noted earlier, this function was denoted by $\cE_{(\threeh,\threeh)}(\tau)$ in  \cite{Green:2005ba} and  was denoted by $\cE_{(0,1)}(\tau)$ in \cite{Green:2014yxa}.  The notation adopted here is in accord with the notation in \cite{Chester:2020vyz}.}
\ie
\label{lapsix}
\left(\Delta_\tau - r(r+1)\right) \cE(r, s_1, s_2,\tau)=-E(s_1,\tau)\, E(s_2,\tau) \, . 
\fe
What is relevant for the $d^6R^4$ interaction is the case with $r=3, s_1=s_2=\threeh$. 
Following \cite{Green:2014yxa}, this equation may be solved in terms of its Fourier modes defined by
	\ie 
	\cE(3,\threeh,\threeh,\tau) = \sum_k  \cF_k(\tau_2) e^{2\pi i k\tau_1}\, .
	\fe
These modes are conveniently written as double sums,
\ie
\cF_k(\tau_2) = \sum_{k_1} \sum_{k_2=k-k_1} f_{k_1,k_2}(\tau_2)\,, \qquad\quad k\neq 0 \,, 
\fe
where $k_1$ and $k_2$ label the mode numbers of the $E(\threeh,\tau)$ factors in the source term in   \eqref{lapsix}.
The zero mode is given by the infinite sum 
\ie
\label{zerom}
\cF_0(\tau_2)  =  \sum_{k_1=0}^\infty f_{k_1,-k_1}( \tau_2) =f_{0,0}( \tau_2) + \sum_{k_1\ne 0} f_{k_1,-k_1}( \tau_2) \, .
\fe
The function $f_{(0,0)}(\tau_2)$ is  a sum of  powers of $\tau_2$ that is given by 
	\ie \label{eq:perty}
	f_{0,0}(\tau_2) = \frac{2 \zeta (3)^2 \tau_2^3 }{3} +\frac{4  \zeta(2) \zeta (3)\tau_2 }{3} +\frac{4 \zeta(4)}{ \tau_2} + {4 \zeta(6) \over 27} \tau_2^{-3}\, ,
	\fe
	where the four terms are interpreted as contributions to string perturbation theory at genus 0 to genus 3.
The terms with $k_1\neq 0$ in  \eqref{zerom} are 	bilinear in K-Bessel functions and have the form 		 
	\ie
	\label{zeromode}
	f_{k_1,-k_1}( \tau_2) = {32 \pi^2 \over 315 |k_1|^3} \sigma_2(|k_1|) \sigma_2(|k_1|) \sum^1_{i,j=0} q_3^{i,j}(\pi |k_1| \tau_2)K_i(2\pi |k_1| \tau_2) K_j(2\pi |k_1| \tau_2)\, ,
	\fe
	where the coefficients $q_3^{i,j} = q_3^{j, i}$ are simple Laurent series of  their arguments.  In the $\tau_2\to \infty$ limit (the weak string coupling limit)  each term behaves as 
 $f_{k_1,-k_1} (\tau_2) \underset {\tau_2\to \infty} {\sim} e^{-4\pi \,k_1 \, \tau_2}$, which is characteristic of an instanton/anti-instanton pair with opposite instanton charges.

	The function $\cF_k(\tau_2)$  with $k \neq 0$ represents an infinite sum of instanton/anti-instanton pairs with total instanton number
	$k$.  We will not reproduce details of the explicit solutions for these modes, which are presented in \cite{Green:2005ba}. 
	
%	\ie
%	q_3^{0,0}(z) &= z \left(-512 z^4+48 z^2-15\right) \, , \cr
%	q_3^{0,1}(z) &= q_3^{1, 0}(z) =  -128 z^4-12 z^2-15 \, , \cr
%	q_3^{1,1}(z) &= 512 z^5+16 z^3+33 z-\frac{15}{z} \, .
%	\fe

\subsubsection{Higher-point interactions}

The modular forms associated with the coefficients of the low-energy expansion of $n$-point MUV amplitudes with $n>4$ that are of dimension 14 (the same order as $d^6R^4$) are again determined using the soft relation  \eqref{eq:soft-amp} together with type IIB supersymmetry.

The  coefficient of the 5-point interaction, which is the modular form  $F_{5}^{(3)}(\tau)$,   $n=5$,  is determined by \eqref{eq:soft-amp} in terms of the four-point interaction,  and is proportional the weight $(1,-1)$ modular form, \footnote{ $\cE^{(3)}_{1,1}(\tau)$ was denoted $\cE^{(3)}_{1}(\tau)$ in \cite{Green:2019rhz} due to the fact that at five points, there is only a single kinematic invariant.}
\ie
 \cE_{1,1}^{(3)}(\tau)  = 2 \cD_0\,  \cE(3,\threeh,\threeh,\tau) \, . 
\fe
For $n = 6$, there are two independent kinematic invariants  $\cO^{(3)}_{6,1} (s_{ij})$ and  $\cO^{(3)}_{6,2} (s_{ij})$ in \eqref{eq:kinn},  which are associated  with two different modular forms of weight $(2,-2)$, which are denoted by $\cE^{(3)}_{2,1}(\tau)$ and $\cE^{(3)}_{2,2}(\tau)$.  The first one  is given by
\ie
\cE^{(3)}_{2,1}(\tau) = 2 \cD_1  \cE_{1,1}^{(3)}(\tau) \, , 
\fe
which obeys inhomogeneous  Laplace equation
\ie
\label{eq:laplace1}
\left(\Delta_{(-)}^{(2)}  -10 \right) \cE^{(3)}_{2,1}(\tau)  = -{15 \over 2} \left( E_0(\threeh, \tau) E_2(\threeh, \tau)  +
{3\over 5} E_1(\threeh, \tau) E_1(\threeh, \tau)  \right) \, .
\fe
The second modular form $\cE^{(3)}_{2,2}(\tau)$ is not related to $ \cE(3,\threeh,\threeh,\tau)$ by \eqref{eq:soft-amp}, and is an independent coefficient satisfying 
\ie
\label{eq:laplace2}
\left(\Delta_{(-)}^{(2)}  -10 \right) \cE^{(3)}_{2,2}(\tau)  = -{5 c_1 \over 2} \left( E_0(\threeh, \tau) E_2(\threeh, \tau)  - E_1(\threeh, \tau) E_1(\threeh, \tau)  \right) \, ,
\fe
where  the overall coefficient $c_1$ was not determined in \cite{Green:2019rhz}. 

Once the six-point coefficients are given, the modular functions associated with higher-point terms $\cO_{n, 1}^{(3)}(s_{ij})$ and $\cO_{n, 2}^{(3)}(s_{ij})$ are then determined from the soft dilaton relation,  \eqref{eq:soft-amp}. They are given by covariant derivatives acting on $ \cE^{(3)}_{2,1}(\tau)$ and $ \cE^{(3)}_{2,2}(\tau)$. In general, the modular forms are given by
\ie
\cO_{n, 1}^{(3)}(s_{ij}): & \qquad \cE^{(3)}_{n-4,1}(\tau) = 2^{n-4} \cD_{n-5} \cdots \cD_0 \cE(3, \threeh, \threeh, \tau)\, , \cr
\cO_{n, 2}^{(3)}(s_{ij}): & \qquad  \cE^{(3)}_{n-4,2}(\tau) = 2^{n-6} \cD_{n-7} \cdots \cD_2  \cE^{(3)}_{2,2}(\tau) \, ,
\fe
where for $\cO_{n, 2}^{(3)}(s_{ij})$ we restrict to the cases with $n \geq 6$.

	\bibliographystyle{ssg}
	\bibliography{U(1)-ref}

\end{document}